\documentclass[hidelinks,onefignum,onetabnum]{siamart251216}

\usepackage[T1]{fontenc}
\usepackage[utf8]{inputenc}
\usepackage{microtype}
\usepackage{textcomp}
\usepackage{cite}

\usepackage{amssymb}
\usepackage{mathtools}
\usepackage{xfrac}

\usepackage{siunitx}
\DeclareSIUnit{\byte}{B}
\DeclareSIUnit{\flop}{FLOP}
\newcommand{\unitPas}{\unit{\pascal\second}}
\newcommand{\unitWmK}{\unit[per-mode=power]{\watt\per\metre\per\kelvin}}

\usepackage{booktabs}
\usepackage{multirow}
\usepackage{array}
\usepackage{longtable}

\usepackage{float}
\usepackage{subcaption}
\usepackage{rotating}
\usepackage{adjustbox}
\usepackage{placeins}
\ifpdf
  \DeclareGraphicsExtensions{.pdf,.png,.jpg,.eps}
\else
  \DeclareGraphicsExtensions{.eps}
\fi

\usepackage{algpseudocode}
\algrenewcommand\algorithmicrequire{\textbf{Input:}}
\algrenewcommand\algorithmicensure{\textbf{Output:}}
\usepackage{listings}

\usepackage{pgfplots}
\pgfplotsset{compat=1.18}
\usepackage{tikz}
\usetikzlibrary{positioning, arrows.meta, shapes.multipart}

\definecolor{codekeyword}{rgb}{0.0, 0.4, 0.65}
\definecolor{codecomment}{rgb}{0.4, 0.45, 0.5}
\definecolor{codestring}{rgb}{0.2, 0.5, 0.25}
\definecolor{codenumber}{rgb}{0.75, 0.75, 0.75}
\definecolor{codebase}{rgb}{0.15, 0.15, 0.15}
\definecolor{lyrA}{rgb}{0.20, 0.50, 0.25}
\definecolor{lyrB}{rgb}{0.02, 0.47, 0.45}
\definecolor{lyrC}{rgb}{0.00, 0.40, 0.65}
\definecolor{lyrD}{rgb}{0.30, 0.28, 0.58}
\headers{Automated Derivation of Lattice Boltzmann Methods}{A.~Kummerl\"ander, F.~Bukreev, and M.~J.~Krause}

\newcommand{\authmark}[1]{\textsuperscript{\ensuremath{#1}}}
\title{Automated Derivation of Lattice Boltzmann Methods for Systems of Conservation Laws}

\author{%
  Adrian Kummerl\"ander\authmark{*,\dagger}%
  \and Fedor Bukreev\authmark{*,\ddagger}%
  \and Mathias J. Krause\authmark{*,\dagger,\ddagger}%
}
\makeatletter
\g@addto@macro\@thanks{%
  \footnotetext[1]{The three authors contributed equally to this work. All
    authors are with the Lattice Boltzmann Research Group (LBRG), Karlsruhe
    Institute of Technology (KIT), 76131 Karlsruhe, Germany
    (\email{kummerlaender@kit.edu}, corresponding author).}%
  \footnotetext[2]{Institute for Applied and Numerical Mathematics (IANM), KIT,
    Englerstra\ss{}e 2.}%
  \footnotetext[3]{Institute of Mechanical Process Engineering and Mechanics
    (MVM), KIT, Stra\ss{}e am Forum 8.}%
}
\makeatother

\ifpdf
\hypersetup{
  pdftitle={Automated Derivation of Lattice Boltzmann Methods for Systems of Conservation Laws (with Supplementary Materials)},
  pdfauthor={A. Kummerlaender, F. Bukreev, and M. J. Krause}
}
\fi

\begin{document}

\maketitle

\begin{abstract}
Multiphysics simulation with \emph{lattice Boltzmann methods} (LBM) requires a scheme hand-derived for each \emph{partial differential equation} (PDE), a labor-intensive, error-prone bottleneck.
We recognize our recently proposed class of LBM schemes~\cite{Bukreev2026} as a discrete-kinetic relaxation approximation of conservation laws and generalize its hand derivation to an automated one for systems of hyperbolic, parabolic, and mixed-type conservation laws.

The derivation splits into three steps:
First, the PDE system is equivalently rearranged into a first-order cascade of conservation laws: every spatial derivative in flux or source becomes an auxiliary variable, recursively for higher derivatives, so all fluxes are algebraic and updates stay local.
Second, the augmented system is approximated by a discrete-velocity kinetic relaxation model with linear, constant-coefficient transport: all nonlinearity resides in a local equilibrium embedding the flux exactly in its first moment, trading the low-Mach truncation for an \emph{a priori} checkable sub-characteristic wave-speed bound.
Third, the relaxation system is discretized by a standard LBM, yielding collide-and-stream algorithms running unchanged on existing solvers.

A symbolic compiler using a domain-specific language encapsulates these steps: unlike existing LBM code generators, which start from the discrete scheme, it automatically derives equilibrium, gradient-tracking cascade, and grid scaling from the declared PDE alone.
We exercise it across twelve PDE systems, including compressible Navier--Stokes--Fourier flow, resistive magnetohydrodynamics, and nonlinear elasticity.
Manufactured-solution verification confirms convergence at or near second order in double precision, retained in single precision by a reference- and equilibrium-shifted formulation.
Targeting OpenLB, the generated GPU kernels reach up to 96\% of the memory-bandwidth roofline.
\end{abstract}



\section{Introduction}

Traditional frameworks for approximating continuous systems of conservation laws on unstructured meshes, most prominently finite volume and finite element methods, evaluate spatial derivatives and interfacial flux tensors by gathering extended, irregularly addressed node neighborhoods.
While robust and geometrically flexible, these indirect, non-local memory access patterns map poorly onto coalesced, throughput-oriented memory systems on contemporary \emph{High-Performance Computing} (HPC) architectures.
As modern HPC nodes commonly lean on thousands of concurrent execution threads within heterogeneous GPU accelerators, the performance of \emph{Partial Differential Equation} (PDE) solvers is increasingly bound by memory bandwidth and synchronization limits rather than raw floating-point throughput.

The classical \emph{Lattice Boltzmann Method} (LBM)~\cite{Krueger2016} circumvents this bottleneck by considering the continuous problem at the mesoscopic level.
By tracking particle distribution functions across a discrete velocity space, the algorithm decomposes into a perfectly parallel collision step and a neighborhood-local streaming step along discrete characteristics.
This locality yields excellent scalability on heterogeneous supercomputers~\cite{Tolke2010, Godenschwager2013, Bauer2021walberla, Kummerlaender2023, Kummerlaender2026c}.
However, standard scalar LBM formulations are structurally rigid.
Constructing a scheme for a new target is fundamentally a moment-matching problem: the discrete velocity moments of the equilibrium distribution must reproduce, order by order, the macroscopic densities and fluxes of the target equations.
This matching is classically realized through high-order Hermite expansions of the equilibrium, custom-tailored to a limited set of target equations~\cite{ShanYuanChen2006, Coreixas2017}, or through entropic equilibria on enlarged velocity sets~\cite{Frapolli2016}.
Because the number of independent moments a lattice can represent isotropically is fixed by its stencil, supporting the moments a given physics demands typically requires case-specific multi-speed lattices.

To overcome these limitations, a class of LBM schemes relying on a point-wise, state-decoupled kinetic formulation was recently advanced in~\cite{Bukreev2026}.
By assigning an independent set of distribution functions to each element of the macroscopic state vector, this approach embeds nonlinear physical fluxes directly in the first-order discrete moments, so a single compact lattice serves diverse conservation laws~\cite{Guillon2024}.
Such point-wise kinetic schemes have so far remained case-specific, requiring manual derivation and custom code for each targeted system~\cite{Boolakee2025, Wissocq2025, Dubois2014, MendozaMunoz2010, Bukreev2026}.
Yet this case-by-case effort is a single mechanical procedure, much as automatic differentiation reduced the hand calculus of derivatives to one graph transformation.
The enabling observation is that the flux-in-first-moment construction is an instance of the discrete-kinetic relaxation approximation of conservation laws~\cite{JinXin1995, Natalini1998, AregbaNatalini2000, Bouchut1999} (Section~\ref{sec:relaxation}): a single map from conservation laws to a lattice scheme, whose validity is fixed a priori by a sub-characteristic condition (Section~\ref{sec:subchar}) rather than rediscovered case by case.

The present work goes beyond these specialized formulations by establishing a generalized and fully automated symbolic compiler for this class of LBM schemes.
Automated code generation for LBM is itself well established: templated GPU solvers such as Sailfish~\cite{Januszewski2014} and TCLB~\cite{LaniewskiWollk2016}, the \texttt{lbmpy}/waLBerla toolchain~\cite{Bauer2021lbmpy, Bauer2021walberla, HennigHolzerRuede2023}, and OpenLB's expression-level kernel optimization and generated adjoints~\cite{Kummerlaender2026c, Ito2026}.
All of these optimize or transform the executable form of a lattice scheme the user must still select: collision model, equilibrium ansatz, stencil, and grid scaling enter as choices.
Our contribution sits one step upstream, deriving the equilibrium, gradient-tracking cascade, and grid scaling automatically from the continuous PDE, and kernel-level generation can consume the schemes we emit.
Driven by a coordinate-free, declarative \emph{Domain-Specific Language} (DSL) built on top of SymPy~\cite{Meurer2017}, our framework automatically deconstructs systems of mixed hyperbolic and parabolic PDEs into point-wise advection-relaxation loops.
When higher-order spatial derivatives or nonlinear coupling tensors are present, the compiler recursively extends the global state vector with auxiliary local gradient-tracking states.
This reduces multi-node differential operators to nested, first-order kinetic interactions, bypassing macroscopic difference stencils to preserve local concurrency.

This paper presents the theoretical foundation, validation, and hardware execution of this multi-physics compiler targeting the platform-transparent framework OpenLB~\cite{Krause2021a,Kummerlaender2026c}: asymptotic convergence across twelve PDE systems and near-roofline GPU-kernel performance.

The remainder of this manuscript is organized as follows: Section~\ref{sec:methodology} analyzes the kinetic formulation, its macroscopic recovery via Chapman--Enskog expansion, and the automated gradient-tracking cascade.
Section~\ref{sec:verification} states the continuous formulation and compiled configuration of the test cases.
Section~\ref{sec:mms} reports the recursive \emph{Method of Manufactured Solutions} (MMS) convergence study, and Section~\ref{sec:performance} evaluates the roofline efficiency of the generated kernels.

\begin{figure}[htbp]
\centering
\resizebox{\textwidth}{!}{%
\begin{tikzpicture}[
  font=\footnotesize,
  stage/.style={rectangle split, rectangle split parts=2, rounded corners=2.5pt,
                rectangle split part fill={#1,#1!7},
                draw=#1!80!black, thick, align=center, text width=28mm, inner sep=4pt},
  vbox/.style={rectangle, rounded corners=2.5pt, fill=#1, draw=#1!80!black,
               thick, inner sep=3pt, minimum height=22mm},
  arr/.style={-{Latex[length=2.2mm]}, thick, black!65},
  brk/.style={black!60, thick},
]
\colorlet{lyrIn}{black!45}
\node[vbox=lyrIn, minimum width=7mm] (b0) {};
\node[rotate=90, font=\bfseries, text=white] at (b0.center) {PDE};
\node[stage=lyrA, anchor=north west] (b1) at ([xshift=10mm]b0.north east) {\textcolor{white}{\textbf{Conservation form}}
  \nodepart{two} {\scriptsize $\partial_t \mathbf{Q} + \nabla \cdot \mathbf{\Phi} = \mathbf{S}$}};
\node[stage=lyrB, anchor=north west, text width=32mm] (b2) at ([xshift=10mm]b1.north east) {\textcolor{white}{\textbf{Augmented system}}
  \nodepart{two} {\scriptsize $\partial_t \mathbf{Q}^{+} + \nabla \cdot \mathbf{\Phi}^{+} = \mathbf{S}^{+}$\\[1pt] $\mathbf{Q}^{+}\!=\!(\mathbf{Q},\mathbf{G})$, $\mathbf{G}\!=\!\nabla\mathbf{Q}$, $\mathbf{\Phi}^{+}$ algebraic}};
\node[stage=lyrC, anchor=north west] (b4) at ([xshift=10mm]b2.north east) {\textcolor{white}{\textbf{Relaxation system}}
  \nodepart{two} {\scriptsize $\partial_t f_{k,i} + \mathbf{\xi}_i \cdot \nabla f_{k,i}$\\[1pt] $\qquad = \tfrac{1}{\varepsilon}(f^{\text{eq}}_{k,i} - f_{k,i})$\\[1pt] $\textstyle\sum_i \mathbf{\xi}_i f^{\text{eq}}_{k,i} = \mathbf{\Phi}_k$}};
\node[stage=lyrD, anchor=north west, text width=40mm] (b5) at ([xshift=10mm]b4.north east) {\textcolor{white}{\textbf{Lattice scheme}}
  \nodepart{two} {\scriptsize $f_{k,i}(\mathbf{x}{+}\mathbf{c}_i,\, t{+}1) =$\\[1pt] $f_{k,i} - \tfrac{1}{\tau}(f_{k,i} - f^{\text{eq}}_{k,i}) + w_i S_k$}};
\node[vbox=lyrIn, minimum width=7mm, anchor=north west] (b6) at ([xshift=10mm]b5.north east) {};
\node[rotate=90, font=\bfseries, text=white] at (b6.center) {Kernels};
\draw[arr] ([yshift=-5mm]b0.north east) -- ([yshift=-5mm]b1.north west);
\draw[arr] ([yshift=-5mm]b1.north east) -- ([yshift=-5mm]b2.north west);
\draw[arr] ([yshift=-5mm]b2.north east) -- ([yshift=-5mm]b4.north west);
\draw[arr] ([yshift=-5mm]b4.north east) -- ([yshift=-5mm]b5.north west);
\draw[arr] ([yshift=-5mm]b5.north east) -- ([yshift=-5mm]b6.north west);
\draw[brk] ([yshift=5.5mm]b0.north west -| b1.west) -- ([yshift=5.5mm]b0.north west -| b2.east)
  node[midway, above=1pt, font=\scriptsize\itshape, text=black!60]{Step 1: reformulation};
\draw[brk] ([yshift=5.5mm]b0.north west -| b1.west) -- ([yshift=4mm]b0.north west -| b1.west);
\draw[brk] ([yshift=5.5mm]b0.north west -| b2.east) -- ([yshift=4mm]b0.north west -| b2.east);
\draw[brk] ([yshift=5.5mm]b0.north west -| b4.west) -- ([yshift=5.5mm]b0.north west -| b4.east)
  node[midway, above=1pt, font=\scriptsize\itshape, text=black!60]{Step 2: relaxation};
\draw[brk] ([yshift=5.5mm]b0.north west -| b4.west) -- ([yshift=4mm]b0.north west -| b4.west);
\draw[brk] ([yshift=5.5mm]b0.north west -| b4.east) -- ([yshift=4mm]b0.north west -| b4.east);
\draw[brk] ([yshift=5.5mm]b0.north west -| b5.west) -- ([yshift=5.5mm]b0.north west -| b5.east)
  node[midway, above=1pt, font=\scriptsize\itshape, text=black!60]{Step 3: discretization};
\draw[brk] ([yshift=5.5mm]b0.north west -| b5.west) -- ([yshift=4mm]b0.north west -| b5.west);
\draw[brk] ([yshift=5.5mm]b0.north west -| b5.east) -- ([yshift=4mm]b0.north west -| b5.east);
\end{tikzpicture}%
}
\caption{A PDE declared in the DSL (Listing~\ref{lst:dsl_euler}) is equivalently rearranged into a first-order cascade of conservation laws by adjoining an auxiliary variable for each spatial derivative (\cref{eq:pde_target}). This exact system is approximated by a discrete-velocity relaxation model whose collision both embeds the physical flux in the equilibrium and relaxes the auxiliary variables onto the gradients they represent (\cref{eq:gradient_advection,eq:feq}), then integrated into the collide-and-stream update (\cref{eq:lbe_update}).}
\label{fig:derivation}
\end{figure}
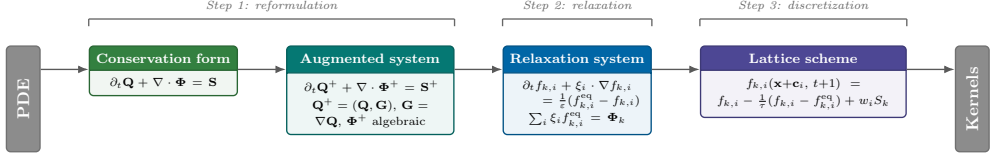

Listing~\ref{lst:dsl_euler} gives the complete DSL specification of the compressible Euler equations: from these few coordinate-free lines the compiler generates a validated OpenLB solver.

\begin{lstlisting}[language=Python, caption={DSL formulation of the compressible Euler equations targeting OpenLB.}, label={lst:dsl_euler}, float=htbp]
from pde2lbm import *

eqs = ConservationLaws(dim=2)

# 1. Define Variables
rho, E = eqs.scalars("rho", "E")
rhoU = eqs.vector("rhoU")

gamma = eqs.register_parameter("gamma")

# 2. Dependent Variables
u = rhoU / rho
p = (gamma - 1) * (E - 0.5 * rho * u.dot(u))

# 3. Equation System
eqs.add([
    Eq(dt(rho)  + div(rhoU)                       , 0),
    Eq(dt(rhoU) + div(outer(rhoU, u) + p * eye(2)), 0),
    Eq(dt(E)    + div((E + p) * u)                , 0)
])

# 4. Track Statistics
eqs.track("rho",  rho)
eqs.track("uSqr", u.dot(u))

# 5. Generate OpenLB solver
eqs.compile(class_name="EulerDynamics")
\end{lstlisting}

\FloatBarrier

\section{Methodology}\label{sec:methodology}

\subsection{Reformulation to Conservation Form}\label{sec:reformulation}

We consider a system of $K$ continuous PDEs specified on a $D$-dimensional spatial domain $\Omega \subset \mathbb{R}^D$.
The target problem class is cast into the general conservation form:
\begin{align} \label{eq:pde_target}
\partial_t \mathbf{Q} + \nabla \cdot \mathbf{\Phi} = \mathbf{S},
\end{align}
where $\mathbf{Q}=[Q_1, Q_2, \dots, Q_K]^T \in \mathbb{R}^K$ denotes the primary macroscopic state vector field.
The physical flux tensor $\mathbf{\Phi} \in \mathbb{R}^{K \times D}$ and the localized source vector $\mathbf{S} \in \mathbb{R}^K$ capture linear or nonlinear couplings in the primary states $\mathbf{Q}$, their spatial gradients $\mathbf{G}=\nabla \mathbf{Q}$, or higher-order derivative tensors.

A broad range of physical systems admits this form, differing only in the functional form of the flux $\mathbf{\Phi}$ and source $\mathbf{S}$.
The derivation that follows takes \cref{eq:pde_target} as its sole entry point and proceeds in three mechanical steps (Figure~\ref{fig:derivation}, Introduction): the conservation law is first equivalently rearranged into a first-order cascade by adjoining an auxiliary variable for each spatial derivative, rendering the fluxes algebraic in the augmented state (Section~\ref{sec:gradient_tracking}), then approximated by a discrete-velocity kinetic relaxation scheme whose collision embeds the flux and relaxes those auxiliaries onto the gradients they represent (Section~\ref{sec:kinetic_scheme}), and finally marched by the standard lattice-Boltzmann integration into a local collide-and-stream update (Section~\ref{sec:discretization}).
Table~\ref{tab:pde_taxonomy} catalogues the twelve systems to which we apply this construction, grouped by the character of their principal part (hyperbolic, parabolic, or mixed), which fixes the refinement scaling (acoustic or diffusive, Section~\ref{sec:consistency}) under which each is verified.
Each system is derived, generated, and validated in detail in Section~\ref{sec:verification}.

\begin{table}[htbp]
\begin{center}
\small
\caption{Taxonomy of the verified PDEs and the scaling under which each is refined.}
\label{tab:pde_taxonomy}
\begin{tabular}{@{}>{\raggedright\arraybackslash}p{2.6cm} >{\raggedright\arraybackslash}p{6cm} >{\raggedright\arraybackslash}p{2.3cm}@{}}
\toprule
\textbf{Category} & \textbf{Example PDEs} & \textbf{Scaling} \\
\midrule
\textbf{I. Hyperbolic} & Inviscid Burgers \newline Compressible Euler \newline Shallow Water \newline Ideal Ultrarelativistic \newline Maxwell (TM mode) \newline Nonlinear Elasticity (finite-strain) & Acoustic \\
\midrule
\textbf{II. Parabolic} & Scalar ADR \newline Allen--Cahn Phase-Field \newline Incompressible Navier--Stokes & Diffusive \\
\midrule
\textbf{III. Mixed} & Compressible Navier--Stokes--Fourier \newline Resistive Magnetohydrodynamics\,$^{\dagger}$ \newline 3D Homogenized Compressible Navier--Stokes--Fourier\,$^{\dagger}$ & Diffusive / Acoustic\,$^{\dagger}$ \\
\bottomrule
\end{tabular}
\end{center}

\smallskip
{\footnotesize $^{\dagger}$\,Physically parabolic (resistive/viscous) but wave-dominated: diffusion is sub-dominant to the fast waves, so these systems are refined acoustically, their lattice diffusion number growing only linearly in $N$ and remaining sub-critical on all reported grids.}
\end{table}

\subsubsection{Recursive Gradient Tracking}\label{sec:gradient_tracking}

When physical fluxes or sources contain spatial derivatives, traditional methods typically rely on multi-node difference stencils.
To preserve locality, the compiler instead scans each declared flux $\mathbf{\Phi}_k$ and source $S_k$ for gradients: the field under a gradient $\nabla V$ is its \emph{parent field} $V$.
In the compressible Navier--Stokes--Fourier system, for instance, the heat flux $-K\nabla T$ has the temperature $T$ as its parent and the viscous stress has the velocity $\mathbf{u}$.
Each such $\nabla V$ is then replaced by an independent auxiliary tracker $\mathbf{G} \approx \nabla V$ appended to the global state vector $\mathbf{Q}$, so no macroscopic difference stencil is ever formed.
A parent that is itself a conserved state is tracked directly, while a derived parent is first expanded over the conserved fields, a reduction the compiler performs automatically (Section~\ref{sec:dsl_pipeline}).
Following this approach~\cite{Bukreev2026}, each gradient $\nabla V$ is resolved via a dedicated linear advection-relaxation equation~\cite{JinLiu1998, BouchutGuarguagliniNatalini2000} for the tracker $\mathbf{G}$:

\begin{align} \label{eq:gradient_advection}
\partial_{t}\mathbf{G} + \nabla \cdot \mathbf{\Phi}_{\mathbf{G}} = \mathbf{S}_{\mathbf{G}},
\end{align}
where the auxiliary flux tensor and relaxation source vector are defined as:
\begin{align}
\mathbf{\Phi}_{\mathbf{G}} = \mathbf{G} \otimes \mathbf{u} - \mathbf{I} \otimes \frac{V}{\tau_{\nabla}}, \quad \mathbf{S}_{\mathbf{G}} = -\frac{\mathbf{G}}{\tau_{\nabla}}.
\end{align}
Here, $\mathbf{u}$ represents a background advection velocity vector, $\mathbf{I}$ is the Kronecker identity tensor, and $\tau_{\nabla}$ is a small gradient-tracker relaxation time.
Pre-scaling the injected field by $1/\tau_{\nabla}$ makes the tracker carry the \emph{physical} gradient directly.

In the limit $\tau_{\nabla} \to 0$ the relaxation penalty dominates, forcing $\partial_{t}\mathbf{G} \to 0$.
Since $\nabla \cdot \big(\mathbf{I} \otimes \tfrac{V}{\tau_{\nabla}}\big) \equiv \tfrac{\nabla V}{\tau_{\nabla}}$, the steady-state balance recovers the spatial derivative locally:
\begin{align} \label{eq:gradient_recovery}
\mathbf{G} = \nabla V + \mathcal{O}(\tau_{\nabla}).
\end{align}
Each tracker is a \emph{separate} linear advection-relaxation system whose equilibrium encodes $\nabla V$, so its $\mathcal{O}(\tau_{\nabla})$ consistency follows from \cref{eq:gradient_recovery} directly, with a relaxation lag whenever the parent field is unsteady.
The tracker time is fixed at $\tau_{\nabla} = g_\tau\,\Delta t$ with a resolution-independent step count $g_\tau$ (supplementary materials), so both offset and lag are $\mathcal{O}(\Delta t)$, their interaction with the two refinement laws quantified in Section~\ref{sec:consistency}.
First derivatives are alternatively available from the non-equilibrium populations, but we adopt prognostic trackers uniformly here and note that route as future work.
For physical systems demanding higher-order derivatives, nested gradient tracking hierarchies are constructed by recursively applying this tracking mechanism to previously generated auxiliary states (e.g., $\mathbf{G}_{2} = \nabla \mathbf{G}_{1} = \nabla(\nabla V)$).
This reduces multi-node macroscopic differential operators to chains of nested, first-order kinetic interactions, maintaining stencil locality throughout the calculation.

\subsection{Relaxation to a Discrete Kinetic Scheme}\label{sec:kinetic_scheme}

\subsubsection{Vector Populations and the Exact-Flux Equilibrium}

To solve \cref{eq:pde_target} while maintaining spatial locality, the continuous system is mapped into a discrete kinetic phase space by allocating an independent set of particle distribution functions $f_{k,i}(\mathbf{x},t)$ for each component $k \in \{1,\dots,K\}$ of the macroscopic state vector.
The formulation is dimension- and stencil-agnostic, accommodating any symmetric lattice whose discrete velocities $\mathbf{c}_i$ and weights $w_i$ satisfy the tensor isotropy conditions:
\begin{align}
\sum_i w_i = 1, \quad \sum_i w_i \mathbf{c}_i = \mathbf{0}, \quad \sum_i w_i \mathbf{c}_i \otimes \mathbf{c}_i = c_s^2 \mathbf{I},
\end{align}
where $\mathbf{I}$ is the identity tensor and $c_s$ is the dimensionless lattice speed of sound.

The primary macroscopic variables are recovered point-wise via the zeroth moment:
\begin{align} \label{eq:moments}
Q_k(\mathbf{x}, t) = \sum_{i=0}^{I-1} f_{k,i}(\mathbf{x}, t).
\end{align}
Classical LBM equilibria truncate a Hermite expansion of the Maxwell--Boltzmann distribution.
We instead embed the analytical physical flux $\mathbf{\Phi}_k$ directly and \emph{exactly} into the first-order discrete moment of the equilibrium distribution function $f_{k,i}^{\text{eq}}$, which is therefore linear in the macroscopic data:
\begin{align} \label{eq:feq}
f_{k,i}^{\text{eq}}(\mathbf{x}, t) = w_i \left( Q_k(\mathbf{x}, t) + \frac{\mathbf{c}_i \cdot \left[ \frac{\Delta t}{\Delta x} \mathbf{\Phi}_k(\mathbf{x}, t) \right]}{c_s^2} \right).
\end{align}
Because no series in $\mathbf{u}/c_s$ is formed, no \emph{low-Mach truncation} ceiling arises from this construction: the minimal lattice serves at high Mach number without enlarging the velocity set (wave speeds are instead limited by the sub-characteristic condition of Section~\ref{sec:subchar}), whereas moment-matched Maxwell--Boltzmann constructions extend the compressible range only by enlarging the velocity set~\cite{Frapolli2016} or abandoning the fixed lattice frame~\cite{Dorschner2018}.
We adopt this per-field, flux-in-first-moment equilibrium in the form of~\cite{Bukreev2026}, where it is exercised in the strongly compressible regime.
Taking the exact discrete moments confirms the linear projection:
\begin{align} \label{eq:eq_moments}
\sum_i f_{k,i}^{\text{eq}} = Q_k, \quad \sum_i f_{k,i}^{\text{eq}} \mathbf{c}_i = \frac{\Delta t}{\Delta x} \mathbf{\Phi}_k.
\end{align}

\subsubsection{Relation to Classical Relaxation Schemes}\label{sec:relaxation}

The construction of \crefrange{eq:feq}{eq:lbe_update} is not merely analogous to a relaxation method but an instance of one.
Introducing the physical lattice velocities $\mathbf{\xi}_i = \mathbf{c}_i \frac{\Delta x}{\Delta t}$, the equilibrium moment constraints~\cref{eq:eq_moments} read
\begin{align} \label{eq:moment_constraints}
\sum_i f_{k,i}^{\text{eq}} = Q_k, \qquad \sum_i \mathbf{\xi}_i\, f_{k,i}^{\text{eq}} = \mathbf{\Phi}_k,
\end{align}
which are precisely the conditions defining the multidimensional discrete-velocity kinetic approximation of conservation laws analysed by Natalini and Aregba-Driollet~\cite{Natalini1998, AregbaNatalini2000}, a class that includes the Jin--Xin relaxation schemes~\cite{JinXin1995}.
In this correspondence the BGK collision is the discretization of a local relaxation toward equilibrium with relaxation time $\varepsilon = (\tau_{\text{LB}} - \tfrac{1}{2})\Delta t$ set by the collide-and-stream relaxation time $\tau_{\text{LB}}$ of \cref{eq:lbe_update}.
The discrete velocities $\mathbf{\xi}_i$ act as the constant characteristic speeds of the underlying linear transport operator.
All nonlinearity resides in the local equilibrium.
The streaming is linear and constant-coefficient.

The canonical scalar instance makes the identification concrete: For a one-dimensional law $\partial_t u + \partial_x f(u) = 0$, the two-velocity ($\xi_{1,2} = \pm a$) scheme has equilibria
\begin{align} \label{eq:maxwellians}
M_{\pm}(u) = \tfrac{1}{2}u \pm \tfrac{1}{2a} f(u),
\end{align}
whose zeroth and first moments return $u$ and $f(u)$ exactly, the one-dimensional form of \cref{eq:feq}.
At full relaxation, $\tau_{\text{LB}} = 1$, the populations are replaced by their equilibria in every step and this two-velocity scheme collapses to the classical Lax--Friedrichs update, the relaxation time thus tuning the stabilizing diffusion of Section~\ref{sec:consistency} continuously below the Lax--Friedrichs level as $\tau_{\text{LB}} \to \tfrac{1}{2}$.
Recognizing the scheme as a discrete kinetic relaxation scheme is what grounds the consistency and stability statements that follow: the asymptotic recovery of \cref{eq:pde_target} (Section~\ref{sec:consistency}) and the admissibility boundary (Section~\ref{sec:subchar}).

The correspondence between LBM and relaxation or finite-difference schemes is well established~\cite{Dubois2008, JunkKlarLuo2005, Graille2014, BellottiGrailleMassot2022, Simonis2020, SimonisKrause2025}, and vectorial schemes placing physical fluxes in the first moments have been constructed by hand for particular settings~\cite{Graille2014, Dubois2014, AregbaNataliniTang2004, MendozaMunoz2010}.
Here the correspondence is used \emph{generatively}: the compiler derives the scheme directly from the supplied conservation law, whose flux Jacobians fix the admissibility condition a priori.

The gradient trackers of Section~\ref{sec:gradient_tracking} are the same relaxation construction with a different target: each tracker equilibrium encodes the gradient of an external field rather than an algebraic flux of a conserved state, so its $\mathcal{O}(\tau_{\nabla})$ consistency follows from the steady-state balance \cref{eq:gradient_recovery} directly.
For scalar diffusion this is classically the Jin--Liu relaxation system~\cite{JinLiu1998}, a diffusive BGK approximation~\cite{BouchutGuarguagliniNatalini2000, LattanzioNatalini2002}, with the difference that here the tracked gradient feeds a flux whose physical diffusivity is prescribed independently of the kinetic relaxation times.

\subsection{Discretization}\label{sec:discretization}

Integrating the discrete-velocity relaxation system along its characteristics over one time step, with the collision treated by the trapezoidal rule and absorbed into the relaxation time, yields the collide-and-stream update, instantiated here with the single-relaxation-time \emph{Bhatnagar--Gross--Krook} (BGK) collision~\cite{Krueger2016}:
\begin{align} \label{eq:lbe_update}
f_{k,i}(\mathbf{x} + \mathbf{c}_i \Delta x, t + \Delta t) = f_{k,i} - \frac{1}{\tau_{\text{LB}}} \left( f_{k,i} - f_{k,i}^{\text{eq}} \right) + \Delta t w_i S_k^{\text{total}}(\mathbf{x}, t + \Delta t/2),
\end{align}
where $\tau_{\text{LB}}$ is the relaxation time parameter, unqualified populations are evaluated at $(\mathbf{x}, t)$, and $S_k^{\text{total}}$ incorporates the target PDE sources, evaluated at the temporal midpoint for second-order source coupling.
This is the standard lattice-Boltzmann time-marching.
The construction is orthogonal to the collision operator: the compiler fixes the equilibria, sources, and per-field relaxation slots, and any collision operator relaxing populations toward these equilibria, such as two-relaxation-time or regularized operators~\cite{Krueger2016, Latt2006}, consumes them unchanged in the target framework.
BGK is the operator we analyse and run.

\subsection{Consistency and Stability}\label{sec:analysis}

\subsubsection{Asymptotic Consistency}\label{sec:consistency}

To evaluate asymptotic consistency, we perform a Chapman--Enskog expansion, perturbing the distribution around its equilibrium state: $f_{k,i} = f_{k,i}^{(0)} + \Delta t f_{k,i}^{(1)} + \mathcal{O}(\Delta t^2)$, where $f_{k,i}^{(0)} \equiv f_{k,i}^{\text{eq}}$ and $\sum_i f_{k,i}^{(1)} = 0$.
A Taylor series expansion of the left-hand side of \cref{eq:lbe_update} yields:
\begin{align} \label{eq:taylor}
\begin{split}
f_{k,i}(\mathbf{x} + \mathbf{c}_i \Delta x, t + \Delta t) ={}& f_{k,i} + \Delta t \left( \partial_t + \frac{\Delta x}{\Delta t} \mathbf{c}_i \cdot \nabla \right) f_{k,i} \\
&\hphantom{f_{k,i}} + \frac{\Delta t^2}{2} \left( \partial_t + \frac{\Delta x}{\Delta t} \mathbf{c}_i \cdot \nabla \right)^2 f_{k,i} + \mathcal{O}(\Delta t^3).
\end{split}
\end{align}
Isolating the first-order non-equilibrium perturbation yields:
\begin{align}
f_{k,i}^{(1)} = -\tau_{\text{LB}} \left( \partial_t f_{k,i}^{(0)} + \frac{\Delta x}{\Delta t} \mathbf{c}_i \cdot \nabla f_{k,i}^{(0)} - w_i S_k^{\text{total}} \right).
\end{align}
Taking the first moment of $f_{k,i}^{(1)}$ with respect to $\mathbf{\xi}_i = \mathbf{c}_i \frac{\Delta x}{\Delta t}$ defines the non-equilibrium deviation flux $\mathbf{\Pi}_k^{(1)} = \sum_i f_{k,i}^{(1)} \mathbf{\xi}_i$.
Using the isotropy conditions (odd moments vanish on the symmetric stencil), the second equilibrium moment is $\sum_i f_{k,i}^{(0)} \mathbf{\xi}_i \otimes \mathbf{\xi}_i = a^2 Q_k \mathbf{I}$ with the squared lattice speed $a^2 = c_s^2 (\Delta x/\Delta t)^2$, giving:
\begin{align} \label{eq:deviation_flux}
\mathbf{\Pi}_k^{(1)} = -\tau_{\text{LB}} \left( \partial_t \mathbf{\Phi}_k + a^2 \nabla Q_k \right).
\end{align}
The standard second-order accounting reduces the prefactor $\tau_{\text{LB}}$ to $\tau_{\text{LB}} - \tfrac{1}{2}$~\cite{Krueger2016}, so the deviation enters the macroscopic balance through the relaxation time $\varepsilon = (\tau_{\text{LB}} - \tfrac{1}{2})\Delta t$ of Section~\ref{sec:relaxation}:
\begin{align} \label{eq:recovered_raw}
\partial_t Q_k + \nabla \cdot \mathbf{\Phi}_k = S_k^{\text{total}} + \nabla \cdot \left( \varepsilon\, \partial_t \mathbf{\Phi}_k + \varepsilon\, a^2\, \nabla Q_k \right) + \mathcal{O}(\varepsilon^2).
\end{align}
The correction is not yet closed: it still contains the time derivative $\partial_t \mathbf{\Phi}_k$.
We take the flux to be an algebraic function of the state, $\mathbf{\Phi} = \mathbf{\Phi}(\mathbf{Q})$.
Gradient-dependent fluxes are reduced to this form by the tracker construction of Section~\ref{sec:gradient_tracking}, and prescribed coefficient fields (the homogenized compressible-flow case of Section~\ref{sec:verification}) add an $\mathcal{O}(\varepsilon)$ closure term that vanishes under refinement.
To leading order the conserved fields then obey $\partial_t \mathbf{Q} = -\nabla\cdot\mathbf{\Phi} + \mathbf{S} + \mathcal{O}(\varepsilon)$, so with the flux Jacobians $\mathbf{A}_\alpha = \partial \mathbf{\Phi}_\alpha / \partial \mathbf{Q}$,
\begin{align} \label{eq:flux_time_closure}
\partial_t \mathbf{\Phi}_\alpha = \mathbf{A}_\alpha\, \partial_t \mathbf{Q} = -\,\mathbf{A}_\alpha \mathbf{A}_\beta\, \partial_\beta \mathbf{Q} + \mathbf{A}_\alpha \mathbf{S} + \mathcal{O}(\varepsilon).
\end{align}
Substituting \cref{eq:flux_time_closure} fuses the temporal and spatial corrections into a single second-order operator, the \emph{relaxation diffusion},
\begin{align} \label{eq:relax_diffusion}
\partial_t Q_k + \nabla \cdot \mathbf{\Phi}_k &= S_k^{\text{total}} + \nabla \cdot \big( \boldsymbol{\nu}^{\text{eff}}\, \nabla \mathbf{Q} \big)_k + \mathcal{O}(\varepsilon^2), \\
\boldsymbol{\nu}^{\text{eff}}(\mathbf{n}) &= \varepsilon\big( a^2 \mathbf{I} - \mathbf{A}_\mathbf{n}^2 \big),
\qquad \mathbf{A}_\mathbf{n} = \textstyle\sum_\alpha n_\alpha \mathbf{A}_\alpha,
\end{align}
anisotropic through the directional flux Jacobian $\mathbf{A}_\mathbf{n}$, the combination that governs a plane wave propagating along the unit direction $\mathbf{n}$ (the source coupling $\mathbf{A}_\alpha \mathbf{S}$ of \cref{eq:flux_time_closure} modifies only the effective source, not the principal symbol, and is absorbed into $S_k^{\text{total}}$).
For a scalar law this collapses to $\nu^{\text{eff}} = \varepsilon\,(a^2 - f'(u)^2)$.
The collision entered this derivation only through the relaxation rate of $\mathbf{\Pi}_k^{(1)}$: any operator preserving \cref{eq:moment_constraints} and relaxing this moment at a single rate recovers \cref{eq:relax_diffusion} with $\varepsilon$ set by that rate, differing only at $\mathcal{O}(\varepsilon^2)$.
Whether $a^2\mathbf{I} - \mathbf{A}_\mathbf{n}^2$ keeps the net operator dissipative is the question taken up in Section~\ref{sec:subchar}.

Under refinement the correction is controlled by $\varepsilon = (\tau_{\text{LB}} - \tfrac{1}{2})\Delta t$ together with the lattice speed $a$.
Throughout this work the relaxation time is parametrized by a resolution-independent rate $\tau_R > 0$ as $\tau_{\text{LB}} = \tfrac{1}{2} + \tau_R\,\Delta t$, so $\varepsilon a^2 = c_s^2\,\tau_R\,\Delta x^2$ contracts under \emph{any} refinement law: the relaxation diffusion is a second-order-consistent numerical stabilizer, never a modelled transport coefficient (at fixed $\tau_{\text{LB}}$ it converges to a finite limit instead, the classical route to modelling physical diffusion~\cite{JunkKlarLuo2005, BouchutGuarguagliniNatalini2000}).
Physical diffusion, where the target carries it, enters through the tracked-gradient fluxes of Section~\ref{sec:gradient_tracking}.

The two refinement laws differ in which lattice number they hold fixed.
\emph{Diffusive} scaling ($\Delta t \propto \Delta x^2$) fixes the lattice diffusion number $D\,\Delta t/\Delta x^2$ of the tracked diffusive flux, the explicit-stability requirement of a parabolic principal part, while the lattice speed grows as $a \propto \Delta x^{-1}$ and the sub-characteristic condition of Section~\ref{sec:subchar} relaxes under refinement.
\emph{Acoustic} scaling ($\Delta t \propto \Delta x$) instead fixes the wave Courant number $\lambda\,\Delta t/\Delta x$ of the physical signal speeds $\lambda$ while $\boldsymbol{\nu}^{\text{eff}} = \mathcal{O}(\Delta x^2) \to 0$, the natural regime for hyperbolic targets, which carry no physical diffusion to sustain.
In either regime the truncation error is $\mathcal{O}(\Delta x^2)$.
The trackers' $\mathcal{O}(\Delta t)$ offset and lag accordingly become $\mathcal{O}(\Delta x^2)$ under diffusive and $\mathcal{O}(\Delta x)$ under acoustic scaling, entering only through the small-diffusivity fluxes and staying below the $\mathcal{O}(\Delta x^2)$ remainder at the reported resolutions.
No single refinement holds both numbers fixed as $\Delta x \to 0$, so each target is refined by whichever timescale binds.

\subsubsection{Sub-Characteristic Stability}\label{sec:subchar}

The relaxation structure that supplies the consistency above also delimits the class of systems the framework can stably represent.
The leading correction $\boldsymbol{\nu}^{\text{eff}}$ of \cref{eq:relax_diffusion} is dissipative only where it is positive semi-definite.
Where it is indefinite, the leading-order correction is anti-diffusive.
For a scalar conservation law with lattice speed $a$ this is the sign of $a^2 - f'(u)^2$, i.e. the classical sub-characteristic condition~\cite{Whitham1974,Liu1987}
\begin{align} \label{eq:subchar_scalar}
a \ge |f'(u)| \qquad \text{for all attained } u,
\end{align}
i.e. the lattice must propagate faster than the fastest physical wave.

For a system the requirement is positive semi-definiteness of $\boldsymbol{\nu}^{\text{eff}}(\mathbf{n}) = \varepsilon(a^2\mathbf{I} - \mathbf{A}_\mathbf{n}^2)$ in every direction.
For systems admitting a symmetrizer (equivalently, a convex entropy)~\cite{FriedrichsLax1971,ChenLevermoreLiu1994}, $\mathbf{A}_\mathbf{n} = \sum_\alpha n_\alpha \mathbf{A}_\alpha(\mathbf{Q})$ is diagonalizable with real spectrum in the symmetrized inner product, and positive semi-definiteness of $a^2\mathbf{I} - \mathbf{A}_\mathbf{n}^2$ is equivalent to the lattice dominating the physical wave speeds in every direction, so that the scalar bound becomes the spectral condition
\begin{align} \label{eq:subchar_system}
\max \left| \operatorname{eig} \sum_\alpha n_\alpha \mathbf{A}_\alpha(\mathbf{Q}) \right| \le a, \qquad \forall\, \mathbf{n}, \ \|\mathbf{n}\| = 1, \text{ and attained } \mathbf{Q}.
\end{align}
Both inequalities are checkable symbolically from the Jacobians of the fluxes the compiler already assembles, yielding an a priori admissibility criterion and, where it fails, a principled rescaling of the lattice speed or time step.

Systems whose principal symbol violates hyperbolicity or whose signal speeds no finite lattice can dominate (elliptic constraints, higher-order dispersive operators) fall outside the admissible class a priori, delimiting the targets the framework represents by construction.

\subsection{Declarative DSL and Automated Compiler}\label{sec:dsl_pipeline}

The reformulation, the equilibrium construction, and the admissibility check each manipulate only the declared fluxes and their Jacobians, and are therefore mechanizable.
We encapsulate them in a symbolic compiler that wraps the derivation in two further stages the mathematics does not require, dedimensionalization and reference shifting, so that a model declared in physical units compiles directly to executable kernels.
Rather than requiring manual, error-prone decomposition of physical fluxes, spatial derivatives, and gradient-tracking variables, the framework exposes a coordinate-free declarative DSL on top of the SymPy \emph{Computer Algebra System} (CAS)~\cite{Meurer2017}.

\begin{figure}[htbp]
\centering
\resizebox{\textwidth}{!}{%
\begin{tikzpicture}[
  font=\footnotesize,
  stage/.style={rectangle split, rectangle split parts=2, rounded corners=2.5pt,
                rectangle split part fill={#1,#1!7},
                draw=#1!80!black, thick, align=center, text width=23mm,
                inner sep=4pt},
  chip/.style={draw=#1!70, fill=#1!3, rounded corners=1.5pt, align=center,
               font=\scriptsize\itshape, inner sep=2pt, text=#1!50!black},
  arr/.style={-{Latex[length=2.2mm]}, thick, black!65},
  node distance=21mm,
]
\node[stage=lyrA] (l1) {\textcolor{white}{\textbf{Layer 1}\\[-1pt] DSL frontend}
  \nodepart{two} {\scriptsize coordinate-free\\ PDE in SI units}};
\node[stage=lyrB, right=of l1] (l2) {\textcolor{white}{\textbf{Layer 2}\\[-1pt] Dedimensionalize}
  \nodepart{two} {\scriptsize Buckingham-$\Pi$\\ $(p,q)$ exponents}};
\node[stage=lyrC, right=of l2] (l3) {\textcolor{white}{\textbf{Layer 3}\\[-1pt] Kinetic mapping}
  \nodepart{two} {\scriptsize point-wise\\ relaxation scheme}};
\node[stage=lyrD, right=of l3] (l4) {\textcolor{white}{\textbf{Layer 4}\\[-1pt] Code generation}
  \nodepart{two} {\scriptsize cross-platform\\ OpenLB operators}};
\draw[arr] (l1) -- node[chip=lyrA]{Physical\\PDE} (l2);
\draw[arr] (l2) -- node[chip=lyrB]{Nondim.\\PDE} (l3);
\draw[arr] (l3) -- node[chip=lyrC]{Kinetic\\Scheme} (l4);
\node[draw=lyrC, dashed, thick, rounded corners=2.5pt, below=8mm of l3,
      align=left, inner sep=5pt, text width=60mm, fill=lyrC!6] (passes)
  {\scriptsize {\bfseries\color{lyrC!70!black} Layer 3 passes} (Algorithm~\ref{alg:dsl_compiler}):\\[1pt]
   1.~\textbf{Split}: terms under a $\nabla\cdot$ to the flux, the remainder to the source\\
   2.~\textbf{Cascade}: nested $\nabla$ resolved into tracker fields, optional strain reduction\\
   3.~\textbf{Assemble}: exact-flux equilibria and relaxation slots per field};
\draw[arr, dashed, lyrC] (l3) -- (passes);
\end{tikzpicture}%
}
\caption{The compiler is a four-layer pipeline. A coordinate-free PDE declared in SI units (Layer~1) is dedimensionalized by a Buckingham-$\Pi$ projection that reads off the scaling (Layer~2), mapped to a local vector-kinetic scheme (Layer~3), and emitted as a platform-transparent OpenLB operator (Layer~4). Each layer consumes the previous intermediate representation and produces the next.}
\label{fig:pipeline}
\end{figure}
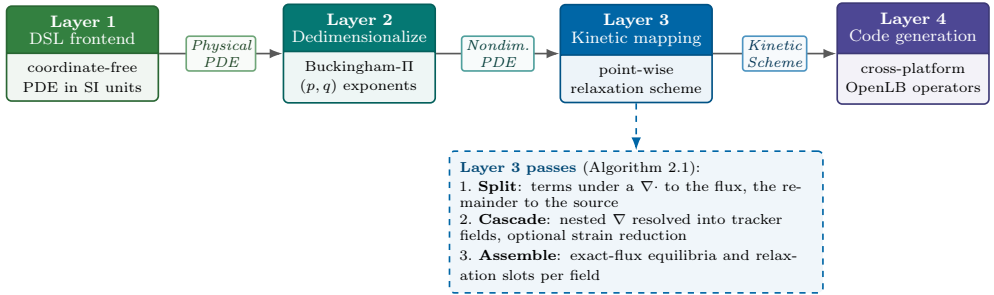

The continuous model of any target system is declared by applying operators directly to symbolic state variables: the temporal derivative ($\mathrm{dt}$), spatial gradient ($\mathrm{grad}$), divergence ($\mathrm{div}$), Laplacian ($\mathrm{laplacian}$), outer tensor product ($\mathrm{outer}$), and identity tensor ($\mathrm{eye}$). For instance, the nonlinear Euler equations of compressible fluid dynamics are declared in a few lines (Listing~\ref{lst:dsl_euler} in the Introduction). The compiler then processes these declared equations through a sequence of symbolic transformation passes prior to C++ code generation, as detailed in Algorithm~\ref{alg:dsl_compiler}.

\begin{algorithm}[tbp]
\caption{The \emph{PDE2LBM} compiler: from conservation law to LBM scheme}
\label{alg:dsl_compiler}
\begin{algorithmic}[1]
\Require conservation laws $\{\partial_t Q + \nabla\cdot\mathbf{\Phi} - S = 0\}$ declared over conserved states, derived quantities, and parameters, in SI units. Lattice $(\mathbf{c}_i, w_i)$ and characteristic references.
\Ensure parameterized collision operator with per-field populations and equilibria
\Statex \textbf{Dedimensionalize} (Layer 2) \hfill \textit{physical PDE $\to$ dimensionless PDE}
\State $(p,q) \gets \Pi\,\mathbf{D}(X)$ for every declared quantity $X$, by Buckingham-$\Pi$ (\cref{eq:pq_projection})
\State rescale states by $dx^{-p} dt^{-q}$, fluxes by $dx^{-(p+1)}dt^{-(q-1)}$, sources by $dx^{-p}dt^{1-q}$
\Statex \textbf{Split} (Layer 3) \hfill \textit{dimensionless PDE $\to (\mathbf{Q}, \mathbf{\Phi}, \mathbf{S})$}
\For{each equation}
  \State $Q_k \gets$ the conserved field under $\partial_t$
  \State $\mathbf{\Phi}_k \gets$ terms under a $\nabla\cdot$
  \State $S_k \gets$ the remainder
\EndFor
\State $\mathbf{\Phi}, \mathbf{S}$ are now algebraic over the states $Q_j$, derived quantities (e.g.\ $\mathbf{u}$, $T = p/\rho R$), parameters, coordinates, and their gradients
\Statex \textbf{Cascade} (Layer 3) \hfill \textit{$(\mathbf{Q}, \mathbf{\Phi}, \mathbf{S}) \to$ gradient-free $(\mathbf{Q}^{+}, \mathbf{\Phi}^{+}, \mathbf{S}^{+})$}
\While{some $\mathbf{\Phi}_k$ or $S_k$ contains a gradient $\nabla V$}\Comment{innermost first}
  \If{$V$ is a conserved state or an existing tracker}
    \State append tracker $\mathbf{G}_V$ to the state vector (\cref{eq:gradient_advection}) and substitute $\nabla V\mapsto\mathbf{G}_V$
  \ElsIf{$V$ is a parameter or coordinate field}
    \State differentiate analytically
  \Else\Comment{$V$ derived}
    \State expand $\nabla V$ over the conserved state, \textbf{or} track $V$ as an auxiliary field
  \EndIf
\EndWhile
\State optional: symmetric velocity-gradient tensor $\to$ strain trackers, $D^2\mapsto D(D{+}1)/2$
\State \textbf{invariant:} every $\mathbf{\Phi}_k$, $S_k$ is algebraic in $\mathbf{Q}^{+}$ (closure premise, Section~\ref{sec:consistency})
\Statex \textbf{Assemble} (Layer 3) \hfill \textit{$(\mathbf{Q}^{+}, \mathbf{\Phi}^{+}, \mathbf{S}^{+}) \to$ kinetic scheme}
\For{each field $k$ of $\mathbf{Q}^{+}$, conserved and tracker}
  \State allocate $f_{k,i}$ with the exact-flux equilibrium \mbox{$f_{k,i}^{\text{eq}} = w_i \big( Q_k + \mathbf{c}_i \cdot \mathbf{\Phi}_k / c_s^2 \big)$} (\cref{eq:feq}, lattice units)
  \State assign the free relaxation slot $\tau_k$ (conserved) or $\tau_{\nabla}$ (tracker)
\EndFor
\Statex \textbf{Emit} (Layer 4) \hfill \textit{kinetic scheme $\to$ OpenLB operator}
\State shift by the reference state $(\mathbf{Q}_0, \mathbf{F}_0)$ (Section~\ref{sec:reference_shift})
\State eliminate common subexpressions and emit the collision operator
\end{algorithmic}
\end{algorithm}

The compiler recursively resolves gradient cascades, which we illustrate by returning to the compressible Navier--Stokes--Fourier system.
Neither its heat-flux gradient $\nabla T$ (with temperature $T = p/(\rho R)$) nor its viscous-stress gradient $\nabla\mathbf{u}$ is the gradient of a conserved variable, so the \emph{Cascade} step of \cref{alg:dsl_compiler} expands them by the chain rule over the conserved fields the scheme can transport ($\mathbf{G}_\rho = \nabla\rho$, $\mathbf{G}_{\rho\mathbf{u}} = \nabla(\rho\mathbf{u})$, $\mathbf{G}_E = \nabla E$). This gives $\nabla\mathbf{u} = \rho^{-1}(\mathbf{G}_{\rho\mathbf{u}} - \mathbf{u}\otimes\mathbf{G}_\rho)$ for the stress and $\nabla T = \rho^{-1}(R^{-1}\mathbf{G}_p - T\,\mathbf{G}_\rho)$ for the heat flux (with the pressure gradient $\mathbf{G}_p=\nabla p$ itself reconstructed from $\mathbf{G}_\rho,\mathbf{G}_{\rho\mathbf{u}},\mathbf{G}_E$), then resolves the resulting conserved-field gradients into trackers.
A single sweep therefore resolves this chain of nested first-order derivatives without ever forming a multi-node stencil. Systems whose fluxes contain still higher derivatives trigger additional passes of the same loop, each registering the gradient of a previously generated auxiliary state, so the cascade terminates at the highest derivative order present in $\mathbf{\Phi}$ and $\mathbf{S}$.

\subsubsection{Automated Physical-Unit Scaling}

In traditional hand-coded LBMs, converting physical variables $Q_{\text{phys}}$ into dimensionless lattice variables $Q_{\text{LBM}}$ requires solving by hand for the exponents $p$ and $q$ in the grid-dependent conversion $Q_{\text{phys}} = Q_{\text{LBM}} \cdot dx^p \, dt^q$. To eliminate this bookkeeping, our symbolic DSL lets the user declare the governing PDE directly in physical (SI) units: each state variable and parameter is registered with its physical dimension (e.g.\ a velocity as $\mathrm{length}/\mathrm{time}$).
The compiler then reduces every quantity to the lattice scales by a Buckingham-$\Pi$ projection (the \emph{Dedimensionalize} step of \cref{alg:dsl_compiler}). Each declared quantity $X$ carries an SI dimension whose exponents over the $b$ base dimensions (length and time first) form a rational vector $\mathbf{D}(X)\in\mathbb{Q}^b$. The lattice carries only two independent scales, the spacings $dx$ and $dt$, so reducing $\mathbf{D}(X)$ to them is a linear map $\pi:\mathbf{D}(X)\mapsto(p,q)$ fixed by three requirements: length and time map to themselves, $\pi(\mathrm{length})=(1,0)$ and $\pi(\mathrm{time})=(0,1)$, and each remaining base dimension is eliminated through a characteristic reference quantity $R_3,\dots,R_b$ already present in the model and held lattice-invariant, $\pi(\mathbf{D}(R_k))=(0,0)$. These conditions make $\{\mathrm{length},\mathrm{time},R_3,\dots,R_b\}$ an alternative basis of dimension space, whose exponent vectors form the columns of the invertible $b\times b$ matrix $B = [\, \mathbf{e}_1,\ \mathbf{e}_2,\ \mathbf{D}(R_3),\ \dots,\ \mathbf{D}(R_b) \,]$, so that the projection becomes the explicit linear map
\begin{equation}\label{eq:pq_projection}
\big(p(X),\,q(X)\big)^{\!\top} = \Pi\,\mathbf{D}(X), \qquad \Pi = [\,I_2 \mid \mathbf{0}\,]\,B^{-1},
\end{equation}
in which $B^{-1}$ re-expresses any SI dimension in the $\{\mathrm{length},\mathrm{time},\text{reference}\}$ basis and $\Pi$ reads off its length and time content. The references need only be dimensionally independent of length, time, and each other, so that $B$ is invertible. The exponents are in general non-zero and field-specific (for example $(p, q) = (1, -1)$ for a velocity), degenerating to $(p, q) = (0, 0)$ for already lattice-dimensionless quantities, including each characteristic reference itself: the density, once mass is eliminated through $\rho_0$, carries no grid-dependent conversion at all.

Fluxes and sources carry their own grid-dependent conversions: a flux is the conserved quantity carried at a velocity, so it inherits the field's own scaling $dx^{-p}\,dt^{-q}$ together with one velocity conversion factor $dt/dx$:
\begin{equation}
\mathbf{F}_{\text{LBM}} = \mathbf{F}_{\text{phys}} \cdot dx^{-p}\,dt^{-q}\,\frac{dt}{dx} = \mathbf{F}_{\text{phys}} \cdot dx^{-(p+1)}\,dt^{-(q-1)},
\end{equation}
while a source, being a rate, picks up one time step instead, $\mathbf{S}_{\text{LBM}} = \mathbf{S}_{\text{phys}} \cdot dx^{-p}\,dt^{1-q}$. Each gradient tracker carries its parent field's exponents shifted by one inverse length, $(p-1,\,q)$, so the gradient cascade is scaled consistently. The compiler injects these transformations automatically: each PDE is written once, in SI form.

Returning to the running example, the same system is declared in SI units, the \emph{dynamic} viscosity ($\unitPas$), thermal conductivity ($\unitWmK$), and specific gas constant each carrying named SI dimensions (the complete DSL specification is given in the supplementary material). Eliminating the mass and temperature dimensions through the density and the gas constant and projecting via \cref{eq:pq_projection}, both the dynamic viscosity and the thermal conductivity collapse to the identical $(p,q)=(2,-1)$ scaling of a $\mathrm{length}^2/\mathrm{time}$ diffusivity, recovering the momentum and thermal diffusion scalings. The result is the compressible Navier--Stokes--Fourier kernel verified at second order in Section~\ref{sec:verification}, generated and run with no hand-derived equilibrium, stencil, or unit conversion.

\subsubsection{Automated Reference and Equilibrium Shifting}\label{sec:reference_shift}

Many target systems carry a small fluctuation $\delta \mathbf{Q}(\mathbf{x}, t)$ on a large background $\mathbf{Q}_0$, $|\delta \mathbf{Q}| \ll |\mathbf{Q}_0|$. Populations formed from absolute state values then push the active variations into the low mantissa bits, and spatial differences suffer catastrophic cancellation, especially in single precision (\texttt{float}).

Each component $Q_k$ is carried by its own substencil whose resting equilibrium $f_{k, i}^{\text{eq}} = w_i Q_k$ distributes $Q_k$ over the lattice weights, so near a background reference state $\mathbf{Q}_0$ even the quiescent populations carry a large static offset $w_i Q_{k, 0} \neq 0$~\cite{Krause2021a}.

The compiler removes this bias at code-generation time through two coordinated shifts. First, a \emph{reference-state shift} moves the macroscopic origin off the large background: the compiler evolves only the fluctuation $\delta\mathbf{Q} = \mathbf{Q} - \mathbf{Q}_0$ about a declared background state $\mathbf{Q}_0$. Second, an \emph{equilibrium-population shift} subtracts the matching background equilibrium from every population: evaluating $\mathbf{Q}_0$ and its flux $\mathbf{F}_{k, 0}$, the compiler stores only the deviation
\begin{equation}
f_{k, i}^* = f_{k, i} - f^{\text{eq}}_{k, i}(\mathbf{Q}_0, \mathbf{F}_0),
\end{equation}
subtracting the background equilibrium of \cref{eq:feq}, so the zeroth-moment state offset $w_i Q_{k, 0}$ and the first-moment flux offset are removed together. The kernels then run entirely on these shifted populations $f_{k, i}^* = \mathcal{O}(\delta Q)$, which vanish at background equilibrium, so the full mantissa is spent on the active fluctuation. This reduces roundoff and improves stability, decisively so under single precision.

\section{Applications}\label{sec:verification}

To validate the generality of the compiler's kinetic mapping, a suite of twelve continuous target systems has been classified in Table~\ref{tab:pde_taxonomy}.
The systems span fluid mechanics, electromagnetics, and solid mechanics.
All are evaluated on a two-dimensional periodic domain $\Omega=[0,1]^2$, except the three-dimensional \emph{Homogenized Compressible Navier--Stokes--Fourier} (HCNSF) case, which uses $\Omega=[0,1]^3$.
Each system is refined under the lattice scaling matched to its dominant transport mechanism (Section~\ref{sec:consistency}): diffusive ($\Delta t \propto \Delta x^2$) for the diffusion-resolving parabolic systems and acoustic ($\Delta t \propto \Delta x$) for the hyperbolic and convection-dominated ones.

\subsection{Inviscid Burgers}
The inviscid Burgers equation is the minimal scalar conservation law with a state-dependent characteristic speed $f'(u) = u$.
It isolates the core mechanism of the framework (the exact nonlinear flux $\tfrac{1}{2}u^2$ embedded directly in the equilibrium first moment) with no gradient trackers, no coupling, and a single field.
The scalar law reads
\begin{align}
\partial_t u + \nabla \cdot \begin{pmatrix*}[l] \tfrac{1}{2}u^2 \\ \tfrac{1}{2}u^2 \end{pmatrix*} = 0,
\end{align}
so that $\nabla \cdot \mathbf{\Phi} = u\left( \partial_{x_0} u + \partial_{x_1} u \right)$, where $u$ is a velocity-like scalar field.
This continuous system maps directly onto \cref{eq:pde_target} with:
\begin{align}
    \mathbf{Q} = \begin{pmatrix*}[l] u \end{pmatrix*}, \quad
    \mathbf{\Phi} = \begin{pmatrix*}[l] \tfrac{1}{2}u^2 \\ \tfrac{1}{2}u^2 \end{pmatrix*}, \quad
    \mathbf{S} = \begin{pmatrix*}[l] 0 \end{pmatrix*}.
\end{align}
Because $\max|f'(u)| = \max|u|$, the manufactured amplitude directly sets the physical wave speeds the lattice must dominate, making this the cleanest setting in which to probe the sub-characteristic bound empirically.

\begin{figure}[htbp]
\centering
\includegraphics[width=0.9\linewidth]{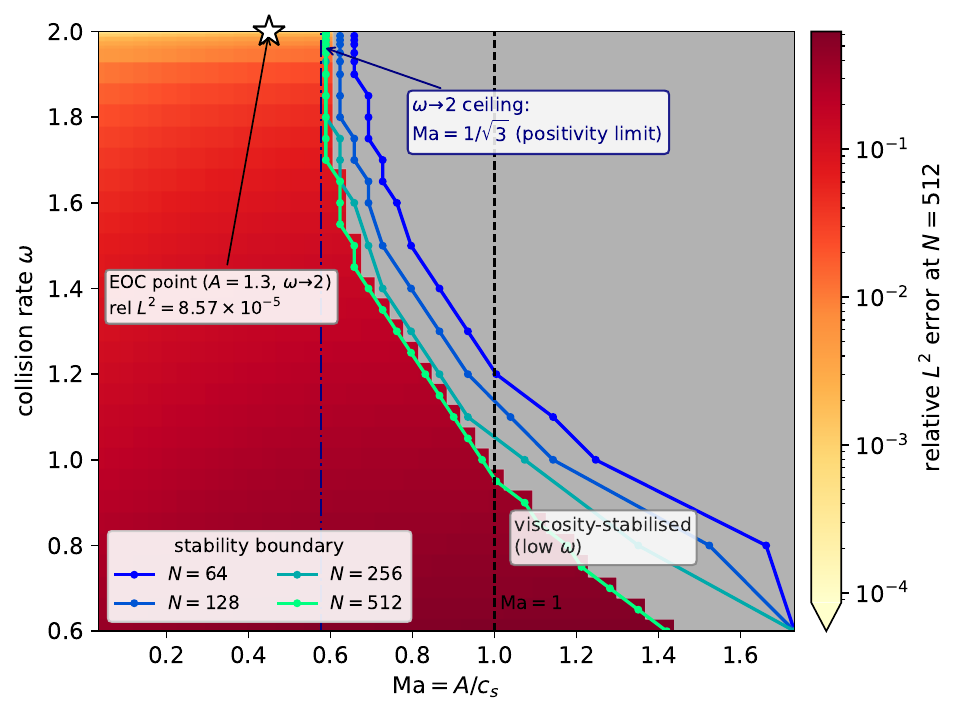}
\caption{Brute-force stability map for the inviscid Burgers scheme, sweeping Mach number $\mathrm{Ma} = A/a$ against collision rate $\omega$. Background shading is the relative $L^2$ error at $N = 512$, and solid curves trace the empirical stability boundary at $N = 64, 128, 256, 512$.
The boundary converges to the equilibrium-monotonicity bound $\mathrm{Ma} = 1/\sqrt{3}$ as $\omega \to 2$ (dash-dotted line), tightening the sub-characteristic condition of Section~\ref{sec:subchar}.
The star marks the verification working point of Table~\ref{tab:mms_results}.}
\label{fig:stability_burgers}
\end{figure}

\paragraph{Empirical Sub-Characteristic Boundary}
Figure~\ref{fig:stability_burgers} reports a brute-force stability sweep of the generated scheme over the manufactured Mach number $\mathrm{Ma} = A/a$ and the collision rate $\omega$.
For each pair the simulation is run to $t = 1.0$ and classified as stable or divergent, while the background shading records the relative $L^2$ error at $N = 512$.
Mapped stability boundaries in the velocity--relaxation plane are an established tool of lattice Boltzmann stability analysis~\cite{SterlingChen1996, WorthingMozerSeeley1997, Simonis2021}.
The empirically resolved boundary tightens under grid refinement toward $\mathrm{Ma} = 1/\sqrt{3}$ at $\omega \to 2$: the bound $\max|f'(u)| \le c_s\,a$ at which the exact-flux equilibria \cref{eq:feq} cease to be monotone in $u$, the monotone-equilibria condition of the relaxation framework~\cite{AregbaNatalini2000}.
It sharpens the sub-characteristic condition of \cref{eq:subchar_scalar} by the factor $c_s$ in the over-relaxation limit, while under heavier relaxation damping the boundary rises toward the per-axis wave-speed ceiling.
The verification working point ($A = 1.3$, $\mathrm{Ma} \approx 0.45$, starred) lies safely within the stable interior, consistent with the a priori criterion of Section~\ref{sec:subchar}.

\subsection{Compressible Euler}
The Euler equations govern the conservation of mass, momentum, and energy in inviscid compressible fluids, where thermal and mechanical energy modes are coupled dynamically through an equation of state.
The system reads
\begin{align}
\begin{aligned}
\partial_t \rho + \nabla \cdot (\rho \mathbf{u}) &= 0, \\
\partial_t (\rho \mathbf{u}) + \nabla \cdot (\rho \mathbf{u} \otimes \mathbf{u} + p\mathbf{I}) &= \mathbf{0}, \\
\partial_t E + \nabla \cdot ((E+p)\mathbf{u}) &= 0,
\end{aligned}
\end{align}
where $\rho$ represents the fluid mass density, $\mathbf{u}$ is the physical velocity vector, $E$ is the total energy density, and $p$ is the static thermal pressure.
LBM formulations for the compressible Euler equations have been developed using multi-speed lattices to decouple fluid velocity from temperature~\cite{Kataoka2004}.
This system maps onto \cref{eq:pde_target} with:
\begin{align}
    \mathbf{Q} = \begin{pmatrix*}[l]
    \rho \\
    \rho \mathbf{u} \\
    E
    \end{pmatrix*}, \quad
    \mathbf{\Phi} = \begin{pmatrix*}[l]
    \rho \mathbf{u} \\
    \rho \mathbf{u} \otimes \mathbf{u} + p\mathbf{I} \\
    (E+p)\mathbf{u}
    \end{pmatrix*}, \quad
    \mathbf{S} = \begin{pmatrix*}[l]
    0 \\
    \mathbf{0} \\
    0
    \end{pmatrix*}.
\end{align}
The pressure is dynamically closed via the ideal gas equation of state $p = (\gamma - 1)(E - \sfrac{1}{2}\rho |\mathbf{u}|^2)$, where $\gamma = 1.4$ is the adiabatic index.

\subsection{Shallow Water}
The \emph{Shallow Water} equations (SWE) describe flow propagation in open channels and coastal zones, modeling depth-averaged fluid thickness and momentum under hydrostatic pressure.
LBM formulations for the shallow water equations identify the fluid height with density~\cite{Salmon1999} and the system is also a code-generation example in \texttt{lbmpy}~\cite{HennigHolzerRuede2023}.
The depth-averaged nonlinear free-surface model simulates channel hydraulics over a static spatial bottom topography field $z_b(\mathbf{x})$:
\begin{align}
\begin{aligned}
\partial_t h + \nabla \cdot (h \mathbf{u}) &= 0, \\
\partial_t (h \mathbf{u}) + \nabla \cdot \left( h \mathbf{u} \otimes \mathbf{u} + \frac{1}{2}g h^2 \mathbf{I} \right) &= -g h \nabla z_b,
\end{aligned}
\end{align}
where $h$ represents the depth-averaged fluid thickness (water column height), $\mathbf{u}$ is the depth-averaged horizontal velocity vector, $g$ is the gravitational acceleration constant, and $z_b$ is the stationary spatial bottom topography profile.
It maps onto \cref{eq:pde_target} with:
\begin{align}
    \mathbf{Q} = \begin{pmatrix*}[l]
    h \\
    h \mathbf{u}
    \end{pmatrix*}, \quad
    \mathbf{\Phi} = \begin{pmatrix*}[l]
    h \mathbf{u} \\
    h \mathbf{u} \otimes \mathbf{u} + \frac{1}{2}g h^2 \mathbf{I}
    \end{pmatrix*}, \quad
    \mathbf{S} = \begin{pmatrix*}[l]
    0 \\
    -g h \nabla z_b
    \end{pmatrix*}.
\end{align}

\subsection{Ideal Ultrarelativistic Fluid}
The ultrarelativistic fluid is the ideal gas of massless particles ($c = 1$, equation of state $p = \sfrac{e}{3}$), whose dynamics is closed by energy--momentum conservation alone: the barotropic relativistic Euler system with sound speed $\sfrac{1}{\sqrt{3}}$~\cite{SmollerTemple1993}.
LBM formulations have been developed for relativistic~\cite{Mendoza2010} and ultrarelativistic~\cite{Mohseni2013} hydrodynamics.
Tracking the energy density $E$ and relativistic momentum density vector $\mathbf{S}$,
\begin{align}
\begin{aligned}
\partial_t E + \nabla \cdot \mathbf{S} &= 0, \\
\partial_t \mathbf{S} + \nabla \cdot \left( \mathbf{S} \otimes \mathbf{v} + p \mathbf{I} \right) &= \mathbf{0},
\end{aligned}
\end{align}
the system maps onto \cref{eq:pde_target} with:
\begin{align}
    \mathbf{Q} = \begin{pmatrix*}[l]
    E \\
    \mathbf{S}
    \end{pmatrix*}, \quad
    \mathbf{\Phi} = \begin{pmatrix*}[l]
    \mathbf{S} \\
    \mathbf{S} \otimes \mathbf{v} + p \mathbf{I}
    \end{pmatrix*}, \quad
    \mathbf{S}_{\text{source}} = \begin{pmatrix*}[l]
    0 \\
    \mathbf{0}
    \end{pmatrix*}.
\end{align}
For the massless gas the primitive recovery is algebraic:
\begin{align}
p = \frac{\sqrt{4E^2 - 3|\mathbf{S}|^2} - E}{3}, \quad \mathbf{v} = \frac{\mathbf{S}}{E + p},
\end{align}
with physicality requiring $E > |\mathbf{S}|$, the null limit $|\mathbf{v}| \to 1$ attained at equality.

\subsection{Maxwell Electromagnetics}
Maxwell's equations are a first-order linear hyperbolic system of conservation laws whose defining operator is the curl, carried here as a pure flux divergence in the equilibrium first moment, realized previously by hand for the full three-dimensional system~\cite{MendozaMunoz2010}.
We treat the two-dimensional \emph{Transverse-Magnetic} (TM) mode in vacuum, with out-of-plane electric field $E_z$ and in-plane magnetic field $(H_x, H_y)$ propagating at light speed $c$~\cite{Hanasoge2011}:
\begin{align}
\begin{aligned}
\partial_t E_z + \nabla \cdot \big( c\,(-H_y,\, H_x) \big) &= 0, \\
\partial_t H_x + \nabla \cdot \big( c\,(0,\, E_z) \big) &= 0, \\
\partial_t H_y + \nabla \cdot \big( c\,(-E_z,\, 0) \big) &= 0,
\end{aligned}
\end{align}
which reproduce Ampère's and Faraday's laws and, on eliminating $\mathbf{H}$, the scalar wave equation $\partial_{tt} E_z = c^2 \nabla^2 E_z$.
The system maps onto \cref{eq:pde_target} with
\begin{align}
    \mathbf{Q} = \begin{pmatrix*}[l] E_z \\ H_x \\ H_y \end{pmatrix*}, \quad
    \mathbf{\Phi} = \begin{pmatrix*}[l] c\,(-H_y,\, H_x) \\ c\,(0,\, E_z) \\ c\,(-E_z,\, 0) \end{pmatrix*}, \quad
    \mathbf{S} = \begin{pmatrix*}[l] 0 \\ 0 \\ 0 \end{pmatrix*}.
\end{align}

\paragraph{Admissibility at Constant Wave Speed}
For Maxwell the admissibility condition of Section~\ref{sec:subchar} applies identically at every state: the flux Jacobians are \emph{constant}, with eigenvalues $\{+c, -c, 0\}$ independent of the field magnitude.
With the naive choice $c = 1$ the wave speed equals the full lattice speed, so the effective diffusion of \cref{eq:relax_diffusion} carries $c_s^2 - c^2 = \tfrac{1}{3} - 1 < 0$: the recovered operator is \emph{anti-diffusive} and the scheme is unstable at every $\tau$.
The remedy is to declare $c$ as a physical quantity of dimension length/time: the dedimensionalizer then scales the flux by the acoustic factor $f = \Delta t / \Delta x$, as for every hyperbolic case.
The lattice wave speed is then $cf$, and the sub-characteristic bound of \cref{eq:subchar_system} becomes the ordinary lattice-Mach constraint
\begin{align}
c\,f \;\le\; c_s = \tfrac{1}{\sqrt{3}}, \qquad \text{i.e.}\qquad f \;\le\; \frac{c_s}{c}.
\end{align}
At the working point $c = 1$, $f = 0.2$ the bound holds with nearly threefold margin, and the scheme runs stably and at second order at the BGK edge ($\tau_{\text{LB}} \approx 0.5$).

\subsection{Nonlinear Finite-Strain Elasticity}
For large-deformation solid mechanics we take the deformation gradient $\mathbf{F} = \partial \mathbf{x}/\partial \mathbf{X}$ and the velocity $\mathbf{v}$ as the conserved state, evolved through the kinematic compatibility relation and the momentum balance in total-Lagrangian form (gradients with respect to the reference coordinate $\mathbf{X}$):
\begin{align}
\begin{aligned}
\partial_t \mathbf{F} - \nabla \mathbf{v} &= \mathbf{0}, \\
\partial_t (\rho_0 \mathbf{v}) - \nabla \cdot \mathbf{P} &= \mathbf{0},
\end{aligned}
\end{align}
where $\rho_0$ is the constant reference-configuration density and $\mathbf{P}$ is the first Piola--Kirchhoff stress of a compressible neo-Hookean solid,
\begin{align}
\mathbf{P} = \mu_s \left( \mathbf{F} - \mathbf{F}^{-T} \right) + \lambda_s \ln(J)\, \mathbf{F}^{-T}, \qquad J = \det \mathbf{F},
\end{align}
with shear modulus $\mu_s$ and first Lam\'e parameter $\lambda_s$.
This system maps onto \cref{eq:pde_target} with
\begin{align}
    \mathbf{Q} = \begin{pmatrix*}[l]
    \mathbf{F} \\
    \rho_0 \mathbf{v}
    \end{pmatrix*}, \quad
    \mathbf{\Phi} = \begin{pmatrix*}[l]
    - \mathbf{v} \otimes \mathbf{I} \\
    - \mathbf{P}
    \end{pmatrix*}, \quad
    \mathbf{S} = \begin{pmatrix*}[l]
    \mathbf{0} \\
    \mathbf{0}
    \end{pmatrix*}.
\end{align}
Prior LBM treatments of elasticity map Cauchy-stress components to single-distribution moments~\cite{Marconi2003} or build vectorial formulations for linear elastic media~\cite{Boolakee2025}.
The same deformation-gradient formulation was arrived at concurrently~\cite{FengChu2026}.

\subsection{Advection-Diffusion-Reaction}
The \emph{Advection-Diffusion-Reaction} (ADR) equation models scalar chemical transport under a prescribed constant velocity field coupled with linear isotropic diffusion and a nonlinear logistic Fisher--KPP reaction source term.
The governing equation reads
\begin{align}
\partial_t c + \nabla \cdot \left( c \mathbf{u}_{\text{adv}} - D \nabla c \right) = R\,c(1-c),
\end{align}
where $c$ is the scalar concentration field, $\mathbf{u}_{\text{adv}}$ is the constant advective velocity vector, $D$ is the isotropic diffusion coefficient, and $R$ is the reaction rate constant.
The LBM has long been applied to such transport, from an early diffusion model~\cite{WolfGladrow1995} to the single-relaxation-time advection-diffusion scheme~\cite{Chopard2009}.
The system maps onto \cref{eq:pde_target} with:
\begin{align}
    \mathbf{Q} = \begin{pmatrix*}[l]
    c \\
\mathbf{G}_c
    \end{pmatrix*}, \quad
    \mathbf{\Phi} = \begin{pmatrix*}[l]
    c \mathbf{u}_{\text{adv}} - D \mathbf{G}_c \\
\mathbf{G}_c \otimes \mathbf{u}_{\text{adv}} - \mathbf{I}\,c/\tau_{\nabla}
    \end{pmatrix*}, \quad
    \mathbf{S} = \begin{pmatrix*}[l]
    R\,c(1-c) \\
- \mathbf{G}_c/\tau_{\nabla}
    \end{pmatrix*}.
\end{align}

\subsection{Allen--Cahn Phase-Field}
The Allen--Cahn equation models non-conservative interface dynamics, coupling diffusion with a bistable cubic reaction.
The governing equation reads
\begin{align}
\partial_t \phi + \nabla \cdot \left( -\gamma \nabla \phi \right) = r\left( \phi - \phi^3 \right),
\end{align}
where $\phi$ is the dimensionless order parameter, $\gamma$ is the interface diffusivity (with units of $\text{length}^2/\text{time}$), and $r$ is the reaction rate.
LBM phase-field models have been developed for interface capturing~\cite{Geier2015}.
The system maps onto \cref{eq:pde_target} with:
\begin{align}
    \mathbf{Q} = \begin{pmatrix*}[l]
    \phi \\
\mathbf{G}_\phi
    \end{pmatrix*}, \quad
    \mathbf{\Phi} = \begin{pmatrix*}[l]
    -\gamma \mathbf{G}_\phi \\
- \mathbf{I}\,\phi/\tau_{\nabla}
    \end{pmatrix*}, \quad
    \mathbf{S} = \begin{pmatrix*}[l]
    r\left( \phi - \phi^3 \right) \\
- \mathbf{G}_\phi/\tau_{\nabla}
    \end{pmatrix*}.
\end{align}

\subsection{Incompressible Navier--Stokes}
The Incompressible \emph{Navier--Stokes Equations} (NSE) model the conservation of momentum and mass in viscous, constant-density fluids.
The equations read
\begin{align}
\begin{aligned}
\nabla \cdot \mathbf{u} &= 0, \\
\partial_t \mathbf{u} + \mathbf{u} \cdot \nabla \mathbf{u} + \frac{1}{\rho_0} \nabla p - \nu \nabla^2 \mathbf{u} &= \mathbf{0},
\end{aligned}
\end{align}
where $\rho_0$ is the constant mass density, $\mathbf{u}$ is the fluid velocity vector, $p$ is the physical pressure, and $\nu = \mu / \rho_0$ is the kinematic viscosity.
Because the divergence-free condition is a kinematic constraint requiring a non-local elliptic Poisson solver, it cannot be directly represented in our explicit local framework (\cref{eq:pde_target}).
Instead, isothermal viscous transport is solved using a weakly compressible formulation~\cite{Krueger2016}, where density fluctuations represent a pressure proxy $p = \rho c_{s,\text{phys}}^2$ with prescribed artificial sound speed $c_{s,\text{phys}}$ in the low-Mach limit, restoring a local conservation form:
\begin{align}
\begin{aligned}
\partial_t \rho + \nabla \cdot (\rho \mathbf{u}) &= 0, \\
\partial_t (\rho \mathbf{u}) + \nabla \cdot \left( \rho \mathbf{u} \otimes \mathbf{u} + \rho c_{s,\text{phys}}^2 \mathbf{I} - \boldsymbol{\tau} \right) &= \mathbf{0},
\end{aligned}
\end{align}
where $\boldsymbol{\tau} = \mu (\nabla \mathbf{u} + (\nabla \mathbf{u})^T)$ is the viscous shear stress tensor with dynamic viscosity coefficient $\mu$.
Closing $\nabla\mathbf{u}\approx\nabla(\rho\mathbf{u})/\rho$ to $\mathcal{O}(\mathrm{Ma}^2)$, this system maps onto \cref{eq:pde_target} with:
\begin{align}
    \mathbf{Q} = \begin{pmatrix*}[l]
    \rho \\
    \rho \mathbf{u} \\
\mathbf{G}_{\rho\mathbf{u}}
    \end{pmatrix*}, \quad
    \mathbf{\Phi} = \begin{pmatrix*}[l]
    \rho \mathbf{u} \\
    \rho \mathbf{u} \otimes \mathbf{u} + \rho c_{s,\text{phys}}^2 \mathbf{I} - \frac{\mu}{\rho}(\mathbf{G}_{\rho\mathbf{u}} + \mathbf{G}_{\rho\mathbf{u}}^T) \\
\mathbf{G}_{\rho\mathbf{u}} \otimes \mathbf{u} - \mathbf{I} \otimes \rho\mathbf{u}/\tau_{\nabla}
    \end{pmatrix*}, \quad
    \mathbf{S} = \begin{pmatrix*}[l]
    0 \\
    \mathbf{0} \\
- \mathbf{G}_{\rho\mathbf{u}}/\tau_{\nabla}
    \end{pmatrix*}.
\end{align}

\subsection{Compressible Navier--Stokes--Fourier}
The \emph{Navier--Stokes--Fourier} (NSF) equations, the running example of Section~\ref{sec:dsl_pipeline}, govern viscous, heat-conducting gas flows: the full compressible system closing the momentum balance with the Newtonian viscous stress and the energy balance with Fourier heat conduction $\mathbf{q}=-K\nabla T$.
This is the first-order Chapman--Enskog closure, as opposed to extended-moment methods such as Grad-13~\cite{Struchtrup2005}.
Prior LBM treatments rely on tailored equilibria~\cite{Alexander1992} or multi-speed lattices~\cite{Frapolli2016}.
The governing equations read:
\begin{align}
\begin{aligned}
\partial_t \rho + \nabla \cdot (\rho \mathbf{u}) &= 0, \\
\partial_t (\rho \mathbf{u}) + \nabla \cdot \left( \rho \mathbf{u} \otimes \mathbf{u} + p \mathbf{I} - \boldsymbol{\tau} \right) &= \mathbf{0}, \\
\partial_t E + \nabla \cdot \left( (E + p)\mathbf{u} - \boldsymbol{\tau} \cdot \mathbf{u} - K \nabla T \right) &= 0,
\end{aligned}
\end{align}
where $\rho$ represents the fluid density, $\mathbf{u}$ is the physical fluid velocity vector, $E$ is the total energy density, $p$ is the ideal-gas pressure of the Euler system, $T = \sfrac{p}{\rho R}$ is the temperature field with specific gas constant $R$, $K$ is the isotropic thermal conductivity, and $\boldsymbol{\tau} = \mu \left( \nabla \mathbf{u} + (\nabla \mathbf{u})^T - \frac{2}{3}(\nabla \cdot \mathbf{u})\mathbf{I} \right)$ is the viscous stress tensor with dynamic viscosity coefficient $\mu$.
The system maps onto \cref{eq:pde_target} with:
\begin{align}
    \mathbf{Q} = \begin{pmatrix*}[l]
    \rho \\
    \rho \mathbf{u} \\
    E \\
\mathbf{G}_\rho \\
\mathbf{G}_{\rho\mathbf{u}} \\
\mathbf{G}_E
    \end{pmatrix*}, \quad
    \mathbf{\Phi} = \begin{pmatrix*}[l]
    \rho \mathbf{u} \\
    \rho \mathbf{u} \otimes \mathbf{u} + p \mathbf{I} - \boldsymbol{\tau} \\
    (E + p)\mathbf{u} - \boldsymbol{\tau} \cdot \mathbf{u} - K \mathbf{G}_{T}\\
\mathbf{G}_\rho \otimes \mathbf{u} - \mathbf{I}\,\rho/\tau_{\nabla} \\
\mathbf{G}_{\rho\mathbf{u}} \otimes \mathbf{u} - \mathbf{I} \otimes \rho\mathbf{u}/\tau_{\nabla} \\
\mathbf{G}_E \otimes \mathbf{u} - \mathbf{I}\,E/\tau_{\nabla}
    \end{pmatrix*}, \quad
    \mathbf{S} = \begin{pmatrix*}[l]
    0 \\
    \mathbf{0} \\
    0 \\
- \mathbf{G}_\rho/\tau_{\nabla} \\
- \mathbf{G}_{\rho\mathbf{u}}/\tau_{\nabla} \\
- \mathbf{G}_E/\tau_{\nabla}
    \end{pmatrix*},
\end{align}
where the trackers and the chain-rule reconstruction of $\nabla\mathbf{u}$ and $\mathbf{G}_T = \nabla T$ are those of the running example (Section~\ref{sec:dsl_pipeline}).

\subsection{Resistive Compressible Magnetohydrodynamics}
The Resistive \emph{Magnetohydrodynamics} (MHD) equations couple viscous, thermally conducting, electrically active fluids to magnetic fields, capturing shear stress, heat flux, and magnetic dissipation.
Existing LBMs represent the magnetic field by vector-valued distribution functions, with the divergence error controlled rather than enforced~\cite{Dellar2002, Dellar2009}.
The system, augmented by the Godunov--Powell divergence-cleaning that restores the symmetrizability assumed in Section~\ref{sec:subchar} and controls the growth of $\nabla\cdot\mathbf{B}$~\cite{Powell1999}, reads:
\begin{align}
\begin{aligned}
\partial_t \rho + \nabla \cdot (\rho \mathbf{u}) &= 0, \\
\partial_t (\rho \mathbf{u}) + \nabla \cdot \left( \rho \mathbf{u} \otimes \mathbf{u} + p_{\text{tot}}\mathbf{I} - \mathbf{B} \otimes \mathbf{B} - \boldsymbol{\tau} \right) &= -(\nabla \cdot \mathbf{B})\,\mathbf{B}, \\
\partial_t E + \nabla \cdot \left( (E + p_{\text{tot}})\mathbf{u} - \mathbf{B}(\mathbf{u} \cdot \mathbf{B}) - \boldsymbol{\tau} \cdot \mathbf{u} + \mathbf{q} + \eta (\mathbf{j} \times \mathbf{B}) \right) &= -(\nabla \cdot \mathbf{B})(\mathbf{u} \cdot \mathbf{B}), \\
\partial_t \mathbf{B} + \nabla \cdot \left( \mathbf{B} \otimes \mathbf{u} - \mathbf{u} \otimes \mathbf{B} - \eta \nabla \mathbf{B} \right) &= -(\nabla \cdot \mathbf{B})\,\mathbf{u},
\end{aligned}
\end{align}
where $\rho$ represents the mass density, $\mathbf{u}$ is the physical velocity vector, $E$ is the total energy density, $\mathbf{B}$ is the magnetic field vector, $\eta$ is the uniform magnetic resistivity, $\boldsymbol{\tau} = \mu \left( \nabla \mathbf{u} + (\nabla \mathbf{u})^T - \frac{2}{3}(\nabla \cdot \mathbf{u})\mathbf{I} \right)$ is the viscous shear stress tensor with dynamic viscosity coefficient $\mu$, $\mathbf{q} = - K \nabla T$ is the thermal heat flux vector with conductivity $K$ and temperature $T = \sfrac{p_{\text{gas}}}{\rho}$, $\mathbf{j} = \nabla \times \mathbf{B}$ is the electrical current density, and $p_{\text{tot}} = p_{\text{gas}} + \sfrac{1}{2}|\mathbf{B}|^2$ is the total pressure with $p_{\text{gas}} = (\gamma - 1)(E - \sfrac{1}{2}\rho |\mathbf{u}|^2 - \sfrac{1}{2}|\mathbf{B}|^2)$.
To evaluate the dissipative gradients without stencils, the system maps onto \cref{eq:pde_target} with:
\begin{equation}
\adjustbox{max width=\linewidth}{$\displaystyle
    \mathbf{Q} = \begin{pmatrix*}[l]
    \rho \\
    \rho \mathbf{u} \\
    E \\
    \mathbf{B} \\
\mathbf{G}_{\mathbf{B}} \\
\mathbf{G}_\rho \\
\mathbf{G}_{\rho\mathbf{u}} \\
\mathbf{G}_E
    \end{pmatrix*}, \;
    \mathbf{\Phi} = \begin{pmatrix*}[l]
    \rho \mathbf{u} \\
    \rho \mathbf{u} \otimes \mathbf{u} + p_{\text{tot}}\mathbf{I} - \mathbf{B} \otimes \mathbf{B} - \boldsymbol{\tau} \\
    (E + p_{\text{tot}})\mathbf{u} - \mathbf{B}(\mathbf{u} \cdot \mathbf{B}) - \boldsymbol{\tau} \cdot \mathbf{u} + \mathbf{q} + \eta (\mathbf{j} \times \mathbf{B}) \\
    \mathbf{B} \otimes \mathbf{u} - \mathbf{u} \otimes \mathbf{B} - \eta \mathbf{G}_{\mathbf{B}} \\
\mathbf{G}_{\mathbf{B}} \otimes \mathbf{u} - \mathbf{I} \otimes \mathbf{B}/\tau_{\nabla} \\
\mathbf{G}_\rho \otimes \mathbf{u} - \mathbf{I}\,\rho/\tau_{\nabla} \\
\mathbf{G}_{\rho\mathbf{u}} \otimes \mathbf{u} - \mathbf{I} \otimes \rho\mathbf{u}/\tau_{\nabla} \\
\mathbf{G}_E \otimes \mathbf{u} - \mathbf{I}\,E/\tau_{\nabla}
    \end{pmatrix*}, \;
    \mathbf{S} = \begin{pmatrix*}[l]
    0 \\
    -(\nabla \cdot \mathbf{B})\,\mathbf{B} \\
    -(\nabla \cdot \mathbf{B})(\mathbf{u} \cdot \mathbf{B}) \\
    -(\nabla \cdot \mathbf{B})\,\mathbf{u} \\
- \mathbf{G}_{\mathbf{B}}/\tau_{\nabla} \\
- \mathbf{G}_\rho/\tau_{\nabla} \\
- \mathbf{G}_{\rho\mathbf{u}}/\tau_{\nabla} \\
- \mathbf{G}_E/\tau_{\nabla}
    \end{pmatrix*}.
$}
\end{equation}
The divergence-cleaning source is itself stencil-free: $\nabla\cdot\mathbf{B} = \mathrm{tr}\,\mathbf{G}_{\mathbf{B}}$ is read off the same tracker, so the Powell correction is generated automatically alongside the resistive term.
This system is treated by hand in the companion work~\cite{Bukreev2026}.
The compiler derives the same scheme from the conservation form alone, differing only in the dissipative closure: conserved-field gradient trackers with chain-rule reconstruction in place of the transported viscous stress.

\subsection{Homogenized Compressible Navier--Stokes--Fourier}
This case exercises two capabilities at once: a three-dimensional lattice (D3Q7) and \emph{prescribed external coefficient fields} that enter the dynamics without being functions of the evolved state.
It models compressible viscous flow interacting with a moving solid through a \emph{homogenized} formulation: a prescribed porosity field $\alpha(\mathbf{x},t)\in[0,1]$ ($\alpha=1$ pure fluid, $\alpha\to 0$ solid) volume-averages the dissipative fluxes, while a velocity coupling drives the momentum toward a prescribed solid velocity $\mathbf{u}_s(\mathbf{x},t)$ in the solid region.
The porosity acts as a numerical volume penalization rather than a physical homogenization: a volume-averaged form of Brinkman penalization~\cite{LiuVasilyev2007}, in the tradition of partially saturated cells and \emph{Homogenized Lattice Boltzmann} (HLBM) fluid--structure coupling~\cite{Noble1998, Krause2017, Kummerlaender2026a}:
\begin{align}
\begin{aligned}
\partial_t \rho + \nabla \cdot (\rho \mathbf{u}) &= 0, \\
\partial_t (\rho \mathbf{u}) + \nabla \cdot \left( \rho \mathbf{u} \otimes \mathbf{u} + p \mathbf{I} - \alpha\,\boldsymbol{\tau} \right) &= \frac{1-\alpha}{\eta}\big( \rho \mathbf{u}_s(\mathbf{x},t) - \rho \mathbf{u} \big), \\
\partial_t E + \nabla \cdot \left( (E + p)\mathbf{u} - \alpha\,\boldsymbol{\tau} \cdot \mathbf{u} + \alpha\,\mathbf{q} \right) &= \frac{1-\alpha}{\eta}\big( \rho \mathbf{u}_s - \rho \mathbf{u} \big) \cdot \mathbf{u},
\end{aligned}
\end{align}
with viscous stress $\boldsymbol{\tau} = \mu\big( \nabla \mathbf{u} + (\nabla \mathbf{u})^T - \tfrac{2}{3}(\nabla \cdot \mathbf{u})\mathbf{I} \big)$, Fourier flux $\mathbf{q} = -\kappa \nabla T$, temperature $T = p/\rho$ (gas constant absorbed, $R=1$), pressure $p = (\gamma - 1)\big( E - \tfrac{1}{2}\rho |\mathbf{u}|^2 \big)$ with adiabatic index $\gamma$, drag time scale $\eta$, and the limit $\alpha \to 1$ recovering the plain compressible NSF system.
Relative to the two-dimensional compressible NSF case of Section~\ref{sec:verification}, the closure is built from a reduced tracker basis that evolves the symmetric velocity gradient $\mathbf{G}_{\mathbf{u}} = \operatorname{sym}\nabla\mathbf{u}$ and the temperature gradient $\mathbf{G}_T = \nabla T$ directly, now in three dimensions and augmented by the homogenization:
\begin{align}
    \mathbf{Q} = \begin{pmatrix*}[l]
    \rho \\
    \rho \mathbf{u} \\
    E \\
\mathbf{G}_{\mathbf{u}} \\
\mathbf{G}_T
    \end{pmatrix*}, \quad
    \mathbf{\Phi} = \begin{pmatrix*}[l]
    \rho \mathbf{u} \\
    \rho \mathbf{u} \otimes \mathbf{u} + p \mathbf{I} - \alpha\boldsymbol{\tau} \\
    (E + p)\mathbf{u} - \alpha\boldsymbol{\tau} \cdot \mathbf{u} - \alpha\kappa \mathbf{G}_T \\
\mathbf{G}_{\mathbf{u}} \otimes \mathbf{u} - \operatorname{sym}(\mathbf{I} \otimes \mathbf{u})/\tau_{\nabla} \\
\mathbf{G}_T \otimes \mathbf{u} - \mathbf{I}\,T/\tau_{\nabla}
    \end{pmatrix*}, \quad
    \mathbf{S} = \begin{pmatrix*}[l]
    0 \\
    \tfrac{1-\alpha}{\eta}(\rho\mathbf{u}_s - \rho\mathbf{u}) \\
    \tfrac{1-\alpha}{\eta}(\rho\mathbf{u}_s - \rho\mathbf{u})\cdot\mathbf{u} \\
- \mathbf{G}_{\mathbf{u}}/\tau_{\nabla} \\
- \mathbf{G}_T/\tau_{\nabla}
    \end{pmatrix*}.
\end{align}

\section{Validation}\label{sec:mms}

To verify the convergence and execution precision of the compiled schemes, we track the global $L_2$ error norm reduction rates across successive grid resolutions ($N \in \{64, \dots, 512\}$; $N \in \{32, \dots, 256\}$ for the three-dimensional case).
For each system, the compiler evaluates the continuous residual of the analytical profiles to construct a manufactured source correction:
\begin{align}
\mathbf{S}_{\text{MMS}}(\mathbf{x}, t) = \partial_t \mathbf{Q}_{\text{exact}} + \nabla \cdot \mathbf{\Phi}(\mathbf{Q}_{\text{exact}}) - \mathbf{S}(\mathbf{Q}_{\text{exact}}).
\end{align}
This source field is injected directly into the mesoscopic update loop via \cref{eq:lbe_update}.
At the final time $t = 1$ we report, per macroscopic component, the discrete relative $L_2$ error $\lVert Q_h - Q_{\text{exact}}\rVert_2 / \lVert Q_{\text{exact}}\rVert_2$ with $\lVert f\rVert_2 = (N_{\text{cells}}^{-1}\sum_{\mathbf{x}} f(\mathbf{x})^2)^{1/2}$, and take the \emph{Empirical Order of Convergence} (EOC) to be the negative of the least-squares slope of $\log_2\lVert e\rVert_2$ against $\log_2 N$ over the full resolution sequence.
The manufactured fields are designed to stress the generated operators rather than for analytical convenience, using large-amplitude fluctuations, broken axis symmetry, independent thermodynamic modes, and exactly satisfied involutions.
The compressible working points are genuinely supersonic, with per-axis mean-flow Mach numbers up to $3$ for the Euler, Navier--Stokes--Fourier, and resistive MHD systems on the same minimal five-velocity lattice.
The per-case design principles are detailed in the supplementary materials.
All cases run on periodic domains, verifying the bulk scheme in isolation.
The generation of boundary operators from the same PDE-level representation is in development.

The resulting convergence matrix across IEEE single and double precisions, each system refined under the lattice scaling matched to its dominant transport mechanism (per the \emph{Scaling} column of Table~\ref{tab:pde_taxonomy}), is documented in Table~\ref{tab:mms_results}.
Every case is fully specified by its refinement law and three resolution-independent scalars: the time-step factor $f$ ($\Delta t = f\,\Delta x$ acoustic, $f\,\Delta x^2$ diffusive), the relaxation rate $\tau_R$ ($\tau_{\text{LB}} = \tfrac{1}{2} + \tau_R\,\Delta t$, Section~\ref{sec:consistency}), and the tracker step count $g_\tau$ ($\tau_\nabla = g_\tau\,\Delta t$, Section~\ref{sec:gradient_tracking}).
The refinement law follows the dominant transport mechanism, $f$ is bounded by the sub-characteristic condition of Section~\ref{sec:subchar}, and $\tau_R$ trades stability margin against the relaxation-diffusion error constant.
No parameter is retuned per grid.
Per-case values are tabulated in the supplementary materials.

\begingroup
\small
\begin{table}[p]
\centering
\scriptsize
\setlength{\tabcolsep}{2pt}
\renewcommand{\arraystretch}{0.95}
\caption{Overview of empirical verification results at the final time $t=1.0$. Errors are relative ($L^\infty$ and $L^2$) at grid resolution $N_{\text{max}}$. Each EOC is the negative least-squares log-log slope over all simulated resolutions for that norm. The \emph{Scaling} column reports the lattice refinement scaling chosen per system (Section~\ref{sec:consistency}). Rows are grouped per physical system, and the \emph{Field} column lists the system's macroscopic components.\label{tab:mms_results}}
\begin{tabular}{>{\raggedright\arraybackslash}p{2.0cm} >{\raggedright\arraybackslash}p{1.2cm} >{\raggedright\arraybackslash}p{1.5cm} l c c c c c}
\toprule
\textbf{System} & \textbf{Scaling} & \textbf{Field} & \textbf{Precision} & \textbf{$N_{\text{max}}$} & \textbf{Rel. $L^\infty$} & \textbf{$L^\infty$ EOC} & \textbf{Rel. $L^2$} & \textbf{$L^2$ EOC} \\
\midrule
\multirow{2}{=}{Inviscid Burgers} & \multirow{2}{=}{Acoustic} & $u$ & Double & 512 & 6.1175e-04 & 1.46 & 8.5671e-05 & 1.87 \\*
 &  &  & Float & 512 & 5.5348e-03 & 1.34 & 8.0850e-04 & 1.73 \\
\midrule
\multirow{6}{=}{Compressible Euler} & \multirow{6}{=}{Acoustic} & $E$ & Double & 512 & 3.3385e-06 & 2.01 & 1.8316e-06 & 2.00 \\*
 &  &  & Float & 512 & 6.7902e-05 & 1.47 & 2.2799e-05 & 1.73 \\*
 &  & $\rho$ & Double & 512 & 1.4401e-06 & 1.98 & 6.8435e-07 & 1.99 \\*
 &  &  & Float & 512 & 6.6678e-05 & 1.34 & 2.4365e-05 & 1.65 \\*
 &  & $\rho\mathbf{u}$ & Double & 512 & 3.2141e-06 & 2.00 & 1.8745e-06 & 2.00 \\*
 &  &  & Float & 512 & 1.0204e-04 & 1.38 & 3.6474e-05 & 1.62 \\
\midrule
\multirow{4}{=}{Shallow Water (SWE)} & \multirow{4}{=}{Acoustic} & $h$ & Double & 512 & 5.5560e-06 & 1.87 & 2.3433e-06 & 1.88 \\*
 &  &  & Float & 512 & 1.5929e-05 & 1.89 & 6.8696e-06 & 1.95 \\*
 &  & $h\mathbf{u}$ & Double & 512 & 2.6837e-05 & 1.76 & 2.0838e-05 & 1.70 \\*
 &  &  & Float & 512 & 5.4059e-05 & 1.81 & 3.6489e-05 & 1.83 \\
\midrule
\multirow{4}{=}{Ideal Ultrarelativistic Fluid} & \multirow{4}{=}{Acoustic} & $E$ & Double & 512 & 4.3449e-06 & 2.00 & 2.1354e-06 & 2.00 \\*
 &  &  & Float & 512 & 1.1342e-05 & 1.59 & 4.7227e-06 & 1.64 \\*
 &  & $\mathbf{S}$ & Double & 512 & 9.6256e-06 & 2.00 & 6.3943e-06 & 2.00 \\*
 &  &  & Float & 512 & 1.3004e-05 & 1.88 & 6.4705e-06 & 2.00 \\
\midrule
\multirow{6}{=}{Maxwell (TM)} & \multirow{6}{=}{Acoustic} & $E_z$ & Double & 512 & 2.9956e-05 & 1.95 & 2.9956e-05 & 1.95 \\*
 &  &  & Float & 512 & 3.7205e-05 & 1.85 & 3.0632e-05 & 1.93 \\*
 &  & $H_x$ & Double & 512 & 1.0951e-04 & 2.00 & 1.0951e-04 & 2.00 \\*
 &  &  & Float & 512 & 1.2184e-04 & 1.95 & 1.1675e-04 & 1.97 \\*
 &  & $H_y$ & Double & 512 & 1.1587e-05 & 2.02 & 1.1587e-05 & 2.02 \\*
 &  &  & Float & 512 & 1.3310e-05 & 1.96 & 5.9351e-06 & 2.33 \\
\midrule
\multirow{4}{=}{Nonlinear Elasticity} & \multirow{4}{=}{Acoustic} & $\mathbf{F}$ & Double & 512 & 3.1874e-05 & 2.00 & 2.8940e-05 & 2.00 \\*
 &  &  & Float & 512 & 9.7364e-05 & 1.48 & 4.2667e-05 & 1.79 \\*
 &  & $\mathbf{v}$ & Double & 512 & 6.4804e-05 & 2.01 & 4.5664e-05 & 1.99 \\*
 &  &  & Float & 512 & 1.2099e-04 & 2.40 & 5.8314e-05 & 2.68 \\
\midrule
\multirow{2}{=}{Scalar ADR} & \multirow{2}{=}{Diffusive} & $c$ & Double & 512 & 2.2901e-04 & 2.50 & 1.2680e-04 & 2.49 \\*
 &  &  & Float & 512 & 1.9430e-04 & 2.55 & 1.0911e-04 & 2.55 \\
\midrule
\multirow{2}{=}{Allen--Cahn Phase-Field} & \multirow{2}{=}{Diffusive} & $\phi$ & Double & 512 & 7.3196e-04 & 2.50 & 7.8653e-04 & 2.51 \\*
 &  &  & Float & 512 & 7.0154e-04 & 2.53 & 5.6359e-04 & 2.66 \\
\midrule
\multirow{4}{=}{Incompressible NSE} & \multirow{4}{=}{Diffusive} & $\rho$ & Double & 512 & 1.3419e-06 & 2.43 & 6.0683e-07 & 2.47 \\*
 &  &  & Float & 512 & 2.6675e-05 & 1.48 & 1.1146e-05 & 1.58 \\*
 &  & $\rho\mathbf{u}$ & Double & 512 & 8.2447e-04 & 2.51 & 8.5394e-04 & 2.49 \\*
 &  &  & Float & 512 & 4.1288e-03 & 2.23 & 4.1154e-03 & 2.23 \\
\midrule
\multirow{6}{=}{Compressible NSF} & \multirow{6}{=}{Diffusive} & $E$ & Double & 512 & 8.3354e-05 & 1.98 & 4.5785e-05 & 2.01 \\*
 &  &  & Float & 512 & 2.4601e-04 & 1.50 & 1.5715e-04 & 1.46 \\*
 &  & $\rho$ & Double & 512 & 7.3384e-05 & 1.91 & 3.7592e-05 & 1.95 \\*
 &  &  & Float & 512 & 2.5013e-04 & 1.36 & 1.5397e-04 & 1.32 \\*
 &  & $\rho\mathbf{u}$ & Double & 512 & 1.1775e-04 & 1.95 & 6.8579e-05 & 1.99 \\*
 &  &  & Float & 512 & 3.7982e-04 & 1.43 & 2.4644e-04 & 1.41 \\
\midrule
\multirow{8}{=}{Resistive MHD} & \multirow{8}{=}{Acoustic} & $\mathbf{B}$ & Double & 512 & 2.4829e-04 & 2.16 & 1.6277e-04 & 2.22 \\*
 &  &  & Float & 512 & 2.8267e-04 & 2.18 & 1.8840e-04 & 2.24 \\*
 &  & $E$ & Double & 512 & 3.6516e-04 & 1.90 & 2.1223e-04 & 1.96 \\*
 &  &  & Float & 512 & 4.0480e-04 & 1.91 & 2.2318e-04 & 2.00 \\*
 &  & $\rho$ & Double & 512 & 2.7365e-04 & 1.76 & 1.6299e-04 & 1.75 \\*
 &  &  & Float & 512 & 3.2867e-04 & 1.70 & 1.9287e-04 & 1.70 \\*
 &  & $\rho\mathbf{u}$ & Double & 512 & 4.6103e-04 & 1.82 & 2.4482e-04 & 1.88 \\*
 &  &  & Float & 512 & 5.2219e-04 & 1.82 & 2.6853e-04 & 1.90 \\
\midrule
\multirow{8}{=}{HCNSF} & \multirow{8}{=}{Acoustic} & $E$ & Double & 256 & 1.7303e-04 & 1.70 & 7.4269e-05 & 1.78 \\*
 &  &  & Float & 256 & 1.7001e-04 & 1.86 & 9.0488e-05 & 1.79 \\*
 &  & $\mathbf{G}_{\mathbf{u}}$ & Double & 256 & 1.0801e-02 & 1.82 & 1.0391e-02 & 1.83 \\*
 &  &  & Float & 256 & 1.1265e-02 & 1.81 & 1.0592e-02 & 1.83 \\*
 &  & $\rho$ & Double & 256 & 2.5489e-04 & 1.87 & 8.4492e-05 & 1.87 \\*
 &  &  & Float & 256 & 3.9397e-04 & 1.87 & 1.3437e-04 & 1.89 \\*
 &  & $\rho\mathbf{u}$ & Double & 256 & 1.3203e-04 & 1.92 & 5.5799e-05 & 1.96 \\*
 &  &  & Float & 256 & 2.2766e-04 & 1.95 & 1.0948e-04 & 1.97 \\
\bottomrule
\end{tabular}
\end{table}

\endgroup

\subsection{Convergence Characteristics}

\paragraph{Second-Order Convergence in Double Precision}
Across the twelve physical systems under double-precision arithmetic, the fitted EOC of the macroscopic state variables is at or near second order, with $L^2$ slopes between $1.70$ and $2.51$.
The fitted slopes near $2.5$ of the three diffusively refined light systems (scalar ADR, Allen--Cahn, incompressible NSE) are not asymptotic superconvergence: their pairwise orders decrease monotonically toward two ($2.8$ down to $2.2$ pairwise).
Physical diffusion is carried by the tracked-gradient fluxes, so under both refinement laws the relaxation diffusion of \cref{eq:relax_diffusion} is a second-order-consistent numerical stabilizer (Section~\ref{sec:consistency}).
For the hyperbolic and convection-dominated systems the observed second order is thus a property of the resolved smooth solutions, not a claim of intrinsic second-order accuracy in the inviscid limit.
In both regimes the clean rate corroborates that the compiler correctly deconstructs the declared PDEs, resolves the automatic gradient cascades, and emits consistent kernels.

\paragraph{Single Precision}
Comparing the single- and double-precision columns isolates the floating-point floor of the method.
Single-precision execution is a standard lever for LBM throughput on GPUs~\cite{Bailey2009, Kuznik2010}, so it matters whether the generated schemes retain their convergence order in \texttt{float}.
The reference- and equilibrium-shifted formulation (Section~\ref{sec:reference_shift}) evolves the fluctuation $\delta\mathbf{Q}$ about the background state directly, so small fluctuations are never formed as the difference of two large numbers.
This removes the catastrophic cancellation that would otherwise destroy the single-precision result, and in several cases is empirically what enables convergence in \texttt{float} at all.
Two residual effects remain when comparing against the double-precision columns.
First, the per-step rounding errors of slowly evolving fields accumulate into a field-dependent relative error floor of order $10^{-5}$--$10^{-4}$ over the $\sim\!10^{5}$ steps of the finest diffusive grids.
Systems whose discretization error stays above it track their double-precision columns (scalar ADR at $2.55$, Allen--Cahn at $2.66$).
The compressible NSF system, whose double error dives below it, tracks double through $N = 256$ and is floor-limited at $N = 512$.
The incompressible NSE density ($1.58$) is floor-limited in the same way.
Second, single precision is unstable near the over-relaxation limit $\omega = 1/\tau_{\text{LB}} \to 2$ where double precision is not.
The affected systems run a uniformly stiffer relaxation configuration in single precision (supplementary materials) and converge cleanly under that single uniform configuration (incompressible NSE momentum at $2.23$, resistive MHD at $2.24$).
The acoustic hyperbolic systems, taking only $\propto N$ steps, erode where the double-precision error is smallest, most visibly the compressible Euler momentum ($1.62$).

\section{Performance}\label{sec:performance}

\begin{figure}[h]
    \centering
    \includegraphics[width=\linewidth]{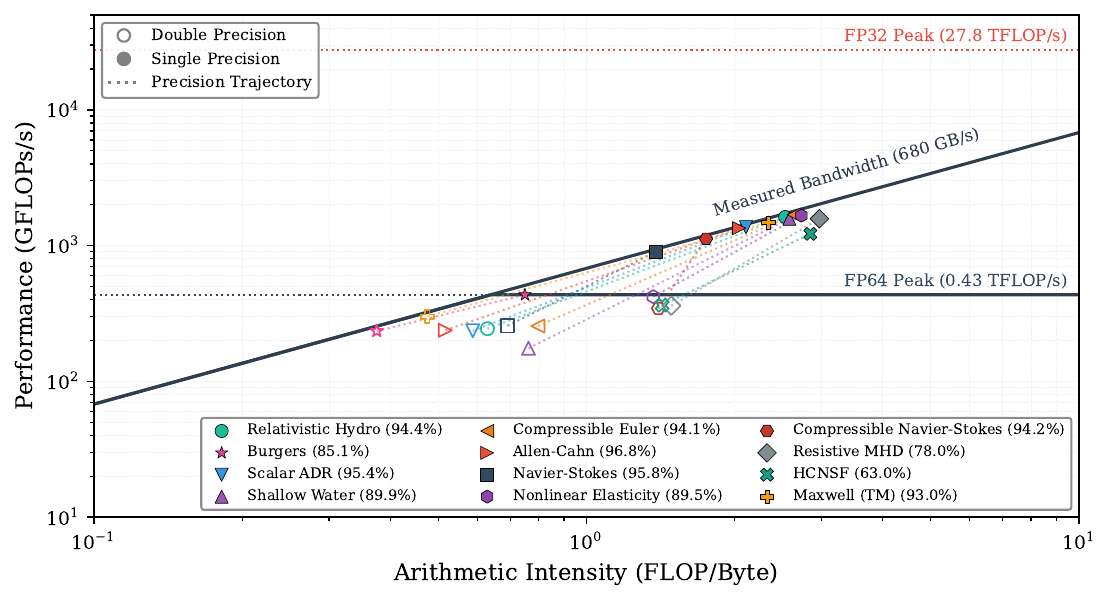}
    \caption{Empirical roofline analysis of the generated LBMs on an NVIDIA RTX A5000.}
    \label{fig:roofline}
\end{figure}

In the final code generation layer (cf. Fig.~\ref{fig:pipeline}) the compiled schemes are emitted as platform-transparent C++ operators for the open-source multiphysics framework OpenLB~\cite{Krause2021a,Kummerlaender2026c}, whose \emph{Common Subexpression Elimination} (CSE) machinery compacts them algebraically: OpenLB extracts each operator's expression tree, reduces it in SymPy, and re-emits it as a per-target template specialization~\cite{Kummerlaender2026c}.
By abstracting memory layout and execution from the continuous equations, this architecture enables compile-time binding to platform-specific data structures, through which OpenLB targets multi-core CPUs and diverse GPU accelerators~\cite{Kummerlaender2023, Kummerlaender2025, Kummerlaender2026c}.
As shown in Figure~\ref{fig:roofline}, the generated cell-level kernels approach the memory-bandwidth roofline on an exemplary NVIDIA RTX A5000, reaching up to $96\%$ of peak in single precision.
The per-cell byte traffic reported in Table~\ref{tab:roofline} is obtained by introspecting the compiled OpenLB dynamics, counting the populations and field components actually read and written per cell update.
The attained throughput is then measured directly from the kernel runtime on a saturating grid, and the saturation column is its ratio to the measured $\qty{680}{\giga\byte\per\second}$ peak (BabelStream Triad~\cite{Deakin2018}).

\begin{sidewaystable}
\centering
\footnotesize
\setlength{\tabcolsep}{2.5pt}
\caption{Performance and per-kernel resource characteristics of the generated LBM kernels across all 12 physical models, measured on an NVIDIA RTX A5000 (\texttt{sm\_86}: 65536 registers and 48 warps per SM, \qty{680.0}{\giga\byte\per\second} peak bandwidth). Throughput columns are for the CSE-optimized build: arithmetic intensity \emph{AI} in FLOP/byte, \emph{Util.\ BW} in \unit{\giga\byte\per\second}, \emph{BW Sat.} relative to peak, \emph{GFLOP/s} attained. Each row reports the faster of the two emitted CSE variants at that precision together with its exact operation count, so \emph{FLOP/cell} can differ between precisions. The resource columns give the binding collide kernel's per-thread register count (\emph{Reg.}), local-memory spill (\emph{Spill}, bytes), and register-limited occupancy (\emph{Occ.}, \%) for the native-CSE build.}
\label{tab:roofline}
\begin{tabular}{l l c l c c c c c c c c c c}
\toprule
\textbf{System} & \textbf{Lattice} & \textbf{Fields} & \textbf{Prec.} & \textbf{FLOP/cell} & \textbf{Byte/cell} & \textbf{AI} & \textbf{MLUP/s} & \textbf{Util.\ BW} & \textbf{BW Sat.} & \textbf{GFLOP/s} & \textbf{Reg.} & \textbf{Spill} & \textbf{Occ.} \\
\midrule
\multirow{2}{*}{Inviscid Burgers} & \multirow{2}{*}{\texttt{D2Q5}} & \multirow{2}{*}{1} & Single & 30 & 40 & 0.75 & 14474 & 579 & 85.1\% & 434 & 36 & -- & 100 \\
 & & & Double & 30 & 80 & 0.38 & 7830 & 626 & 92.1\% & 235 & 36 & -- & 100 \\
\hline
\multirow{2}{*}{Compressible Euler} & \multirow{2}{*}{\texttt{D2Q5}} & \multirow{2}{*}{4} & Single & 422 & 160 & 2.64 & 4000 & 640 & 94.1\% & 1688 & 78 & -- & 52 \\
 & & & Double & 255 & 320 & 0.80 & 1003 & 321 & 47.2\% & 256 & 78 & -- & 52 \\
\hline
\multirow{2}{*}{Shallow Water (SWE)} & \multirow{2}{*}{\texttt{D2Q5}} & \multirow{2}{*}{3} & Single & 310 & 120 & 2.58 & 5093 & 611 & 89.9\% & 1579 & 56 & -- & 75 \\
 & & & Double & 183 & 240 & 0.76 & 958 & 230 & 33.8\% & 175 & 56 & -- & 75 \\
\hline
\multirow{2}{*}{Ideal Ultrarelativistic Fluid} & \multirow{2}{*}{\texttt{D2Q5}} & \multirow{2}{*}{3} & Single & 304 & 120 & 2.53 & 5349 & 642 & 94.4\% & 1626 & 54 & -- & 75 \\
 & & & Double & 151 & 240 & 0.63 & 1618 & 388 & 57.1\% & 244 & 54 & -- & 75 \\
\hline
\multirow{2}{*}{Maxwell (TM)} & \multirow{2}{*}{\texttt{D2Q5}} & \multirow{2}{*}{3} & Single & 281 & 120 & 2.34 & 5269 & 632 & 93.0\% & 1481 & 56 & -- & 75 \\
 & & & Double & 114 & 240 & 0.48 & 2634 & 632 & 93.0\% & 300 & 56 & -- & 75 \\
\hline
\multirow{2}{*}{Nonlinear Elasticity} & \multirow{2}{*}{\texttt{D2Q5}} & \multirow{2}{*}{6} & Single & 656 & 240 & 2.73 & 2536 & 609 & 89.5\% & 1664 & 56 & -- & 75 \\
 & & & Double & 656 & 480 & 1.37 & 641 & 308 & 45.2\% & 420 & 56 & -- & 75 \\
\hline
\multirow{2}{*}{Scalar ADR} & \multirow{2}{*}{\texttt{D2Q5}} & \multirow{2}{*}{3} & Single & 253 & 120 & 2.11 & 5405 & 649 & 95.4\% & 1367 & 58 & -- & 67 \\
 & & & Double & 141 & 240 & 0.59 & 1673 & 402 & 59.0\% & 236 & 58 & -- & 67 \\
\hline
\multirow{2}{*}{Allen--Cahn Phase-Field} & \multirow{2}{*}{\texttt{D2Q5}} & \multirow{2}{*}{3} & Single & 245 & 120 & 2.04 & 5487 & 658 & 96.8\% & 1344 & 64 & -- & 67 \\
 & & & Double & 124 & 240 & 0.52 & 1918 & 460 & 67.7\% & 238 & 64 & -- & 67 \\
\hline
\multirow{2}{*}{Incompressible NSE} & \multirow{2}{*}{\texttt{D2Q5}} & \multirow{2}{*}{7} & Single & 387 & 280 & 1.38 & 2328 & 652 & 95.8\% & 901 & 126 & -- & 33 \\
 & & & Double & 387 & 560 & 0.69 & 662 & 371 & 54.6\% & 256 & 126 & -- & 33 \\
\hline
\multirow{2}{*}{Compressible NSF} & \multirow{2}{*}{\texttt{D2Q5}} & \multirow{2}{*}{12} & Single & 840 & 480 & 1.75 & 1335 & 641 & 94.2\% & 1121 & 206 & -- & 19 \\
 & & & Double & 1344 & 960 & 1.40 & 256 & 246 & 36.2\% & 345 & 210 & -- & 19 \\
\hline
\multirow{2}{*}{Resistive MHD} & \multirow{2}{*}{\texttt{D2Q5}} & \multirow{2}{*}{18} & Single & 2141 & 720 & 2.97 & 736 & 530 & 78.0\% & 1577 & 255 & 32 & 17 \\
 & & & Double & 2141 & 1440 & 1.49 & 168 & 242 & 35.7\% & 361 & 255 & 16 & 17 \\
\hline
\multirow{2}{*}{HCNSF} & \multirow{2}{*}{\texttt{D3Q7}} & \multirow{2}{*}{14} & Single & 2234 & 784 & 2.85 & 546 & 428 & 63.0\% & 1220 & 255 & 104 & 17 \\
 & & & Double & 2234 & 1568 & 1.42 & 163 & 255 & 37.5\% & 363 & 255 & 104 & 17 \\
\bottomrule
\end{tabular}
\end{sidewaystable}

\looseness=-1 The per-kernel resource columns of Table~\ref{tab:roofline} account for the two single-precision bandwidth outliers.
The two largest-state kernels, the 18-field resistive MHD and the 14-field HCNSF system, both reach the 255-register ceiling of the \texttt{sm\_86} architecture and spill to local memory ($32$ and $104$ bytes per thread).
They are precisely the two whose bandwidth saturation falls below the $85$--$97\%$ band, at $78\%$ and $63\%$.
Occupancy alone does not explain this: the 12-field compressible NSF kernel runs at comparable occupancy ($19\%$ against $17\%$) but, fitting within the register file without spilling, still sustains $94\%$ of peak.
Of the two, the HCNSF system saturates the least: its 14-field state on the \texttt{D3Q7} lattice (seven populations per field against five on \texttt{D2Q5}) is the largest live-quantity count in the suite, so at the shared register ceiling it spills the hardest, and the uncounted local-memory traffic of that spill displaces the useful bandwidth.
In double precision the three largest-state kernels become compute-bound: compressible NSF, resistive MHD, and HCNSF sustain $345$, $361$, and $\qty{363}{\giga\flop\per\second}$ against the $\qty{434}{\giga\flop\per\second}$ FP64 peak ($79$--$84\%$), so their low bandwidth-saturation figures (all under $38\%$) reflect the $1{:}64$ FP64 throughput of the architecture rather than a memory-system inefficiency.
Of the remaining double-precision points, Burgers and Maxwell stay at the bandwidth roofline and nonlinear elasticity at the FP64 ceiling.

\section{Conclusion}\label{sec:conclusion}

We introduced a symbolic compiler that automates the translation of systems of continuous conservation laws into optimized, local lattice Boltzmann schemes, work previously restricted to manual and case-specific derivations.
Driven by point-wise advection-relaxation cascades, it maps nonlinear physical fluxes directly onto the first-order discrete moments of independent distribution functions and recursively deconstructs higher-order derivatives into auxiliary states, eliminating non-local finite-difference stencils and preserving local concurrency.
Asymptotic spatial convergence at or near second order of the generated models was verified by MMS across twelve PDEs spanning fluid dynamics, resistive MHD, phase-field interface dynamics, electromagnetics, and nonlinear finite-strain elasticity.
The reference- and equilibrium-shifted formulation makes single precision usable at close to the double-precision convergence order, and the arithmetically minimized device kernels generated for the platform-transparent OpenLB framework reach up to 96\% of the memory-bandwidth roofline.

The present framework is the first in which LBMs themselves are derived automatically from continuous conservation laws rather than supplied as input, in contrast to prior top-down constructions of individual schemes~\cite{MendozaMunoz2010, Graille2014, Dubois2014, Simonis2023}.
Within this class, retargeting LBM to a new system is thereby reduced from a manual derivation to a declarative specification, pairing the rapid PDE retargeting that historically distinguished general-purpose finite-volume and finite-element frameworks with the structural locality and efficiency inherent to LBM.

Building on this, boundary condition generation from the PDE-level representation is in development as the prerequisite for validation against physical, non-manufactured benchmarks, as already demonstrated manually for MHD~\cite{Bukreev2026}.

\section*{Code Availability}

All generated schemes are specified by Sections~\ref{sec:methodology} and \ref{sec:verification} alongside the component equations, manufactured solutions, and numerical configurations in the supplement.
The \emph{PDE2LBM} compiler is available upon reasonable request.

\section*{Acknowledgement}

Google Gemini and Anthropic Claude assisted in drafting and revising the manuscript.
All content was reviewed, verified, and approved by the authors, who take full responsibility for it.

\bibliographystyle{siamplain}
\bibliography{bibliography}

\clearpage

\makeatletter
\def\siamprelabel{SM}%
\def\siampretitle{Supplementary Materials: }%
\makeatother
\setcounter{section}{0}%
\setcounter{figure}{0}%
\setcounter{table}{0}%
\setcounter{equation}{0}%
\renewcommand{\thepage}{\arabic{page}}%

\phantomsection
\addcontentsline{toc}{section}{Supplementary Materials}
\begin{center}
  {\normalfont\scshape\large Supplementary Materials}
\end{center}
\vskip 6pt

\section{Empirical Convergence Profile Plots}\label{sec:appendix_plots}

In this appendix, we present the \emph{Empirical Order of Convergence} (EOC) profiles for all twelve verified PDE targets.
Each plot displays the relative $L^1$, $L^2$, and $L^\infty$ error norms of the macroscopic state components across consecutive grid resolutions, in both double and single precision, together with a second-order reference slope.
Each system is refined under the lattice scaling matched to its dominant transport mechanism (diffusive for the diffusion-resolving parabolic systems, acoustic for the hyperbolic and convection-dominated ones).
The numerical configuration underlying these runs is tabulated in Appendix~\ref{sec:appendix_recipes}.

\newcommand{\eocfig}[2]{%
    \begin{figure}[p]
        \centering
        \includegraphics[width=0.9\textwidth]{figure/#1.pdf}
        \caption{EOC profile for the #2.}
        \label{fig:#1}
    \end{figure}%
}

\eocfig{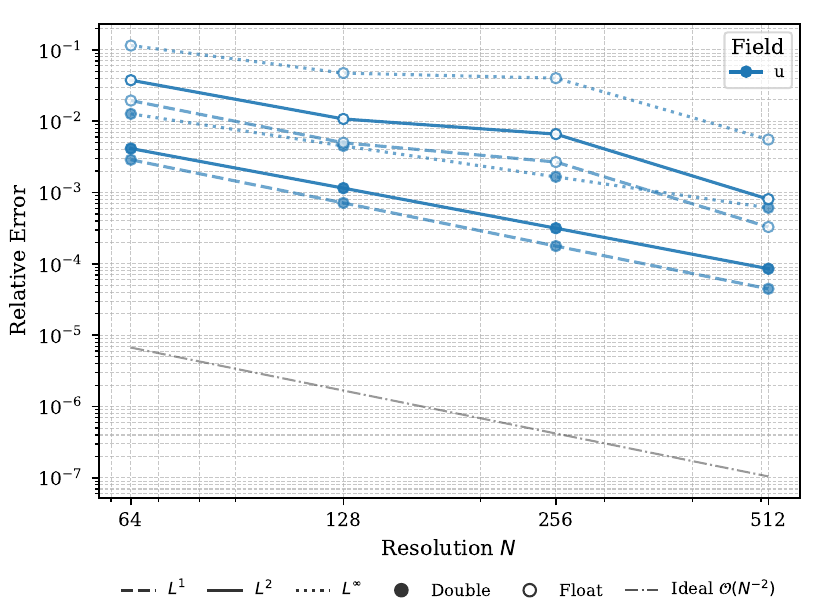}{inviscid Burgers system}
\eocfig{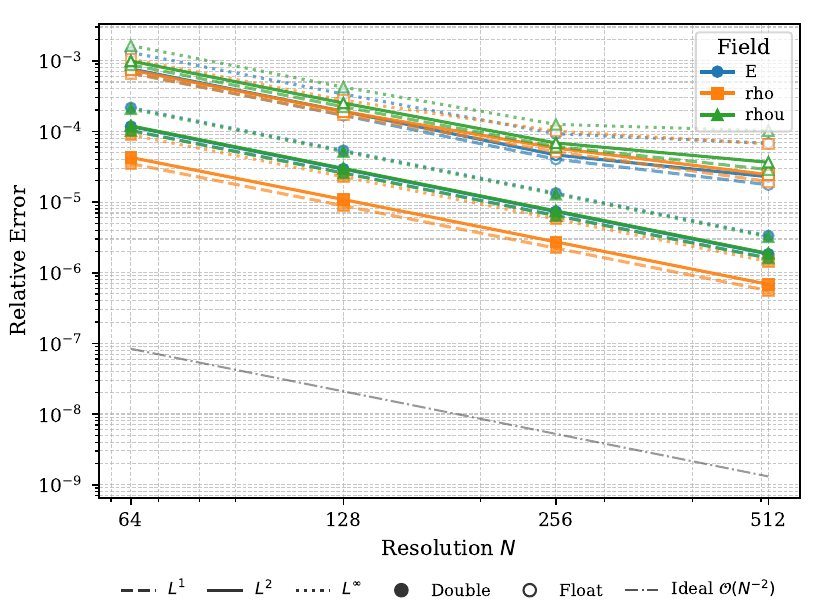}{compressible Euler system}
\eocfig{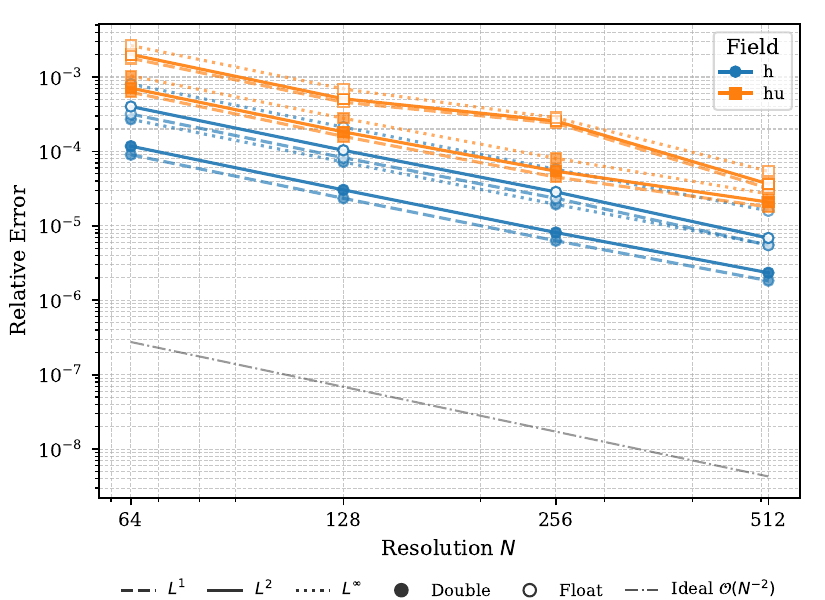}{shallow water equations}
\eocfig{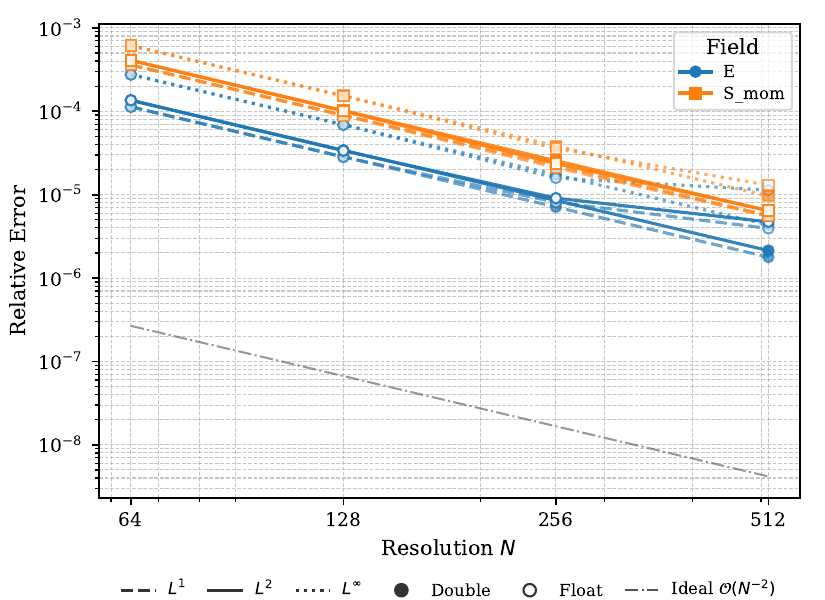}{ideal ultrarelativistic fluid}
\eocfig{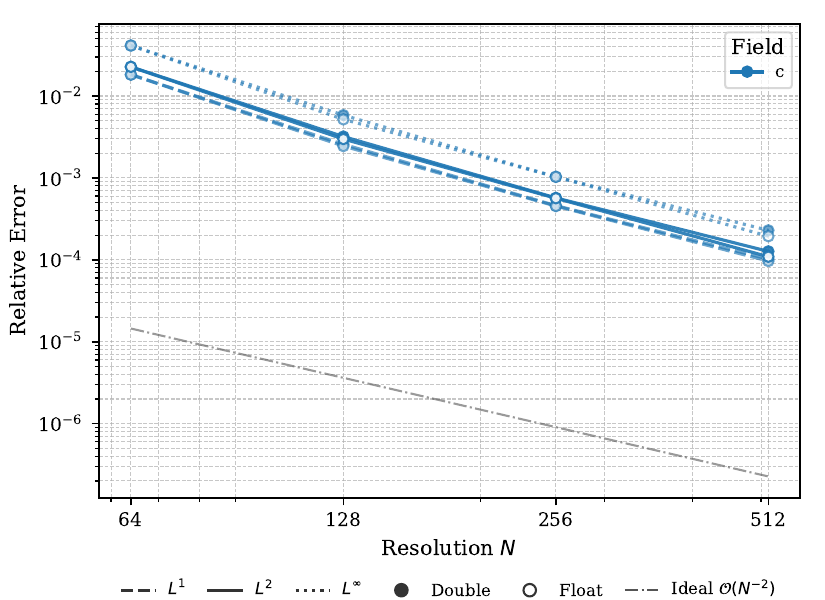}{scalar advection-diffusion-reaction system}
\eocfig{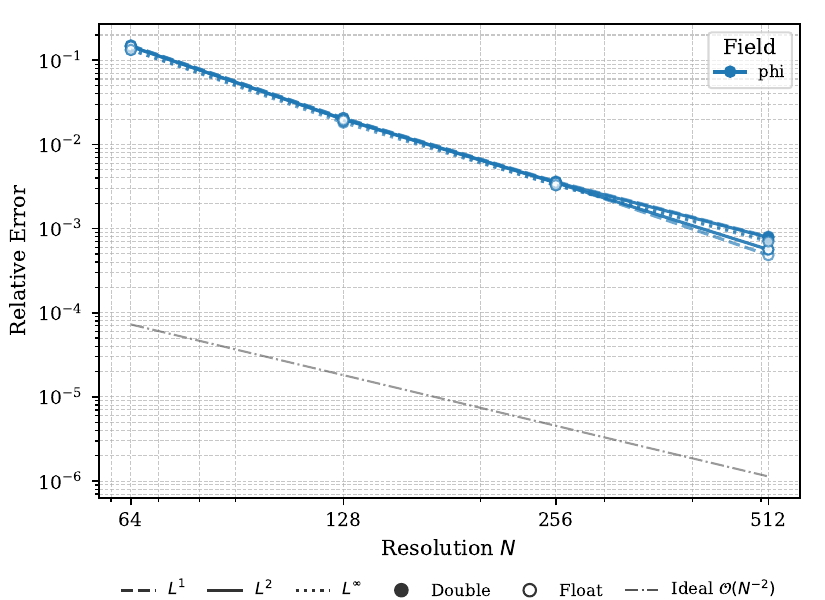}{Allen--Cahn phase-field equation}
\eocfig{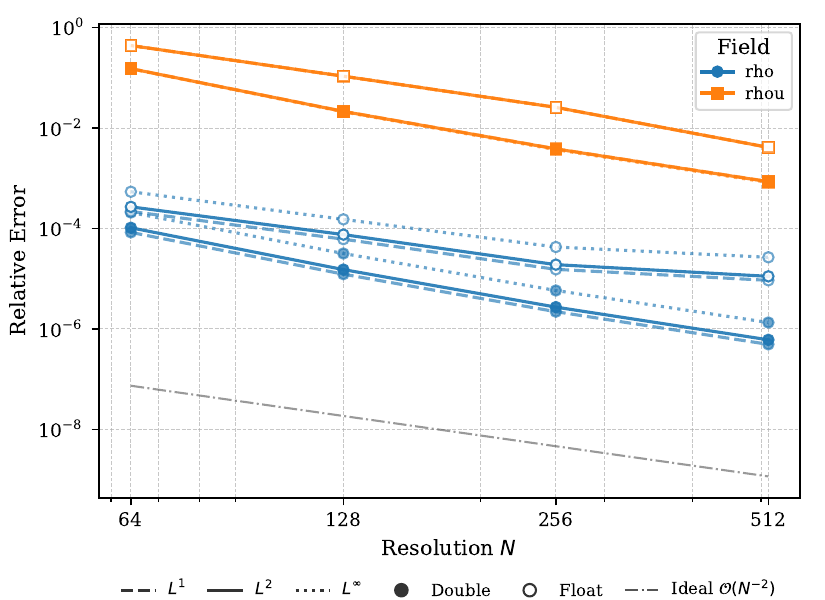}{incompressible Navier--Stokes equations}
\eocfig{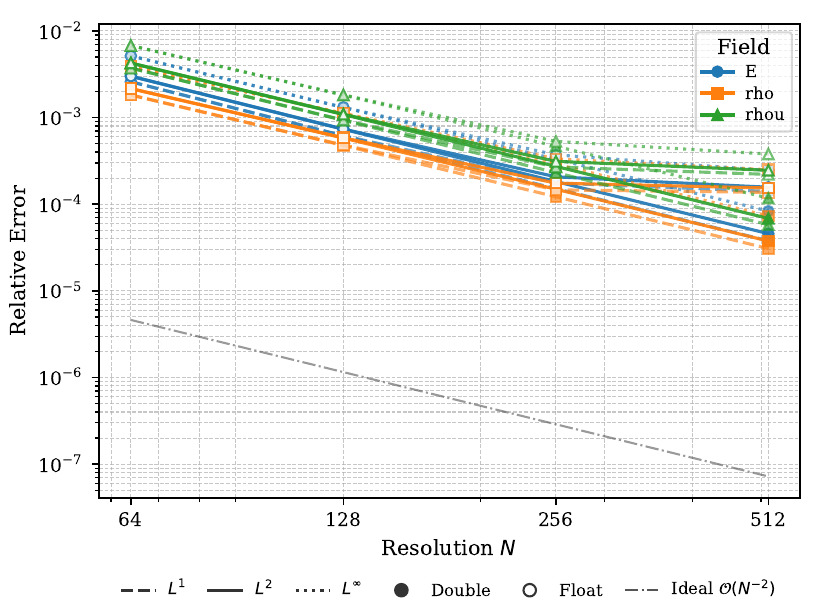}{compressible Navier--Stokes--Fourier equations}
\eocfig{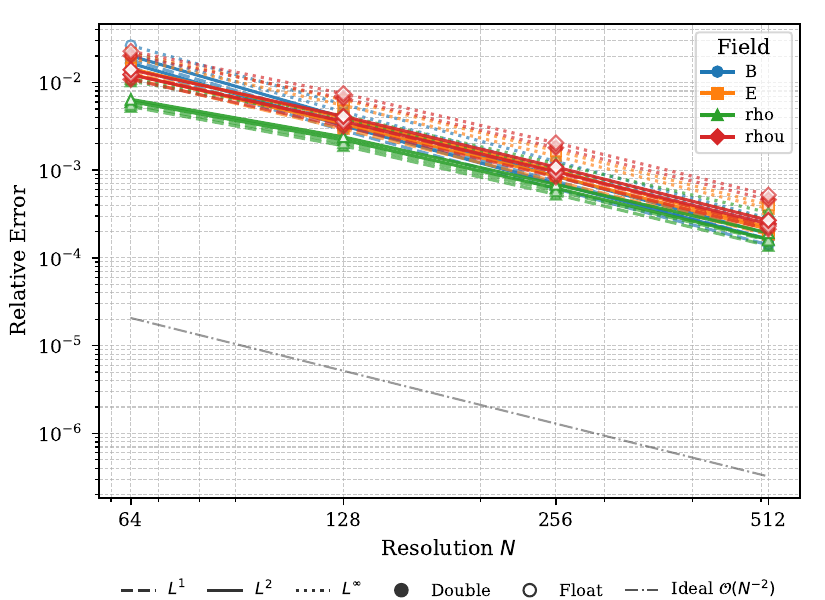}{resistive magnetohydrodynamics system}
\eocfig{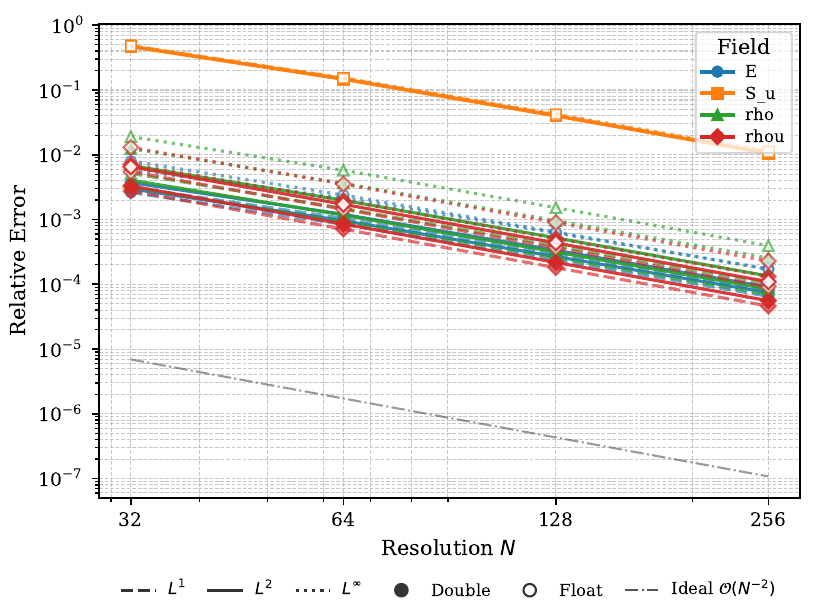}{three-dimensional homogenized compressible Navier--Stokes--Fourier system}
\eocfig{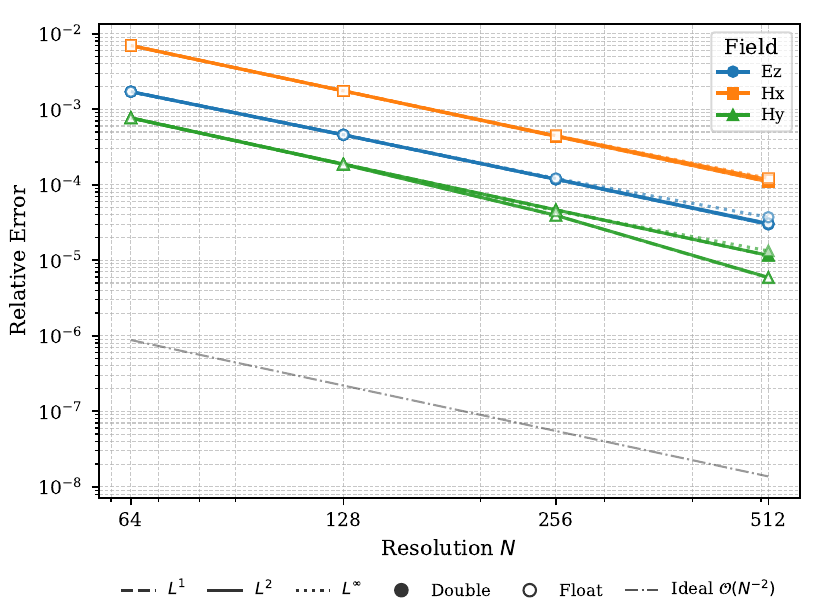}{Maxwell electromagnetics (TM mode)}
\eocfig{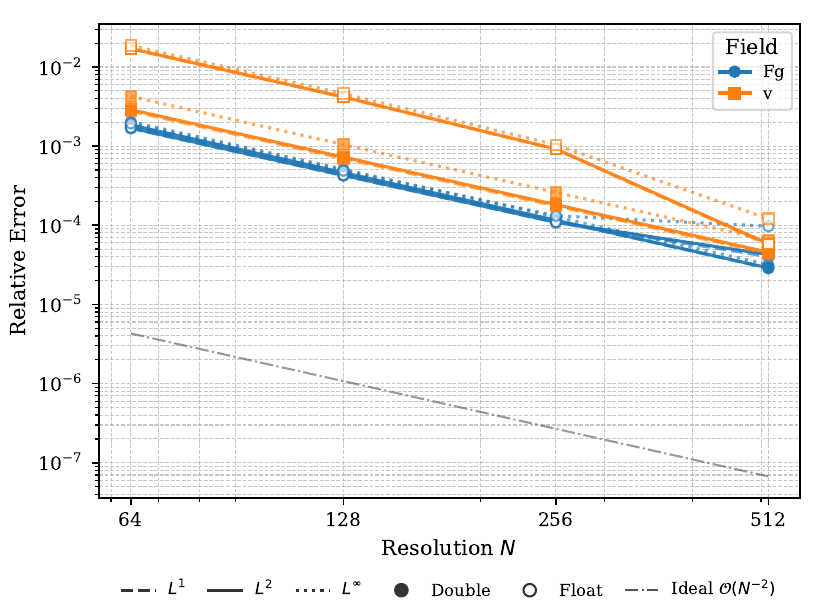}{nonlinear (neo-Hookean) elasticity system}

\clearpage
\section{Numerical Configuration of the Convergence Study}\label{sec:appendix_recipes}

This appendix documents the complete numerical configuration behind the convergence results reported in the main article, so that every run is reproducible from the tabulated values.
All cases share one study harness parametrized by the refinement scaling and three scalar knobs.
The time step follows the per-system scaling law, $\Delta t = f\,\Delta x$ under acoustic and $\Delta t = f\,\Delta x^2$ under diffusive refinement, with time-step factor $f$.
The relaxation time of the physical fields is parametrized by a resolution-independent rate $\tau_R$ as $\tau_{\text{LB}} = \tfrac{1}{2} + \tau_R\,\Delta t$, so the bare relaxation diffusion contracts as $\varepsilon a^2 = c_s^2\,\tau_R\,\Delta x^2$ under either scaling.
The gradient-tracker fields relax on the tracker time $\tau_\nabla = g_\tau\,\Delta t$, and their populations collide at the fixed lattice rate $1/(g_\tau + 1)$.
The two-dimensional cases run the D2Q5 stencil and the three-dimensional HCNSF case runs D3Q7.
Table~\ref{tab:stencils} lists the discrete velocity sets, weights, and lattice speeds of sound.
All populations are initialized at the (shifted) equilibrium of the exact manufactured fields at $t = 0$.

\begin{table}[htbp]
\centering\footnotesize
\caption{Discrete velocity sets of the two stencils used in the study.\label{tab:stencils}}
\begin{tabular}{l l c c c}
\toprule
\textbf{Lattice} & \textbf{Velocities} $\mathbf{c}_i$ & \textbf{Count} & $w_i$ & $c_s^2$ \\
\midrule
\multirow{2}{*}{D2Q5} & $(0,0)$ & 1 & $\tfrac{1}{3}$ & \multirow{2}{*}{$\tfrac{1}{3}$} \\*
 & $(\pm 1,0),\ (0,\pm 1)$ & 4 & $\tfrac{1}{6}$ & \\
\midrule
\multirow{2}{*}{D3Q7} & $(0,0,0)$ & 1 & $\tfrac{1}{4}$ & \multirow{2}{*}{$\tfrac{1}{4}$} \\*
 & $(\pm 1,0,0),\ (0,\pm 1,0),\ (0,0,\pm 1)$ & 6 & $\tfrac{1}{8}$ & \\
\bottomrule
\end{tabular}
\end{table}

Table~\ref{tab:recipes} lists the per-case values.
Every configuration is uniform across the reported resolutions in both precisions: no parameter is retuned per grid, so each fitted convergence order measures a single fixed scheme family.

\begin{table}[htbp]
\centering\footnotesize
\setlength{\tabcolsep}{4pt}
\caption{Per-case numerical configuration of the convergence study. The parametrization of the time-step factor $f$, relaxation rate $\tau_R$, and tracker relaxation $g_\tau$ is defined in the text, and tracker-free systems are marked with a dash. Every value is uniform across all reported grids of its case ($N = 64$ to $512$, and $N = 32$ to $256$ for the 3D HCNSF case): no parameter is retuned per resolution in either precision.\label{tab:recipes}}
\begin{tabular}{l l l l l c}
\toprule
\textbf{System} & \textbf{Scaling} & \textbf{Precision} & $f$ & $\tau_R$ & $g_\tau$ \\
\midrule
\multirow{2}{*}{Inviscid Burgers} & \multirow{2}{*}{Acoustic} & Double & 0.2 & 0.2 & -- \\*
 &  & Float & 0.2 & 2 & -- \\
\midrule
\multirow{2}{*}{Compressible Euler} & \multirow{2}{*}{Acoustic} & Double & 0.05 & 0.2 & -- \\*
 &  & Float & 0.05 & 5 & -- \\
\midrule
\multirow{2}{*}{Shallow Water (SWE)} & \multirow{2}{*}{Acoustic} & Double & 0.2 & 0.5 & -- \\*
 &  & Float & 0.2 & 2 & -- \\
\midrule
\multirow{2}{*}{Ideal Ultrarelativistic Fluid} & \multirow{2}{*}{Acoustic} & Double & 0.2 & 0.2 & -- \\*
 &  & Float & 0.2 & 0.2 & -- \\
\midrule
\multirow{2}{*}{Maxwell (TM)} & \multirow{2}{*}{Acoustic} & Double & 0.2 & 0.2 & -- \\*
 &  & Float & 0.2 & 0.2 & -- \\
\midrule
\multirow{2}{*}{Nonlinear Elasticity} & \multirow{2}{*}{Acoustic} & Double & 0.1 & 0.2 & -- \\*
 &  & Float & 0.15 & 2 & -- \\
\midrule
\multirow{2}{*}{Scalar ADR} & \multirow{2}{*}{Diffusive} & Double & 1.9 & 50 & 6 \\*
 &  & Float & 1.9 & 50 & 6 \\
\midrule
\multirow{2}{*}{Allen--Cahn Phase-Field} & \multirow{2}{*}{Diffusive} & Double & 1.9 & 50 & 6 \\*
 &  & Float & 1.9 & 50 & 6 \\
\midrule
\multirow{2}{*}{Incompressible NSE} & \multirow{2}{*}{Diffusive} & Double & 2 & 50 & 6 \\*
 &  & Float & 2 & 90 & 6 \\
\midrule
\multirow{2}{*}{Compressible NSF} & \multirow{2}{*}{Diffusive} & Double & 2 & 60 & 6 \\*
 &  & Float & 2 & 60 & 6 \\
\midrule
\multirow{2}{*}{Resistive MHD} & \multirow{2}{*}{Acoustic} & Double & 0.025 & 140 & 4 \\*
 &  & Float & 0.025 & 160 & 4 \\
\midrule
\multirow{2}{*}{HCNSF} & \multirow{2}{*}{Acoustic} & Double & 0.05 & 10 & 3 \\*
 &  & Float & 0.05 & 20 & 3 \\
\bottomrule
\end{tabular}
\end{table}

\clearpage
\section{Domain-Specific Language Example}\label{sec:appendix_dsl}

Complementing the compressible Euler specification in the main article, Listing~\ref{lst:dsl_cns} gives the complete DSL specification of the compressible Navier--Stokes--Fourier system declared in SI units.
Two non-dimensionalization steps are declared.
First, characteristic references (the density and the gas constant) eliminate the mass and temperature base dimensions, from which the compiler derives every lattice grid-scaling exponent automatically, with the dynamic viscosity ($\unitPas$) and thermal conductivity ($\unitWmK$) reducing to the diffusive ($\mathrm{length}^2/\mathrm{time}$) scaling.
Second, the realistic-SI magnitudes (density $\qty{1.225}{\kilogram\per\cubic\metre}$, sound speed $\qty{340}{\metre\per\second}$) are declared as characteristic symbols, and the manufactured solution is then authored as (scale)\,$\times$\,(dimensionless), so the value non-dimensionalization divides the scales out through the products and places the working point at the $O(1)$ lattice scale.

\begin{lstlisting}[language=Python, caption={Formulation of the Navier--Stokes--Fourier equations in SI units.}, label={lst:dsl_cns}]
from pde2lbm import *
from sympy.physics.units import (length, time, velocity, mass,
        temperature, energy, pressure, power, momentum)

cns = ConservationLaws(dim=2)

# 1. States and parameters with named SI dimensions
rho  = cns.register_state("rho",  dim=mass/length**3)        # density
rhou = cns.register_state("rhou", shape=2, dim=momentum/length**3)
E    = cns.register_state("E",    dim=energy/length**3)      # energy

gamma = cns.register_parameter("gamma")
mu = cns.register_parameter("mu", dim=pressure*time)         # Pa s
K  = cns.register_parameter("K",  dim=power/(length*temperature))  # W/(m K)
R  = cns.register_parameter("R",  dim=energy/(mass*temperature))   # gas const

# 2a. Eliminate mass and temperature base dimensions
cns.set_characteristic(mass, rho)
cns.set_characteristic(temperature, R)

# 2b. Value non-dimensionalization: realistic SI working scales as characteristic symbols
rho0 = cns.characteristic_symbol(mass,     1.225, name="rho0")  # air at ~1 atm
c_s  = cns.characteristic_symbol(velocity, 340.0, name="c_s")   # sound speed

# 3. Dependent variables
u = cns.register_derived("u", rhou / rho)       # registers u_0, u_1
p = (gamma - 1)*(E - 0.5*rho*u.dot(u))
T = p/(rho*R)                                   # temperature

# 4. Newtonian stress + Fourier flux
gu  = cgrad(u)
tau = mu*(gu + gu.T - (2.0/3.0)*trace(gu)*eye(2))
q   = -K*cgrad(T)

# 5. Equation system
cns.add([
  Eq(dt(rho)  + div(rhou), 0),
  Eq(dt(rhou) + div(outer(rhou,u) + p*eye(2) - tau), 0),
  Eq(dt(E)    + div((E+p)*u - tau*u + q), 0),
])

# 6. Generate OpenLB solver
cns.compile(class_name="CompressibleNSF")
\end{lstlisting}

\clearpage
\allowdisplaybreaks

\section{Fully Expanded Compiled Component Equations}\label{sec:appendix_expanded_pdes}

In this appendix, we present the component-level mathematical expressions of the macroscopic state vectors ($\mathbf{Q}$), physical flux tensors ($\mathbf{\Phi}$), and local source or sink vectors ($\mathbf{S}$) as automatically compiled and executed point-wise on the lattice by our symbolic backend. Each element is listed individually. Compound conserved symbols such as $\rho u_0$ are atomic state components, so ratios like $\rho u_0^2/\rho$ are already in final form: no cancellation against the separate state $\rho$ is available or intended.

\subsection{Inviscid Burgers Equation}\label{sec:appendix_pde_burgers}
The 2D continuous system has a state vector of dimension $N = 1$.

\paragraph{Macroscopic State Components ($Q_i$)}
{\footnotesize
\begin{align*}
Q_{0} = u
\end{align*}
}
\paragraph{Flux Tensor Components ($\Phi_{i,j}$)}
{\footnotesize
\begin{align*}
\Phi_{0,0} &= \frac{u^{2}}{2} \\
\Phi_{0,1} &= \frac{u^{2}}{2}
\end{align*}
}
\paragraph{Source Vector Components ($S_i$)}
All components of the source vector are identically zero: $\mathbf{S}_{\text{burgers}} = \mathbf{0}$.

\subsection{Compressible Euler Equations}\label{sec:appendix_pde_compressible_euler}
The 2D continuous system has a state vector of dimension $N = 4$.

\paragraph{Macroscopic State Components ($Q_i$)}
{\footnotesize
\begin{align*}
Q_{0} = \rho , \quad Q_{1} = \rho u_0 , \quad Q_{2} = \rho u_1 \\
Q_{3} = E
\end{align*}
}
\paragraph{Flux Tensor Components ($\Phi_{i,j}$)}
{\footnotesize
\begin{align*}
\Phi_{0,0} &= \rho u_0 \\
\Phi_{0,1} &= \rho u_1 \\
\Phi_{1,0} &= E \gamma - E - \frac{\gamma \rho u_0^{2}}{2 \rho} - \frac{\gamma \rho u_1^{2}}{2 \rho} + \frac{3 \rho u_0^{2}}{2 \rho} \\
&\quad + \frac{\rho u_1^{2}}{2 \rho} \\
\Phi_{1,1} &= \frac{\rho u_0 \rho u_1}{\rho} \\
\Phi_{2,0} &= \frac{\rho u_0 \rho u_1}{\rho} \\
\Phi_{2,1} &= E \gamma - E - \frac{\gamma \rho u_0^{2}}{2 \rho} - \frac{\gamma \rho u_1^{2}}{2 \rho} + \frac{\rho u_0^{2}}{2 \rho} \\
&\quad + \frac{3 \rho u_1^{2}}{2 \rho} \\
\Phi_{3,0} &= \frac{E \gamma \rho u_0}{\rho} - \frac{\gamma \rho u_0^{3}}{2 \rho^{2}} - \frac{\gamma \rho u_0 \rho u_1^{2}}{2 \rho^{2}} \\
&\quad + \frac{\rho u_0^{3}}{2 \rho^{2}} + \frac{\rho u_0 \rho u_1^{2}}{2 \rho^{2}} \\
\Phi_{3,1} &= \frac{E \gamma \rho u_1}{\rho} - \frac{\gamma \rho u_0^{2} \rho u_1}{2 \rho^{2}} - \frac{\gamma \rho u_1^{3}}{2 \rho^{2}} \\
&\quad + \frac{\rho u_0^{2} \rho u_1}{2 \rho^{2}} + \frac{\rho u_1^{3}}{2 \rho^{2}}
\end{align*}
}
\paragraph{Source Vector Components ($S_i$)}
All components of the source vector are identically zero: $\mathbf{S}_{\text{compressible\_euler}} = \mathbf{0}$.

\subsection{Shallow Water Equations}\label{sec:appendix_pde_shallow_water}
The 2D continuous system has a state vector of dimension $N = 3$.

\paragraph{Macroscopic State Components ($Q_i$)}
{\footnotesize
\begin{align*}
Q_{0} = h , \quad Q_{1} = h u_0 , \quad Q_{2} = h u_1
\end{align*}
}
\paragraph{Flux Tensor Components ($\Phi_{i,j}$)}
{\footnotesize
\begin{align*}
\Phi_{0,0} &= h u_0 \\
\Phi_{0,1} &= h u_1 \\
\Phi_{1,0} &= \frac{g h^{2}}{2} + \frac{h u_0^{2}}{h} \\
\Phi_{1,1} &= \frac{h u_0 h u_1}{h} \\
\Phi_{2,0} &= \frac{h u_0 h u_1}{h} \\
\Phi_{2,1} &= \frac{g h^{2}}{2} + \frac{h u_1^{2}}{h}
\end{align*}
}
\paragraph{Source Vector Components ($S_i$)}
{\footnotesize
\begin{align*}
S_{0} &= 0 \\
S_{1} &= - \frac{\pi g h \cos{(2 \pi x_0 )} \cos{(2 \pi x_1 )}}{50} \\
S_{2} &= \frac{\pi g h \sin{(2 \pi x_0 )} \sin{(2 \pi x_1 )}}{50}
\end{align*}
}

\subsection{Ideal Ultrarelativistic Fluid}\label{sec:appendix_pde_relativistic_hydro}
The 2D continuous system has a state vector of dimension $N = 3$.

\paragraph{Macroscopic State Components ($Q_i$)}
{\footnotesize
\begin{align*}
Q_{0} = E , \quad Q_{1} = S_0 , \quad Q_{2} = S_1
\end{align*}
}
\paragraph{Flux Tensor Components ($\Phi_{i,j}$)}
{\footnotesize
\begin{align*}
\Phi_{0,0} &= S_0 \\
\Phi_{0,1} &= S_1 \\
\Phi_{1,0} &= - \frac{E}{3} + \frac{3 S_0^{2}}{2 E + \sqrt{4 E^{2} - 3 S_0^{2} - 3 S_1^{2}}} + \frac{\sqrt{4 E^{2} - 3 S_0^{2} - 3 S_1^{2}}}{3} \\
\Phi_{1,1} &= \frac{3 S_0 S_1}{2 E + \sqrt{4 E^{2} - 3 S_0^{2} - 3 S_1^{2}}} \\
\Phi_{2,0} &= \frac{3 S_0 S_1}{2 E + \sqrt{4 E^{2} - 3 S_0^{2} - 3 S_1^{2}}} \\
\Phi_{2,1} &= - \frac{E}{3} + \frac{3 S_1^{2}}{2 E + \sqrt{4 E^{2} - 3 S_0^{2} - 3 S_1^{2}}} + \frac{\sqrt{4 E^{2} - 3 S_0^{2} - 3 S_1^{2}}}{3}
\end{align*}
}
\paragraph{Source Vector Components ($S_i$)}
All components of the source vector are identically zero: $\mathbf{S}_{\text{relativistic\_hydro}} = \mathbf{0}$.

\subsection{Maxwell Electromagnetics (TM Mode)}\label{sec:appendix_pde_maxwell}
The 2D continuous system has a state vector of dimension $N = 3$.

\paragraph{Macroscopic State Components ($Q_i$)}
{\footnotesize
\begin{align*}
Q_{0} = E_z , \quad Q_{1} = H_x , \quad Q_{2} = H_y
\end{align*}
}
\paragraph{Flux Tensor Components ($\Phi_{i,j}$)}
{\footnotesize
\begin{align*}
\Phi_{0,0} &= - H_y c \\
\Phi_{0,1} &= H_x c \\
\Phi_{1,0} &= 0 \\
\Phi_{1,1} &= E_z c \\
\Phi_{2,0} &= - E_z c \\
\Phi_{2,1} &= 0
\end{align*}
}
\paragraph{Source Vector Components ($S_i$)}
All components of the source vector are identically zero: $\mathbf{S}_{\text{maxwell}} = \mathbf{0}$.

\subsection{Nonlinear Finite-Strain Elasticity}\label{sec:appendix_pde_nonlinear_elasticity}
The 2D continuous system has a state vector of dimension $N = 6$.

\paragraph{Macroscopic State Components ($Q_i$)}
{\footnotesize
\begin{align*}
Q_{0} = F_{00} , \quad Q_{1} = F_{01} , \quad Q_{2} = F_{10} \\
Q_{3} = F_{11} , \quad Q_{4} = v_0 , \quad Q_{5} = v_1
\end{align*}
}
\paragraph{Flux Tensor Components ($\Phi_{i,j}$)}
with the shared subexpressions
{\footnotesize
\begin{align*}
\chi_{0} &= - v_0 \\
\chi_{1} &= - v_1 \\
\chi_{2} &= \frac{\mu_s}{\rho_{0}} \\
\chi_{3} &= F_{00} F_{11} \\
\chi_{4} &= F_{01} F_{10} \\
\chi_{5} &= \frac{1}{\chi_{3} \rho_{0} - \chi_{4} \rho_{0}} \\
\chi_{6} &= \log{(\chi_{3} - \chi_{4} )} \\
\chi_{7} &= \chi_{5} \chi_{6} \lambda_s \\
\chi_{8} &= \chi_{5} \mu_s
\end{align*}
}
the flux components evaluate to
{\footnotesize
\begin{align*}
\Phi_{0,0} &= \chi_{0} \\
\Phi_{0,1} &= 0 \\
\Phi_{1,0} &= 0 \\
\Phi_{1,1} &= \chi_{0} \\
\Phi_{2,0} &= \chi_{1} \\
\Phi_{2,1} &= 0 \\
\Phi_{3,0} &= 0 \\
\Phi_{3,1} &= \chi_{1} \\
\Phi_{4,0} &= - F_{00} \chi_{2} + F_{11} \chi_{5} \mu_s - F_{11} \chi_{7} \\
\Phi_{4,1} &= - F_{01} \chi_{2} + F_{10} \chi_{5} \chi_{6} \lambda_s - F_{10} \chi_{8} \\
\Phi_{5,0} &= F_{01} \chi_{5} \chi_{6} \lambda_s - F_{01} \chi_{8} - F_{10} \chi_{2} \\
\Phi_{5,1} &= F_{00} \chi_{5} \mu_s - F_{00} \chi_{7} - F_{11} \chi_{2}
\end{align*}
}
\paragraph{Source Vector Components ($S_i$)}
All components of the source vector are identically zero: $\mathbf{S}_{\text{nonlinear\_elasticity}} = \mathbf{0}$.

\subsection{Scalar Advection-Diffusion-Reaction}\label{sec:appendix_pde_adr}
The 2D continuous system has a state vector of dimension $N = 3$.

\paragraph{Macroscopic State Components ($Q_i$)}
{\footnotesize
\begin{align*}
Q_{0} = c , \quad Q_{1} = G_{c,0} , \quad Q_{2} = G_{c,1}
\end{align*}
}
\paragraph{Flux Tensor Components ($\Phi_{i,j}$)}
{\footnotesize
\begin{align*}
\Phi_{0,0} &= - D G_{c,0} + c u_0 \\
\Phi_{0,1} &= - D G_{c,1} + c u_1 \\
\Phi_{1,0} &= G_{c,0} u_0 - \frac{c}{\tau_{\nabla}} \\
\Phi_{1,1} &= G_{c,0} u_1 \\
\Phi_{2,0} &= G_{c,1} u_0 \\
\Phi_{2,1} &= G_{c,1} u_1 - \frac{c}{\tau_{\nabla}}
\end{align*}
}
\paragraph{Source Vector Components ($S_i$)}
{\footnotesize
\begin{align*}
S_{0} &= - R c^{2} + R c \\
S_{1} &= - \frac{G_{c,0}}{\tau_{\nabla}} \\
S_{2} &= - \frac{G_{c,1}}{\tau_{\nabla}}
\end{align*}
}

\subsection{Allen--Cahn Phase-Field Equation}\label{sec:appendix_pde_allen_cahn}
The 2D continuous system has a state vector of dimension $N = 3$.

\paragraph{Macroscopic State Components ($Q_i$)}
{\footnotesize
\begin{align*}
Q_{0} = \phi , \quad Q_{1} = G_{\phi,0} , \quad Q_{2} = G_{\phi,1}
\end{align*}
}
\paragraph{Flux Tensor Components ($\Phi_{i,j}$)}
{\footnotesize
\begin{align*}
\Phi_{0,0} &= - G_{\phi,0} \gamma \\
\Phi_{0,1} &= - G_{\phi,1} \gamma \\
\Phi_{1,0} &= - \frac{\phi}{\tau_{\nabla}} \\
\Phi_{1,1} &= 0 \\
\Phi_{2,0} &= 0 \\
\Phi_{2,1} &= - \frac{\phi}{\tau_{\nabla}}
\end{align*}
}
\paragraph{Source Vector Components ($S_i$)}
{\footnotesize
\begin{align*}
S_{0} &= - \phi^{3} r + \phi r \\
S_{1} &= - \frac{G_{\phi,0}}{\tau_{\nabla}} \\
S_{2} &= - \frac{G_{\phi,1}}{\tau_{\nabla}}
\end{align*}
}

\subsection{Weakly Compressible Navier--Stokes Equations (Incompressible Limit)}\label{sec:appendix_pde_navier_stokes}
The 2D continuous system has a state vector of dimension $N = 7$.

\paragraph{Macroscopic State Components ($Q_i$)}
{\footnotesize
\begin{align*}
Q_{0} = \rho , \quad Q_{1} = \rho u_0 , \quad Q_{2} = \rho u_1 \\
Q_{3} = G_{\rho u, 0, 0} , \quad Q_{4} = G_{\rho u, 0, 1} , \quad Q_{5} = G_{\rho u, 1, 0} \\
Q_{6} = G_{\rho u, 1, 1}
\end{align*}
}
\paragraph{Flux Tensor Components ($\Phi_{i,j}$)}
{\footnotesize
\begin{align*}
\Phi_{0,0} &= \rho u_0 \\
\Phi_{0,1} &= \rho u_1 \\
\Phi_{1,0} &= - \frac{2 G_{\rho u, 0, 0} \mu}{\rho} + c_{s,\text{phys}}^{2} \rho + \frac{\rho u_0^{2}}{\rho} \\
\Phi_{1,1} &= - \frac{G_{\rho u, 0, 1} \mu}{\rho} - \frac{G_{\rho u, 1, 0} \mu}{\rho} + \frac{\rho u_0 \rho u_1}{\rho} \\
\Phi_{2,0} &= - \frac{G_{\rho u, 0, 1} \mu}{\rho} - \frac{G_{\rho u, 1, 0} \mu}{\rho} + \frac{\rho u_0 \rho u_1}{\rho} \\
\Phi_{2,1} &= - \frac{2 G_{\rho u, 1, 1} \mu}{\rho} + c_{s,\text{phys}}^{2} \rho + \frac{\rho u_1^{2}}{\rho} \\
\Phi_{3,0} &= \frac{G_{\rho u, 0, 0} \rho u_0}{\rho} - \frac{\rho u_0}{\tau_{\nabla}} \\
\Phi_{3,1} &= \frac{G_{\rho u, 0, 0} \rho u_1}{\rho} \\
\Phi_{4,0} &= \frac{G_{\rho u, 0, 1} \rho u_0}{\rho} \\
\Phi_{4,1} &= \frac{G_{\rho u, 0, 1} \rho u_1}{\rho} - \frac{\rho u_0}{\tau_{\nabla}} \\
\Phi_{5,0} &= \frac{G_{\rho u, 1, 0} \rho u_0}{\rho} - \frac{\rho u_1}{\tau_{\nabla}} \\
\Phi_{5,1} &= \frac{G_{\rho u, 1, 0} \rho u_1}{\rho} \\
\Phi_{6,0} &= \frac{G_{\rho u, 1, 1} \rho u_0}{\rho} \\
\Phi_{6,1} &= \frac{G_{\rho u, 1, 1} \rho u_1}{\rho} - \frac{\rho u_1}{\tau_{\nabla}}
\end{align*}
}
\paragraph{Source Vector Components ($S_i$)}
{\footnotesize
\begin{align*}
S_{0} &= 0 \\
S_{1} &= 0 \\
S_{2} &= 0 \\
S_{3} &= - \frac{G_{\rho u, 0, 0}}{\tau_{\nabla}} \\
S_{4} &= - \frac{G_{\rho u, 0, 1}}{\tau_{\nabla}} \\
S_{5} &= - \frac{G_{\rho u, 1, 0}}{\tau_{\nabla}} \\
S_{6} &= - \frac{G_{\rho u, 1, 1}}{\tau_{\nabla}}
\end{align*}
}

\subsection{Compressible Navier--Stokes--Fourier Equations}\label{sec:appendix_pde_compressible_navier_stokes}
The 2D continuous system has a state vector of dimension $N = 12$.

\paragraph{Macroscopic State Components ($Q_i$)}
{\footnotesize
\begin{align*}
Q_{0} = \rho , \quad Q_{1} = \rho u_0 , \quad Q_{2} = \rho u_1 \\
Q_{3} = E , \quad Q_{4} = G_{\rho u, 1, 0} , \quad Q_{5} = G_{\rho u, 1, 1} \\
Q_{6} = G_{E,0} , \quad Q_{7} = G_{E,1} , \quad Q_{8} = G_{\rho,0} \\
Q_{9} = G_{\rho,1} , \quad Q_{10} = G_{\rho u, 0, 0} , \quad Q_{11} = G_{\rho u, 0, 1}
\end{align*}
}
\paragraph{Flux Tensor Components ($\Phi_{i,j}$)}
{\footnotesize
\begin{align*}
\Phi_{0,0} &= \rho u_0 \\
\Phi_{0,1} &= \rho u_1 \\
\Phi_{1,0} &= E \gamma - E + \frac{4 G_{\rho,0} \mu \rho u_0}{3 \rho^{2}} - \frac{2 G_{\rho,1} \mu \rho u_1}{3 \rho^{2}} \\
&\quad - \frac{4 G_{\rho u, 0, 0} \mu}{3 \rho} + \frac{2 G_{\rho u, 1, 1} \mu}{3 \rho} - \frac{\gamma \rho u_0^{2}}{2 \rho} \\
&\quad - \frac{\gamma \rho u_1^{2}}{2 \rho} + \frac{3 \rho u_0^{2}}{2 \rho} + \frac{\rho u_1^{2}}{2 \rho} \\
\Phi_{1,1} &= \frac{G_{\rho,0} \mu \rho u_1}{\rho^{2}} + \frac{G_{\rho,1} \mu \rho u_0}{\rho^{2}} - \frac{G_{\rho u, 0, 1} \mu}{\rho} \\
&\quad - \frac{G_{\rho u, 1, 0} \mu}{\rho} + \frac{\rho u_0 \rho u_1}{\rho} \\
\Phi_{2,0} &= \frac{G_{\rho,0} \mu \rho u_1}{\rho^{2}} + \frac{G_{\rho,1} \mu \rho u_0}{\rho^{2}} - \frac{G_{\rho u, 0, 1} \mu}{\rho} \\
&\quad - \frac{G_{\rho u, 1, 0} \mu}{\rho} + \frac{\rho u_0 \rho u_1}{\rho} \\
\Phi_{2,1} &= E \gamma - E - \frac{2 G_{\rho,0} \mu \rho u_0}{3 \rho^{2}} + \frac{4 G_{\rho,1} \mu \rho u_1}{3 \rho^{2}} \\
&\quad + \frac{2 G_{\rho u, 0, 0} \mu}{3 \rho} - \frac{4 G_{\rho u, 1, 1} \mu}{3 \rho} - \frac{\gamma \rho u_0^{2}}{2 \rho} \\
&\quad - \frac{\gamma \rho u_1^{2}}{2 \rho} + \frac{\rho u_0^{2}}{2 \rho} + \frac{3 \rho u_1^{2}}{2 \rho} \\
\Phi_{3,0} &= \frac{E G_{\rho,0} K \gamma}{R \rho^{2}} - \frac{E G_{\rho,0} K}{R \rho^{2}} + \frac{E \gamma \rho u_0}{\rho} \\
&\quad - \frac{G_{E,0} K \gamma}{R \rho} + \frac{G_{E,0} K}{R \rho} - \frac{G_{\rho,0} K \gamma \rho u_0^{2}}{R \rho^{3}} \\
&\quad - \frac{G_{\rho,0} K \gamma \rho u_1^{2}}{R \rho^{3}} + \frac{G_{\rho,0} K \rho u_0^{2}}{R \rho^{3}} \\
&\quad + \frac{G_{\rho,0} K \rho u_1^{2}}{R \rho^{3}} + \frac{4 G_{\rho,0} \mu \rho u_0^{2}}{3 \rho^{3}} \\
&\quad + \frac{G_{\rho,0} \mu \rho u_1^{2}}{\rho^{3}} + \frac{G_{\rho,1} \mu \rho u_0 \rho u_1}{3 \rho^{3}} \\
&\quad + \frac{G_{\rho u, 0, 0} K \gamma \rho u_0}{R \rho^{2}} - \frac{G_{\rho u, 0, 0} K \rho u_0}{R \rho^{2}} \\
&\quad - \frac{4 G_{\rho u, 0, 0} \mu \rho u_0}{3 \rho^{2}} - \frac{G_{\rho u, 0, 1} \mu \rho u_1}{\rho^{2}} \\
&\quad + \frac{G_{\rho u, 1, 0} K \gamma \rho u_1}{R \rho^{2}} - \frac{G_{\rho u, 1, 0} K \rho u_1}{R \rho^{2}} \\
&\quad - \frac{G_{\rho u, 1, 0} \mu \rho u_1}{\rho^{2}} + \frac{2 G_{\rho u, 1, 1} \mu \rho u_0}{3 \rho^{2}} \\
&\quad - \frac{\gamma \rho u_0^{3}}{2 \rho^{2}} - \frac{\gamma \rho u_0 \rho u_1^{2}}{2 \rho^{2}} + \frac{\rho u_0^{3}}{2 \rho^{2}} \\
&\quad + \frac{\rho u_0 \rho u_1^{2}}{2 \rho^{2}} \\
\Phi_{3,1} &= \frac{E G_{\rho,1} K \gamma}{R \rho^{2}} - \frac{E G_{\rho,1} K}{R \rho^{2}} + \frac{E \gamma \rho u_1}{\rho} \\
&\quad - \frac{G_{E,1} K \gamma}{R \rho} + \frac{G_{E,1} K}{R \rho} + \frac{G_{\rho,0} \mu \rho u_0 \rho u_1}{3 \rho^{3}} \\
&\quad - \frac{G_{\rho,1} K \gamma \rho u_0^{2}}{R \rho^{3}} - \frac{G_{\rho,1} K \gamma \rho u_1^{2}}{R \rho^{3}} \\
&\quad + \frac{G_{\rho,1} K \rho u_0^{2}}{R \rho^{3}} + \frac{G_{\rho,1} K \rho u_1^{2}}{R \rho^{3}} \\
&\quad + \frac{G_{\rho,1} \mu \rho u_0^{2}}{\rho^{3}} + \frac{4 G_{\rho,1} \mu \rho u_1^{2}}{3 \rho^{3}} \\
&\quad + \frac{2 G_{\rho u, 0, 0} \mu \rho u_1}{3 \rho^{2}} + \frac{G_{\rho u, 0, 1} K \gamma \rho u_0}{R \rho^{2}} \\
&\quad - \frac{G_{\rho u, 0, 1} K \rho u_0}{R \rho^{2}} - \frac{G_{\rho u, 0, 1} \mu \rho u_0}{\rho^{2}} \\
&\quad - \frac{G_{\rho u, 1, 0} \mu \rho u_0}{\rho^{2}} + \frac{G_{\rho u, 1, 1} K \gamma \rho u_1}{R \rho^{2}} \\
&\quad - \frac{G_{\rho u, 1, 1} K \rho u_1}{R \rho^{2}} - \frac{4 G_{\rho u, 1, 1} \mu \rho u_1}{3 \rho^{2}} \\
&\quad - \frac{\gamma \rho u_0^{2} \rho u_1}{2 \rho^{2}} - \frac{\gamma \rho u_1^{3}}{2 \rho^{2}} \\
&\quad + \frac{\rho u_0^{2} \rho u_1}{2 \rho^{2}} + \frac{\rho u_1^{3}}{2 \rho^{2}} \\
\Phi_{4,0} &= \frac{G_{\rho u, 1, 0} \rho u_0}{\rho} - \frac{\rho u_1}{\tau_{\nabla}} \\
\Phi_{4,1} &= \frac{G_{\rho u, 1, 0} \rho u_1}{\rho} \\
\Phi_{5,0} &= \frac{G_{\rho u, 1, 1} \rho u_0}{\rho} \\
\Phi_{5,1} &= \frac{G_{\rho u, 1, 1} \rho u_1}{\rho} - \frac{\rho u_1}{\tau_{\nabla}} \\
\Phi_{6,0} &= - \frac{E}{\tau_{\nabla}} + \frac{G_{E,0} \rho u_0}{\rho} \\
\Phi_{6,1} &= \frac{G_{E,0} \rho u_1}{\rho} \\
\Phi_{7,0} &= \frac{G_{E,1} \rho u_0}{\rho} \\
\Phi_{7,1} &= - \frac{E}{\tau_{\nabla}} + \frac{G_{E,1} \rho u_1}{\rho} \\
\Phi_{8,0} &= \frac{G_{\rho,0} \rho u_0}{\rho} - \frac{\rho}{\tau_{\nabla}} \\
\Phi_{8,1} &= \frac{G_{\rho,0} \rho u_1}{\rho} \\
\Phi_{9,0} &= \frac{G_{\rho,1} \rho u_0}{\rho} \\
\Phi_{9,1} &= \frac{G_{\rho,1} \rho u_1}{\rho} - \frac{\rho}{\tau_{\nabla}} \\
\Phi_{10,0} &= \frac{G_{\rho u, 0, 0} \rho u_0}{\rho} - \frac{\rho u_0}{\tau_{\nabla}} \\
\Phi_{10,1} &= \frac{G_{\rho u, 0, 0} \rho u_1}{\rho} \\
\Phi_{11,0} &= \frac{G_{\rho u, 0, 1} \rho u_0}{\rho} \\
\Phi_{11,1} &= \frac{G_{\rho u, 0, 1} \rho u_1}{\rho} - \frac{\rho u_0}{\tau_{\nabla}}
\end{align*}
}
\paragraph{Source Vector Components ($S_i$)}
{\footnotesize
\begin{align*}
S_{0} &= 0 \\
S_{1} &= 0 \\
S_{2} &= 0 \\
S_{3} &= 0 \\
S_{4} &= - \frac{G_{\rho u, 1, 0}}{\tau_{\nabla}} \\
S_{5} &= - \frac{G_{\rho u, 1, 1}}{\tau_{\nabla}} \\
S_{6} &= - \frac{G_{E,0}}{\tau_{\nabla}} \\
S_{7} &= - \frac{G_{E,1}}{\tau_{\nabla}} \\
S_{8} &= - \frac{G_{\rho,0}}{\tau_{\nabla}} \\
S_{9} &= - \frac{G_{\rho,1}}{\tau_{\nabla}} \\
S_{10} &= - \frac{G_{\rho u, 0, 0}}{\tau_{\nabla}} \\
S_{11} &= - \frac{G_{\rho u, 0, 1}}{\tau_{\nabla}}
\end{align*}
}

\subsection{Resistive Compressible Magnetohydrodynamics}\label{sec:appendix_pde_mhd}
The 2D continuous system has a state vector of dimension $N = 18$.

\paragraph{Macroscopic State Components ($Q_i$)}
{\footnotesize
\begin{align*}
Q_{0} = \rho , \quad Q_{1} = \rho u_0 , \quad Q_{2} = \rho u_1 \\
Q_{3} = E , \quad Q_{4} = B_0 , \quad Q_{5} = B_1 \\
Q_{6} = G_{\rho,0} , \quad Q_{7} = G_{\rho,1} , \quad Q_{8} = G_{\rho u, 1, 0} \\
Q_{9} = G_{\rho u, 1, 1} , \quad Q_{10} = G_{E,0} , \quad Q_{11} = G_{E,1} \\
Q_{12} = G_{B, 1, 0} , \quad Q_{13} = G_{B, 1, 1} , \quad Q_{14} = G_{\rho u, 0, 0} \\
Q_{15} = G_{\rho u, 0, 1} , \quad Q_{16} = G_{B, 0, 0} , \quad Q_{17} = G_{B, 0, 1}
\end{align*}
}
\paragraph{Flux Tensor Components ($\Phi_{i,j}$)}
{\footnotesize
\begin{align*}
\Phi_{0,0} &= \rho u_0 \\
\Phi_{0,1} &= \rho u_1 \\
\Phi_{1,0} &= - \frac{B_0^{2} \gamma}{2} - \frac{B_1^{2} \gamma}{2} + B_1^{2} + E \gamma - E + \frac{4 G_{\rho,0} \mu \rho u_0}{3 \rho^{2}} \\
&\quad - \frac{2 G_{\rho,1} \mu \rho u_1}{3 \rho^{2}} - \frac{4 G_{\rho u, 0, 0} \mu}{3 \rho} + \frac{2 G_{\rho u, 1, 1} \mu}{3 \rho} \\
&\quad - \frac{\gamma \rho u_0^{2}}{2 \rho} - \frac{\gamma \rho u_1^{2}}{2 \rho} + \frac{3 \rho u_0^{2}}{2 \rho} \\
&\quad + \frac{\rho u_1^{2}}{2 \rho} \\
\Phi_{1,1} &= - B_0 B_1 + \frac{G_{\rho,0} \mu \rho u_1}{\rho^{2}} + \frac{G_{\rho,1} \mu \rho u_0}{\rho^{2}} \\
&\quad - \frac{G_{\rho u, 0, 1} \mu}{\rho} - \frac{G_{\rho u, 1, 0} \mu}{\rho} + \frac{\rho u_0 \rho u_1}{\rho} \\
\Phi_{2,0} &= - B_0 B_1 + \frac{G_{\rho,0} \mu \rho u_1}{\rho^{2}} + \frac{G_{\rho,1} \mu \rho u_0}{\rho^{2}} \\
&\quad - \frac{G_{\rho u, 0, 1} \mu}{\rho} - \frac{G_{\rho u, 1, 0} \mu}{\rho} + \frac{\rho u_0 \rho u_1}{\rho} \\
\Phi_{2,1} &= - \frac{B_0^{2} \gamma}{2} + B_0^{2} - \frac{B_1^{2} \gamma}{2} + E \gamma - E - \frac{2 G_{\rho,0} \mu \rho u_0}{3 \rho^{2}} \\
&\quad + \frac{4 G_{\rho,1} \mu \rho u_1}{3 \rho^{2}} + \frac{2 G_{\rho u, 0, 0} \mu}{3 \rho} - \frac{4 G_{\rho u, 1, 1} \mu}{3 \rho} \\
&\quad - \frac{\gamma \rho u_0^{2}}{2 \rho} - \frac{\gamma \rho u_1^{2}}{2 \rho} + \frac{\rho u_0^{2}}{2 \rho} \\
&\quad + \frac{3 \rho u_1^{2}}{2 \rho} \\
\Phi_{3,0} &= - \frac{B_0^{2} G_{\rho,0} K \gamma}{2 \rho^{2}} + \frac{B_0^{2} G_{\rho,0} K}{2 \rho^{2}} \\
&\quad - \frac{B_0^{2} \gamma \rho u_0}{2 \rho} - \frac{B_0 B_1 \rho u_1}{\rho} + \frac{B_0 G_{B, 0, 0} K \gamma}{\rho} \\
&\quad - \frac{B_0 G_{B, 0, 0} K}{\rho} - \frac{B_1^{2} G_{\rho,0} K \gamma}{2 \rho^{2}} + \frac{B_1^{2} G_{\rho,0} K}{2 \rho^{2}} \\
&\quad - \frac{B_1^{2} \gamma \rho u_0}{2 \rho} + \frac{B_1^{2} \rho u_0}{\rho} + B_1 G_{B, 0, 1} \eta \\
&\quad + \frac{B_1 G_{B, 1, 0} K \gamma}{\rho} - \frac{B_1 G_{B, 1, 0} K}{\rho} - B_1 G_{B, 1, 0} \eta \\
&\quad + \frac{E G_{\rho,0} K \gamma}{\rho^{2}} - \frac{E G_{\rho,0} K}{\rho^{2}} + \frac{E \gamma \rho u_0}{\rho} \\
&\quad - \frac{G_{E,0} K \gamma}{\rho} + \frac{G_{E,0} K}{\rho} - \frac{G_{\rho,0} K \gamma \rho u_0^{2}}{\rho^{3}} \\
&\quad - \frac{G_{\rho,0} K \gamma \rho u_1^{2}}{\rho^{3}} + \frac{G_{\rho,0} K \rho u_0^{2}}{\rho^{3}} \\
&\quad + \frac{G_{\rho,0} K \rho u_1^{2}}{\rho^{3}} + \frac{4 G_{\rho,0} \mu \rho u_0^{2}}{3 \rho^{3}} \\
&\quad + \frac{G_{\rho,0} \mu \rho u_1^{2}}{\rho^{3}} + \frac{G_{\rho,1} \mu \rho u_0 \rho u_1}{3 \rho^{3}} \\
&\quad + \frac{G_{\rho u, 0, 0} K \gamma \rho u_0}{\rho^{2}} - \frac{G_{\rho u, 0, 0} K \rho u_0}{\rho^{2}} \\
&\quad - \frac{4 G_{\rho u, 0, 0} \mu \rho u_0}{3 \rho^{2}} - \frac{G_{\rho u, 0, 1} \mu \rho u_1}{\rho^{2}} \\
&\quad + \frac{G_{\rho u, 1, 0} K \gamma \rho u_1}{\rho^{2}} - \frac{G_{\rho u, 1, 0} K \rho u_1}{\rho^{2}} \\
&\quad - \frac{G_{\rho u, 1, 0} \mu \rho u_1}{\rho^{2}} + \frac{2 G_{\rho u, 1, 1} \mu \rho u_0}{3 \rho^{2}} \\
&\quad - \frac{\gamma \rho u_0^{3}}{2 \rho^{2}} - \frac{\gamma \rho u_0 \rho u_1^{2}}{2 \rho^{2}} + \frac{\rho u_0^{3}}{2 \rho^{2}} \\
&\quad + \frac{\rho u_0 \rho u_1^{2}}{2 \rho^{2}} \\
\Phi_{3,1} &= - \frac{B_0^{2} G_{\rho,1} K \gamma}{2 \rho^{2}} + \frac{B_0^{2} G_{\rho,1} K}{2 \rho^{2}} \\
&\quad - \frac{B_0^{2} \gamma \rho u_1}{2 \rho} + \frac{B_0^{2} \rho u_1}{\rho} - \frac{B_0 B_1 \rho u_0}{\rho} \\
&\quad + \frac{B_0 G_{B, 0, 1} K \gamma}{\rho} - \frac{B_0 G_{B, 0, 1} K}{\rho} - B_0 G_{B, 0, 1} \eta + B_0 G_{B, 1, 0} \eta \\
&\quad - \frac{B_1^{2} G_{\rho,1} K \gamma}{2 \rho^{2}} + \frac{B_1^{2} G_{\rho,1} K}{2 \rho^{2}} \\
&\quad - \frac{B_1^{2} \gamma \rho u_1}{2 \rho} + \frac{B_1 G_{B, 1, 1} K \gamma}{\rho} - \frac{B_1 G_{B, 1, 1} K}{\rho} \\
&\quad + \frac{E G_{\rho,1} K \gamma}{\rho^{2}} - \frac{E G_{\rho,1} K}{\rho^{2}} + \frac{E \gamma \rho u_1}{\rho} \\
&\quad - \frac{G_{E,1} K \gamma}{\rho} + \frac{G_{E,1} K}{\rho} + \frac{G_{\rho,0} \mu \rho u_0 \rho u_1}{3 \rho^{3}} \\
&\quad - \frac{G_{\rho,1} K \gamma \rho u_0^{2}}{\rho^{3}} - \frac{G_{\rho,1} K \gamma \rho u_1^{2}}{\rho^{3}} \\
&\quad + \frac{G_{\rho,1} K \rho u_0^{2}}{\rho^{3}} + \frac{G_{\rho,1} K \rho u_1^{2}}{\rho^{3}} \\
&\quad + \frac{G_{\rho,1} \mu \rho u_0^{2}}{\rho^{3}} + \frac{4 G_{\rho,1} \mu \rho u_1^{2}}{3 \rho^{3}} \\
&\quad + \frac{2 G_{\rho u, 0, 0} \mu \rho u_1}{3 \rho^{2}} + \frac{G_{\rho u, 0, 1} K \gamma \rho u_0}{\rho^{2}} \\
&\quad - \frac{G_{\rho u, 0, 1} K \rho u_0}{\rho^{2}} - \frac{G_{\rho u, 0, 1} \mu \rho u_0}{\rho^{2}} \\
&\quad - \frac{G_{\rho u, 1, 0} \mu \rho u_0}{\rho^{2}} + \frac{G_{\rho u, 1, 1} K \gamma \rho u_1}{\rho^{2}} \\
&\quad - \frac{G_{\rho u, 1, 1} K \rho u_1}{\rho^{2}} - \frac{4 G_{\rho u, 1, 1} \mu \rho u_1}{3 \rho^{2}} \\
&\quad - \frac{\gamma \rho u_0^{2} \rho u_1}{2 \rho^{2}} - \frac{\gamma \rho u_1^{3}}{2 \rho^{2}} \\
&\quad + \frac{\rho u_0^{2} \rho u_1}{2 \rho^{2}} + \frac{\rho u_1^{3}}{2 \rho^{2}} \\
\Phi_{4,0} &= - G_{B, 0, 0} \eta \\
\Phi_{4,1} &= \frac{B_0 \rho u_1}{\rho} - \frac{B_1 \rho u_0}{\rho} - G_{B, 0, 1} \eta \\
\Phi_{5,0} &= - \frac{B_0 \rho u_1}{\rho} + \frac{B_1 \rho u_0}{\rho} - G_{B, 1, 0} \eta \\
\Phi_{5,1} &= - G_{B, 1, 1} \eta \\
\Phi_{6,0} &= \frac{G_{\rho,0} \rho u_0}{\rho} - \frac{\rho}{\tau_{\nabla}} \\
\Phi_{6,1} &= \frac{G_{\rho,0} \rho u_1}{\rho} \\
\Phi_{7,0} &= \frac{G_{\rho,1} \rho u_0}{\rho} \\
\Phi_{7,1} &= \frac{G_{\rho,1} \rho u_1}{\rho} - \frac{\rho}{\tau_{\nabla}} \\
\Phi_{8,0} &= \frac{G_{\rho u, 1, 0} \rho u_0}{\rho} - \frac{\rho u_1}{\tau_{\nabla}} \\
\Phi_{8,1} &= \frac{G_{\rho u, 1, 0} \rho u_1}{\rho} \\
\Phi_{9,0} &= \frac{G_{\rho u, 1, 1} \rho u_0}{\rho} \\
\Phi_{9,1} &= \frac{G_{\rho u, 1, 1} \rho u_1}{\rho} - \frac{\rho u_1}{\tau_{\nabla}} \\
\Phi_{10,0} &= - \frac{E}{\tau_{\nabla}} + \frac{G_{E,0} \rho u_0}{\rho} \\
\Phi_{10,1} &= \frac{G_{E,0} \rho u_1}{\rho} \\
\Phi_{11,0} &= \frac{G_{E,1} \rho u_0}{\rho} \\
\Phi_{11,1} &= - \frac{E}{\tau_{\nabla}} + \frac{G_{E,1} \rho u_1}{\rho} \\
\Phi_{12,0} &= - \frac{B_1}{\tau_{\nabla}} + \frac{G_{B, 1, 0} \rho u_0}{\rho} \\
\Phi_{12,1} &= \frac{G_{B, 1, 0} \rho u_1}{\rho} \\
\Phi_{13,0} &= \frac{G_{B, 1, 1} \rho u_0}{\rho} \\
\Phi_{13,1} &= - \frac{B_1}{\tau_{\nabla}} + \frac{G_{B, 1, 1} \rho u_1}{\rho} \\
\Phi_{14,0} &= \frac{G_{\rho u, 0, 0} \rho u_0}{\rho} - \frac{\rho u_0}{\tau_{\nabla}} \\
\Phi_{14,1} &= \frac{G_{\rho u, 0, 0} \rho u_1}{\rho} \\
\Phi_{15,0} &= \frac{G_{\rho u, 0, 1} \rho u_0}{\rho} \\
\Phi_{15,1} &= \frac{G_{\rho u, 0, 1} \rho u_1}{\rho} - \frac{\rho u_0}{\tau_{\nabla}} \\
\Phi_{16,0} &= - \frac{B_0}{\tau_{\nabla}} + \frac{G_{B, 0, 0} \rho u_0}{\rho} \\
\Phi_{16,1} &= \frac{G_{B, 0, 0} \rho u_1}{\rho} \\
\Phi_{17,0} &= \frac{G_{B, 0, 1} \rho u_0}{\rho} \\
\Phi_{17,1} &= - \frac{B_0}{\tau_{\nabla}} + \frac{G_{B, 0, 1} \rho u_1}{\rho}
\end{align*}
}
\paragraph{Source Vector Components ($S_i$)}
{\footnotesize
\begin{align*}
S_{0} &= 0 \\
S_{1} &= - B_0 G_{B, 0, 0} - B_0 G_{B, 1, 1} \\
S_{2} &= - B_1 G_{B, 0, 0} - B_1 G_{B, 1, 1} \\
S_{3} &= - \frac{B_0 G_{B, 0, 0} \rho u_0}{\rho} - \frac{B_0 G_{B, 1, 1} \rho u_0}{\rho} - \frac{B_1 G_{B, 0, 0} \rho u_1}{\rho} \\
&\quad - \frac{B_1 G_{B, 1, 1} \rho u_1}{\rho} \\
S_{4} &= - \frac{G_{B, 0, 0} \rho u_0}{\rho} - \frac{G_{B, 1, 1} \rho u_0}{\rho} \\
S_{5} &= - \frac{G_{B, 0, 0} \rho u_1}{\rho} - \frac{G_{B, 1, 1} \rho u_1}{\rho} \\
S_{6} &= - \frac{G_{\rho,0}}{\tau_{\nabla}} \\
S_{7} &= - \frac{G_{\rho,1}}{\tau_{\nabla}} \\
S_{8} &= - \frac{G_{\rho u, 1, 0}}{\tau_{\nabla}} \\
S_{9} &= - \frac{G_{\rho u, 1, 1}}{\tau_{\nabla}} \\
S_{10} &= - \frac{G_{E,0}}{\tau_{\nabla}} \\
S_{11} &= - \frac{G_{E,1}}{\tau_{\nabla}} \\
S_{12} &= - \frac{G_{B, 1, 0}}{\tau_{\nabla}} \\
S_{13} &= - \frac{G_{B, 1, 1}}{\tau_{\nabla}} \\
S_{14} &= - \frac{G_{\rho u, 0, 0}}{\tau_{\nabla}} \\
S_{15} &= - \frac{G_{\rho u, 0, 1}}{\tau_{\nabla}} \\
S_{16} &= - \frac{G_{B, 0, 0}}{\tau_{\nabla}} \\
S_{17} &= - \frac{G_{B, 0, 1}}{\tau_{\nabla}}
\end{align*}
}

\subsection{3D Homogenized Compressible Navier--Stokes--Fourier Equations}\label{sec:appendix_pde_hcfnse}
The 3D continuous system has a state vector of dimension $N = 14$.

\paragraph{Macroscopic State Components ($Q_i$)}
{\footnotesize
\begin{align*}
Q_{0} = \rho , \quad Q_{1} = \rho u_0 , \quad Q_{2} = \rho u_1 \\
Q_{3} = \rho u_2 , \quad Q_{4} = E , \quad Q_{5} = G_{u,1,1} \\
Q_{6} = G_{u,2,2} , \quad Q_{7} = G^{\text{aux}}_{0} , \quad Q_{8} = G^{\text{aux}}_{1} \\
Q_{9} = G^{\text{aux}}_{2} , \quad Q_{10} = G_{u,0,0} , \quad Q_{11} = S_{u,01} \\
Q_{12} = S_{u,02} , \quad Q_{13} = S_{u,12}
\end{align*}
}
\paragraph{Flux Tensor Components ($\Phi_{i,j}$)}
with the shared subexpressions
{\footnotesize
\begin{align*}
\chi_{0} &= \rho u_0^{2} \\
\chi_{1} &= \frac{1}{\rho} \\
\chi_{2} &= \frac{3 \chi_{1}}{2} \\
\chi_{3} &= \alpha \mu \\
\chi_{4} &= \frac{4 \chi_{3}}{3} \\
\chi_{5} &= - E \\
\chi_{6} &= E \gamma \\
\chi_{7} &= \rho u_2^{2} \\
\chi_{8} &= \frac{\chi_{1}}{2} \\
\chi_{9} &= \chi_{7} \chi_{8} \\
\chi_{10} &= \chi_{0} \chi_{8} \\
\chi_{11} &= - \chi_{10} \gamma \\
\chi_{12} &= \rho u_1^{2} \\
\chi_{13} &= \chi_{12} \chi_{8} \\
\chi_{14} &= - \chi_{13} \gamma \\
\chi_{15} &= - \chi_{9} \gamma \\
\chi_{16} &= \frac{2 \chi_{3}}{3} \\
\chi_{17} &= G_{u,2,2} \chi_{16} + \chi_{11} + \chi_{14} + \chi_{15} + \chi_{5} + \chi_{6} + \chi_{9} \\
\chi_{18} &= G_{u,1,1} \chi_{16} + \chi_{13} \\
\chi_{19} &= 2 \chi_{3} \\
\chi_{20} &= - S_{u,01} \chi_{19} + \chi_{1} \rho u_0 \rho u_1 \\
\chi_{21} &= - S_{u,02} \chi_{19} + \chi_{1} \rho u_0 \rho u_2 \\
\chi_{22} &= G_{u,0,0} \chi_{16} + \chi_{10} \\
\chi_{23} &= - S_{u,12} \chi_{19} + \chi_{1} \rho u_1 \rho u_2 \\
\chi_{24} &= \alpha \kappa \\
\chi_{25} &= \frac{1}{\rho^{2}} \\
\chi_{26} &= \rho u_0^{3} \\
\chi_{27} &= \frac{\chi_{25}}{2} \\
\chi_{28} &= \chi_{1} \rho u_0 \\
\chi_{29} &= G_{u,0,0} \chi_{28} \\
\chi_{30} &= \chi_{1} \rho u_1 \\
\chi_{31} &= S_{u,01} \chi_{30} \\
\chi_{32} &= \chi_{1} \rho u_2 \\
\chi_{33} &= S_{u,02} \chi_{32} \\
\chi_{34} &= \chi_{27} \rho u_0 \\
\chi_{35} &= \rho u_1^{3} \\
\chi_{36} &= G_{u,1,1} \chi_{30} \\
\chi_{37} &= S_{u,01} \chi_{28} \\
\chi_{38} &= S_{u,12} \chi_{32} \\
\chi_{39} &= \chi_{27} \rho u_1 \\
\chi_{40} &= \rho u_2^{3} \\
\chi_{41} &= G_{u,2,2} \chi_{32} \\
\chi_{42} &= S_{u,02} \chi_{28} \\
\chi_{43} &= S_{u,12} \chi_{30} \\
\chi_{44} &= \chi_{27} \rho u_2 \\
\chi_{45} &= \frac{1}{\tau_{\nabla}} \\
\chi_{46} &= \chi_{1} \chi_{45} \\
\chi_{47} &= \chi_{27} \chi_{45} \\
\chi_{48} &= \chi_{0} \chi_{47} \\
\chi_{49} &= \chi_{12} \chi_{47} \\
\chi_{50} &= \chi_{47} \chi_{7} \\
\chi_{51} &= E \chi_{46} - \chi_{46} \chi_{6} + \chi_{48} \gamma - \chi_{48} + \chi_{49} \gamma - \chi_{49} + \chi_{50} \gamma - \chi_{50} \\
\chi_{52} &= \chi_{45} \chi_{8} \\
\chi_{53} &= - \chi_{52} \rho u_1 \\
\chi_{54} &= - \chi_{52} \rho u_0 \\
\chi_{55} &= - \chi_{52} \rho u_2
\end{align*}
}
the flux components evaluate to
{\footnotesize
\begin{align*}
\Phi_{0,0} &= \rho u_0 \\
\Phi_{0,1} &= \rho u_1 \\
\Phi_{0,2} &= \rho u_2 \\
\Phi_{1,0} &= - G_{u,0,0} \chi_{4} + \chi_{0} \chi_{2} + \chi_{17} + \chi_{18} \\
\Phi_{1,1} &= \chi_{20} \\
\Phi_{1,2} &= \chi_{21} \\
\Phi_{2,0} &= \chi_{20} \\
\Phi_{2,1} &= - G_{u,1,1} \chi_{4} + \chi_{12} \chi_{2} + \chi_{17} + \chi_{22} \\
\Phi_{2,2} &= \chi_{23} \\
\Phi_{3,0} &= \chi_{21} \\
\Phi_{3,1} &= \chi_{23} \\
\Phi_{3,2} &= - G_{u,2,2} \chi_{4} + \chi_{11} + \chi_{14} + \chi_{15} + \chi_{18} + \chi_{2} \chi_{7} + \chi_{22} + \chi_{5} + \chi_{6} \\
\Phi_{4,0} &= E \chi_{1} \gamma \rho u_0 - G^{\text{aux}}_{0} \chi_{24} + \frac{2 G_{u,1,1} \alpha \chi_{1} \mu \rho u_0}{3} \\
&\quad + \frac{2 G_{u,2,2} \alpha \chi_{1} \mu \rho u_0}{3} + \frac{\chi_{12} \chi_{25} \rho u_0}{2} - \chi_{12} \chi_{34} \gamma \\
&\quad - \chi_{19} \chi_{31} - \chi_{19} \chi_{33} + \frac{\chi_{25} \chi_{26}}{2} + \frac{\chi_{25} \chi_{7} \rho u_0}{2} \\
&\quad - \chi_{26} \chi_{27} \gamma - \chi_{29} \chi_{4} - \chi_{34} \chi_{7} \gamma \\
\Phi_{4,1} &= E \chi_{1} \gamma \rho u_1 - G^{\text{aux}}_{1} \chi_{24} + \frac{2 G_{u,0,0} \alpha \chi_{1} \mu \rho u_1}{3} \\
&\quad + \frac{2 G_{u,2,2} \alpha \chi_{1} \mu \rho u_1}{3} + \frac{\chi_{0} \chi_{25} \rho u_1}{2} - \chi_{0} \chi_{39} \gamma \\
&\quad - \chi_{19} \chi_{37} - \chi_{19} \chi_{38} + \frac{\chi_{25} \chi_{35}}{2} + \frac{\chi_{25} \chi_{7} \rho u_1}{2} \\
&\quad - \chi_{27} \chi_{35} \gamma - \chi_{36} \chi_{4} - \chi_{39} \chi_{7} \gamma \\
\Phi_{4,2} &= E \chi_{1} \gamma \rho u_2 - G^{\text{aux}}_{2} \chi_{24} + \frac{2 G_{u,0,0} \alpha \chi_{1} \mu \rho u_2}{3} \\
&\quad + \frac{2 G_{u,1,1} \alpha \chi_{1} \mu \rho u_2}{3} + \frac{\chi_{0} \chi_{25} \rho u_2}{2} - \chi_{0} \chi_{44} \gamma \\
&\quad + \frac{\chi_{12} \chi_{25} \rho u_2}{2} - \chi_{12} \chi_{44} \gamma - \chi_{19} \chi_{42} - \chi_{19} \chi_{43} \\
&\quad + \frac{\chi_{25} \chi_{40}}{2} - \chi_{27} \chi_{40} \gamma - \chi_{4} \chi_{41} \\
\Phi_{5,0} &= G_{u,1,1} \chi_{28} \\
\Phi_{5,1} &= - \chi_{30} \chi_{45} + \chi_{36} \\
\Phi_{5,2} &= G_{u,1,1} \chi_{32} \\
\Phi_{6,0} &= G_{u,2,2} \chi_{28} \\
\Phi_{6,1} &= G_{u,2,2} \chi_{30} \\
\Phi_{6,2} &= - \chi_{32} \chi_{45} + \chi_{41} \\
\Phi_{7,0} &= G^{\text{aux}}_{0} \chi_{28} + \chi_{51} \\
\Phi_{7,1} &= G^{\text{aux}}_{0} \chi_{30} \\
\Phi_{7,2} &= G^{\text{aux}}_{0} \chi_{32} \\
\Phi_{8,0} &= G^{\text{aux}}_{1} \chi_{28} \\
\Phi_{8,1} &= G^{\text{aux}}_{1} \chi_{30} + \chi_{51} \\
\Phi_{8,2} &= G^{\text{aux}}_{1} \chi_{32} \\
\Phi_{9,0} &= G^{\text{aux}}_{2} \chi_{28} \\
\Phi_{9,1} &= G^{\text{aux}}_{2} \chi_{30} \\
\Phi_{9,2} &= G^{\text{aux}}_{2} \chi_{32} + \chi_{51} \\
\Phi_{10,0} &= - \chi_{28} \chi_{45} + \chi_{29} \\
\Phi_{10,1} &= G_{u,0,0} \chi_{30} \\
\Phi_{10,2} &= G_{u,0,0} \chi_{32} \\
\Phi_{11,0} &= \chi_{37} + \chi_{53} \\
\Phi_{11,1} &= \chi_{31} + \chi_{54} \\
\Phi_{11,2} &= S_{u,01} \chi_{32} \\
\Phi_{12,0} &= \chi_{42} + \chi_{55} \\
\Phi_{12,1} &= S_{u,02} \chi_{30} \\
\Phi_{12,2} &= \chi_{33} + \chi_{54} \\
\Phi_{13,0} &= S_{u,12} \chi_{28} \\
\Phi_{13,1} &= \chi_{43} + \chi_{55} \\
\Phi_{13,2} &= \chi_{38} + \chi_{53}
\end{align*}
}
\paragraph{Source Vector Components ($S_i$)}
{\footnotesize
\begin{align*}
S_{0} &= 0 \\
S_{1} &= - \frac{\alpha u_{s,0} \rho}{\eta} + \frac{\alpha \rho u_0}{\eta} + \frac{u_{s,0} \rho}{\eta} - \frac{\rho u_0}{\eta} \\
S_{2} &= - \frac{\alpha u_{s,1} \rho}{\eta} + \frac{\alpha \rho u_1}{\eta} + \frac{u_{s,1} \rho}{\eta} - \frac{\rho u_1}{\eta} \\
S_{3} &= - \frac{\alpha u_{s,2} \rho}{\eta} + \frac{\alpha \rho u_2}{\eta} + \frac{u_{s,2} \rho}{\eta} - \frac{\rho u_2}{\eta} \\
S_{4} &= - \frac{\alpha u_{s,0} \rho u_0}{\eta} - \frac{\alpha u_{s,1} \rho u_1}{\eta} - \frac{\alpha u_{s,2} \rho u_2}{\eta} \\
&\quad + \frac{\alpha \rho u_0^{2}}{\eta \rho} + \frac{\alpha \rho u_1^{2}}{\eta \rho} + \frac{\alpha \rho u_2^{2}}{\eta \rho} \\
&\quad + \frac{u_{s,0} \rho u_0}{\eta} + \frac{u_{s,1} \rho u_1}{\eta} + \frac{u_{s,2} \rho u_2}{\eta} - \frac{\rho u_0^{2}}{\eta \rho} \\
&\quad - \frac{\rho u_1^{2}}{\eta \rho} - \frac{\rho u_2^{2}}{\eta \rho} \\
S_{5} &= - \frac{G_{u,1,1}}{\tau_{\nabla}} \\
S_{6} &= - \frac{G_{u,2,2}}{\tau_{\nabla}} \\
S_{7} &= - \frac{G^{\text{aux}}_{0}}{\tau_{\nabla}} \\
S_{8} &= - \frac{G^{\text{aux}}_{1}}{\tau_{\nabla}} \\
S_{9} &= - \frac{G^{\text{aux}}_{2}}{\tau_{\nabla}} \\
S_{10} &= - \frac{G_{u,0,0}}{\tau_{\nabla}} \\
S_{11} &= - \frac{S_{u,01}}{\tau_{\nabla}} \\
S_{12} &= - \frac{S_{u,02}}{\tau_{\nabla}} \\
S_{13} &= - \frac{S_{u,12}}{\tau_{\nabla}}
\end{align*}
}

\clearpage
\allowdisplaybreaks

\section{Method of Manufactured Solutions (MMS) Exact Fields}\label{sec:appendix_mms}

In this appendix, we document the analytical expressions of the \emph{Method of Manufactured Solutions} (MMS) exact fields ($\mathbf{Q}^{\text{exact}}$) for all twelve verified physical PDE targets. The volume forcing terms are derived from these fields symbolically at compile time as the continuous residual $\mathbf{S}_{\text{mms}} = \partial_t \mathbf{Q}^{\text{exact}} + \nabla \cdot \mathbf{\Phi}(\mathbf{Q}^{\text{exact}}) - \mathbf{S}(\mathbf{Q}^{\text{exact}})$. We list the conserved macroscopic components. The gradient-tracker components carried by some systems are initialized from the spatial derivatives of these fields and are omitted here. Cases declared with SI-valued scales author their fields as (scale)\,$\times$\,(dimensionless), and the value non-dimensionalization divides the scales out to reach the $O(1)$ lattice working point tabulated per case.

\subsection{Design Principles for the Manufactured Solutions}\label{sec:appendix_mms_design}

\begin{enumerate}
\item \emph{Smoothness.} All fields are $C^\infty$ trigonometric--exponential profiles, so the measured order reflects the truncation error of the scheme.
\item \emph{Large amplitudes.} Fluctuations reach a finite fraction of the mean state, so the nonlinearities are exercised.
\item \emph{Independent modes.} Fields coupled by a constitutive law carry independent fluctuation modes rather than being slaved to one another.
\item \emph{Broken axis symmetry.} Per-axis wavenumbers, Mach numbers, or phase patterns are distinct.
\item \emph{Off-resonance forcing.} Temporal frequencies avoid the free eigenfrequencies, and the evaluation instant falls at a generic phase.
\item \emph{Involutions by construction.} Differential constraints carried by the physics are satisfied analytically by the exact fields.
\end{enumerate}

\subsection{Inviscid Burgers Equation MMS}\label{sec:appendix_mms_burgers}
\paragraph{Manufactured solution (physical form)}
The exact manufactured profile is a smooth periodic wave of amplitude $A = \max|u|$:
\begin{align}
u_{\text{exact}}(\mathbf{x}, t) = A\,\sin(2\pi x_0)\sin(2\pi x_1)\cos t.
\end{align}

\begin{table}[H]
\centering\footnotesize
\caption{Symbol values for Inviscid Burgers Equation MMS.}\label{tab:mmsvals_burgers}
\begin{tabular}{@{}lll@{}}
\toprule
Symbol & Value & Quantity \\
\midrule
$A$ & \num{1.3} & manufactured-solution amplitude \\
\bottomrule
\end{tabular}
\end{table}

\paragraph{Compiler-expanded fields ($Q_i^{\text{exact}}$)}
{\footnotesize
\begin{align*}
Q_{0}^{\text{exact}} &= A \sin{(2 \pi x_0 )} \sin{(2 \pi x_1 )} \cos{(t )}
\end{align*}
}

\subsection{Compressible Euler Equations MMS}\label{sec:appendix_mms_compressible_euler}
\paragraph{Manufactured solution (physical form)}
The manufactured solution superimposes smooth fluctuations on a supersonic advection background at per-axis Mach number $3.0$. The reference state has density $\rho_0$ and sound speed $c_s = \sqrt{\gamma p_0/\rho_0}$, with reference pressure $p_0 = \rho_0 c_s^2/\gamma$ and background velocity $u_0 = v_0 = 3\,c_s$. With wavenumber $k = 2\pi$ the fluctuations are
\begin{align}
\begin{aligned}
\Delta\rho &= 0.3\,\rho_0\,\sin(k x_0)\cos(k x_1)\cos t, \\
\Delta u   &= 0.3\,c_s\,\cos(k x_0)\sin(k x_1)\sin t, \\
\Delta v   &= 0.3\,c_s\,\sin(k x_0)\sin(k x_1)\sin t, \\
\Delta p   &= c_s^2\,\Delta\rho + 0.2\,p_0\,\cos(k x_0)\sin(k x_1)\sin t,
\end{aligned}
\end{align}
so the pressure carries an independent mode in addition to the linear response $c_s^2\Delta\rho$. The conserved state is assembled as $\rho_{\text{exact}} = \rho_0 + \Delta\rho$, $(\rho\mathbf{u})_{\text{exact}} = \rho_{\text{exact}}(\mathbf{u}_0 + \Delta\mathbf{u})$, $p_{\text{exact}} = p_0 + \Delta p$, and $E_{\text{exact}} = p_{\text{exact}}/(\gamma-1) + \tfrac12\rho_{\text{exact}}|\mathbf{u}_{\text{exact}}|^2$.

\begin{table}[H]
\centering\footnotesize
\caption{Symbol values for Compressible Euler Equations MMS.}\label{tab:mmsvals_compressible_euler}
\begin{tabular}{@{}lll@{}}
\toprule
Symbol & Value & Quantity \\
\midrule
$\rho_{0}$ & \qty{1.225}{\kilogram\per\cubic\metre} & reference density \\
$c_{s}$ & \qty{340.3}{\metre\per\second} & sound-speed scale \\
$\gamma$ & \num{1.4} & adiabatic index \\
\bottomrule
\end{tabular}
\end{table}

\paragraph{Compiler-expanded fields ($Q_i^{\text{exact}}$)}
{\footnotesize
\begin{align*}
Q_{0}^{\text{exact}} &= \frac{3 \rho_{0} \sin{(2 \pi x_0 )} \cos{(t )} \cos{(2 \pi x_1 )}}{10} + \rho_{0} \\
Q_{1}^{\text{exact}} &= (\frac{3 c_{s} \sin{(t )} \sin{(2 \pi x_1 )} \cos{(2 \pi x_0 )}}{10} \\
&\quad + 3 c_{s}) (\frac{3 \rho_{0} \sin{(2 \pi x_0 )} \cos{(t )} \cos{(2 \pi x_1 )}}{10} + \rho_{0}) \\
Q_{2}^{\text{exact}} &= (\frac{3 c_{s} \sin{(t )} \sin{(2 \pi x_0 )} \sin{(2 \pi x_1 )}}{10} \\
&\quad + 3 c_{s}) (\frac{3 \rho_{0} \sin{(2 \pi x_0 )} \cos{(t )} \cos{(2 \pi x_1 )}}{10} + \rho_{0}) \\
Q_{3}^{\text{exact}} &= \frac{5 c_{s}^{2} \rho_{0} \sin{(t )} \sin{(2 \pi x_1 )} \cos{(2 \pi x_0 )}}{14} \\
&\quad + \frac{3 c_{s}^{2} \rho_{0} \sin{(2 \pi x_0 )} \cos{(t )} \cos{(2 \pi x_1 )}}{4} + \frac{25 c_{s}^{2} \rho_{0}}{14} \\
&\quad + (\frac{3 \rho_{0} \sin{(2 \pi x_0 )} \cos{(t )} \cos{(2 \pi x_1 )}}{20} \\
&\quad + \frac{\rho_{0}}{2}) ((\frac{3 c_{s} \sin{(t )} \sin{(2 \pi x_0 )} \sin{(2 \pi x_1 )}}{10} + 3 c_{s})^{2} \\
&\quad + (\frac{3 c_{s} \sin{(t )} \sin{(2 \pi x_1 )} \cos{(2 \pi x_0 )}}{10} + 3 c_{s})^{2})
\end{align*}
}

\subsection{Shallow Water Equations MMS}\label{sec:appendix_mms_shallow_water}
\paragraph{Manufactured solution (physical form)}
The manufactured fields evolve the water depth $h$ and discharge $h\mathbf{u}$ over a static bathymetry $z_b(\mathbf{x}) = 0.01\sin(2\pi x_0)\cos(2\pi x_1)$. Depth fluctuations are $40\%$ of the mean depth, giving a subcritical Froude number $\approx 0.3$ with $h \in [0.06,\,0.14]$ so $h > 0$ everywhere:
\begin{align}
\begin{aligned}
h_{\text{exact}} &= 0.04\,\sin(2\pi x_0)\sin(2\pi x_1)\cos t + 0.1, \\
(hu_0)_{\text{exact}} &= 0.02\,\cos(2\pi x_0)\cos(2\pi x_1)\sin t + 0.03, \\
(hu_1)_{\text{exact}} &= 0.02\,\sin(2\pi x_0)\sin(2\pi x_1)\sin t + 0.03.
\end{aligned}
\end{align}
The characteristic gravity-wave speed $c_s = \sqrt{g H}$ at reference depth $H = 0.1$ sets the convective time scale and reduces the realistic-SI gravity to the working point. The values are listed below.

\begin{table}[H]
\centering\footnotesize
\caption{Symbol values for Shallow Water Equations MMS.}\label{tab:mmsvals_shallow_water}
\begin{tabular}{@{}lll@{}}
\toprule
Symbol & Value & Quantity \\
\midrule
$c_{s}$ & \qty{0.9905}{\metre\per\second} & sound-speed scale \\
$g$ & \qty{9.81}{\metre\per\second\squared} & gravitational acceleration \\
\bottomrule
\end{tabular}
\end{table}

\paragraph{Compiler-expanded fields ($Q_i^{\text{exact}}$)}
{\footnotesize
\begin{align*}
Q_{0}^{\text{exact}} &= \frac{\sin{(2 \pi x_0 )} \sin{(2 \pi x_1 )} \cos{(t )}}{25} + \frac{1}{10} \\
Q_{1}^{\text{exact}} &= \frac{\sin{(t )} \cos{(2 \pi x_0 )} \cos{(2 \pi x_1 )}}{50} + \frac{3}{100} \\
Q_{2}^{\text{exact}} &= \frac{\sin{(t )} \sin{(2 \pi x_0 )} \sin{(2 \pi x_1 )}}{50} + \frac{3}{100}
\end{align*}
}

\subsection{Ideal Ultrarelativistic Fluid MMS}\label{sec:appendix_mms_relativistic_hydro}
\paragraph{Manufactured solution (physical form)}
The manufactured fields for the ideal ultrarelativistic fluid are smooth fluctuations about a uniform background, kept inside the physicality region $E > |\mathbf{S}|$ (three-velocities up to $|\mathbf{v}| \approx 0.79$, Lorentz factor $\Gamma \approx 1.6$ where $E$ is low and $|\mathbf{S}|$ high):
\begin{align}
\begin{aligned}
E_{\text{exact}}(\mathbf{x}, t) &= 0.4\,\sin(2\pi x_0)\sin(2\pi x_1)\cos t + 3.0, \\
S_{0,\text{exact}}(\mathbf{x}, t) &= 0.4\,\cos(2\pi x_0)\cos(2\pi x_1)\sin t + 1.2, \\
S_{1,\text{exact}}(\mathbf{x}, t) &= 0.4\,\sin(2\pi x_0)\sin(2\pi x_1)\sin t + 1.2,
\end{aligned}
\end{align}
with conserved energy density $E$ and momentum $\mathbf{S}$ in natural units ($c = 1$).

\paragraph{Compiler-expanded fields ($Q_i^{\text{exact}}$)}
{\footnotesize
\begin{align*}
Q_{0}^{\text{exact}} &= \frac{2 \sin{(2 \pi x_0 )} \sin{(2 \pi x_1 )} \cos{(t )}}{5} + 3 \\
Q_{1}^{\text{exact}} &= \frac{2 \sin{(t )} \cos{(2 \pi x_0 )} \cos{(2 \pi x_1 )}}{5} + \frac{6}{5} \\
Q_{2}^{\text{exact}} &= \frac{2 \sin{(t )} \sin{(2 \pi x_0 )} \sin{(2 \pi x_1 )}}{5} + \frac{6}{5}
\end{align*}
}

\subsection{Maxwell Electromagnetics (TM Mode) MMS}\label{sec:appendix_mms_maxwell}
\paragraph{Manufactured solution (physical form)}
The manufactured solution is a TM pattern with distinct wavenumbers $k_x = 2\pi$ and $k_y = 4\pi$, amplitude $A = \tfrac{1}{2}$, and temporal frequency $\omega = 4$, with the electric and magnetic fields in the physical $90^\circ$ quadrature ($E_z \sim \cos\omega t$, $\mathbf{H} \sim \sin\omega t$):
\begin{align}
\begin{aligned}
E_{z,\text{exact}}(\mathbf{x}, t) &= A \sin(k_x x_0)\sin(k_y x_1)\cos(\omega t), \\
H_{x,\text{exact}}(\mathbf{x}, t) &= A \sin(k_x x_0)\cos(k_y x_1)\sin(\omega t), \\
H_{y,\text{exact}}(\mathbf{x}, t) &= -A \frac{k_x}{k_y} \cos(k_x x_0)\sin(k_y x_1)\sin(\omega t).
\end{aligned}
\end{align}
The field satisfies $\nabla \cdot \mathbf{H} = 0$ analytically.

\begin{table}[H]
\centering\footnotesize
\caption{Symbol values for Maxwell Electromagnetics (TM Mode) MMS.}\label{tab:mmsvals_maxwell}
\begin{tabular}{@{}lll@{}}
\toprule
Symbol & Value & Quantity \\
\midrule
$c$ & \num{1} & electromagnetic wave speed \\
\bottomrule
\end{tabular}
\end{table}

\paragraph{Compiler-expanded fields ($Q_i^{\text{exact}}$)}
{\footnotesize
\begin{align*}
Q_{0}^{\text{exact}} &= \frac{\sin{(2 \pi x_0 )} \sin{(4 \pi x_1 )} \cos{(4 t )}}{2} \\
Q_{1}^{\text{exact}} &= \frac{\sin{(4 t )} \sin{(2 \pi x_0 )} \cos{(4 \pi x_1 )}}{2} \\
Q_{2}^{\text{exact}} &= - \frac{\sin{(4 t )} \sin{(4 \pi x_1 )} \cos{(2 \pi x_0 )}}{4}
\end{align*}
}

\subsection{Nonlinear Finite-Strain Elasticity MMS}\label{sec:appendix_mms_nonlinear_elasticity}
\paragraph{Manufactured solution (physical form)}
The case is the deformation-gradient (total-Lagrangian) formulation of compressible neo-Hookean elastodynamics of the main article, with conserved state $(\mathbf{F}, \mathbf{v})$. The manufactured fields derive from a smooth, non-separable displacement potential $\mathbf{w}$ of amplitude $A = 0.06$ via $\mathbf{F} = \mathbf{I} + \nabla\mathbf{w}$ (curl-free, so the compatibility involution holds exactly) and $\mathbf{v} = \partial_t\mathbf{w}$, reaching a $37\%$ finite strain with $\det\mathbf{F}\ge 0.68 > 0$ everywhere:
\begin{align}
\begin{aligned}
w_0 &= A\big(\sin(2\pi x_0)\sin(2\pi x_1)\cos t + \tfrac{1}{2}\sin(2\pi x_0)\cos(2\pi x_1)\sin t \\
    &\qquad + \tfrac{1}{4}\cos(2\pi(x_0+x_1))\cos t\big), \\
w_1 &= A\big(\cos(2\pi x_0)\cos(2\pi x_1)\sin t + \tfrac{1}{2}\cos(2\pi x_0)\sin(2\pi x_1)\cos t \\
    &\qquad + \tfrac{1}{4}\sin(2\pi(x_0-x_1))\sin t\big).
\end{aligned}
\end{align}
The material values (aluminium-like) are listed below. The rest state $\mathbf{F} = \mathbf{I}$, $\mathbf{v} = 0$ gives $\mathbf{P}(\mathbf{I}) = 0$, so the reference state is the rest configuration.

\begin{table}[H]
\centering\footnotesize
\caption{Symbol values for Nonlinear Finite-Strain Elasticity MMS.}\label{tab:mmsvals_nonlinear_elasticity}
\begin{tabular}{@{}lll@{}}
\toprule
Symbol & Value & Quantity \\
\midrule
$\rho_{0}$ & \num{2700} & reference density \\
$\mu_s$ & \num{2.6e+10} & shear modulus \\
$\lambda_s$ & \num{5.2e+10} & first Lame parameter \\
\bottomrule
\end{tabular}
\end{table}

\paragraph{Compiler-expanded fields ($Q_i^{\text{exact}}$)}
{\footnotesize
\begin{align*}
Q_{0}^{\text{exact}} &= \frac{3 \pi \sin{(t )} \cos{(2 \pi x_0 )} \cos{(2 \pi x_1 )}}{50} \\
&\quad + \frac{3 \pi \sin{(2 \pi x_1 )} \cos{(t )} \cos{(2 \pi x_0 )}}{25} - \frac{3 \pi \sin{(\pi (2 x_0 + 2 x_1) )} \cos{(t )}}{100} \\
&\quad + 1 \\
Q_{1}^{\text{exact}} &= - \frac{3 \pi \sin{(t )} \sin{(2 \pi x_0 )} \sin{(2 \pi x_1 )}}{50} \\
&\quad + \frac{3 \pi \sin{(2 \pi x_0 )} \cos{(t )} \cos{(2 \pi x_1 )}}{25} - \frac{3 \pi \sin{(\pi (2 x_0 + 2 x_1) )} \cos{(t )}}{100} \\
Q_{2}^{\text{exact}} &= - \frac{3 \pi \sin{(t )} \sin{(2 \pi x_0 )} \cos{(2 \pi x_1 )}}{25} + \frac{3 \pi \sin{(t )} \cos{(\pi (2 x_0 - 2 x_1) )}}{100} \\
&\quad - \frac{3 \pi \sin{(2 \pi x_0 )} \sin{(2 \pi x_1 )} \cos{(t )}}{50} \\
Q_{3}^{\text{exact}} &= - \frac{3 \pi \sin{(t )} \sin{(2 \pi x_1 )} \cos{(2 \pi x_0 )}}{25} - \frac{3 \pi \sin{(t )} \cos{(\pi (2 x_0 - 2 x_1) )}}{100} \\
&\quad + \frac{3 \pi \cos{(t )} \cos{(2 \pi x_0 )} \cos{(2 \pi x_1 )}}{50} + 1 \\
Q_{4}^{\text{exact}} &= - \frac{3 \sin{(t )} \sin{(2 \pi x_0 )} \sin{(2 \pi x_1 )}}{50} - \frac{3 \sin{(t )} \cos{(\pi (2 x_0 + 2 x_1) )}}{200} \\
&\quad + \frac{3 \sin{(2 \pi x_0 )} \cos{(t )} \cos{(2 \pi x_1 )}}{100} \\
Q_{5}^{\text{exact}} &= - \frac{3 \sin{(t )} \sin{(2 \pi x_1 )} \cos{(2 \pi x_0 )}}{100} + \frac{3 \sin{(\pi (2 x_0 - 2 x_1) )} \cos{(t )}}{200} \\
&\quad + \frac{3 \cos{(t )} \cos{(2 \pi x_0 )} \cos{(2 \pi x_1 )}}{50}
\end{align*}
}

\subsection{Scalar Advection-Diffusion-Reaction MMS}\label{sec:appendix_mms_adr}
\paragraph{Manufactured solution (physical form)}
The exact manufactured verification profile is defined as:
\begin{align}
c_{\text{exact}}(\mathbf{x}, t) = 0.5 \sin(2\pi(x_0 - u_{0}t))\cos(2\pi(x_1 - u_{1}t))\exp(-t) + 0.5.
\end{align}
System parameters are listed below.

\begin{table}[H]
\centering\footnotesize
\caption{Symbol values for Scalar Advection-Diffusion-Reaction MMS.}\label{tab:mmsvals_adr}
\begin{tabular}{@{}lll@{}}
\toprule
Symbol & Value & Quantity \\
\midrule
$u_0$ & \num{0.1} & advection velocity (x) \\
$u_1$ & \num{-0.1} & advection velocity (y) \\
$D$ & \num{0.01} & diffusivity \\
$R$ & \num{1} & reaction rate \\
\bottomrule
\end{tabular}
\end{table}

\paragraph{Compiler-expanded fields ($Q_i^{\text{exact}}$)}
{\footnotesize
\begin{align*}
Q_{0}^{\text{exact}} &= \frac{1}{2} + \frac{e^{- t} \sin{(\pi (- 2 t u_0 + 2 x_0) )} \cos{(\pi (- 2 t u_1 + 2 x_1) )}}{2}
\end{align*}
}

\subsection{Allen--Cahn Phase-Field Equation MMS}\label{sec:appendix_mms_allen_cahn}
\paragraph{Manufactured solution (physical form)}
The exact manufactured verification profile is a smooth periodic order parameter whose amplitude is kept below unity:
\begin{align}
\phi_{\text{exact}}(\mathbf{x}, t) = 0.7 \sin(2\pi x_0)\cos(2\pi x_1)\cos(t).
\end{align}
System parameters are listed below.

\begin{table}[H]
\centering\footnotesize
\caption{Symbol values for Allen--Cahn Phase-Field Equation MMS.}\label{tab:mmsvals_allen_cahn}
\begin{tabular}{@{}lll@{}}
\toprule
Symbol & Value & Quantity \\
\midrule
$\gamma$ & \num{0.01} & interface mobility times width squared \\
$r$ & \num{1} & reaction rate \\
\bottomrule
\end{tabular}
\end{table}

\paragraph{Compiler-expanded fields ($Q_i^{\text{exact}}$)}
{\footnotesize
\begin{align*}
Q_{0}^{\text{exact}} &= \frac{7 \sin{(2 \pi x_0 )} \cos{(t )} \cos{(2 \pi x_1 )}}{10}
\end{align*}
}

\subsection{Weakly Compressible Navier--Stokes Equations (Incompressible Limit) MMS}\label{sec:appendix_mms_navier_stokes}
\paragraph{Manufactured solution (physical form)}
The manufactured solution is the two-dimensional Taylor--Green vortex, an exact solution of the incompressible Navier--Stokes equations, used here to verify the weakly compressible (artificial-compressibility) scheme in its incompressible limit. The working point is weakly compressible at Mach number $\text{Ma} = U/c_s = 0.1$ and Reynolds number $\text{Re} = U L/\nu = 10$, with reference density $\rho_0$, artificial sound speed $c_s$, velocity amplitude $U$, and kinematic viscosity $\nu$ ($\mu = \rho_0\nu$). With wavenumber $k = 2\pi$ and the exact viscous decay $\mathcal{D}(t) = \exp(-2(\nu/c_s)k^2 t)$ in characteristic time, the velocity and pressure are
\begin{align}
\begin{aligned}
u &= -U\,\cos(k x_0)\sin(k x_1)\,\mathcal{D}(t), \\
v &=  U\,\sin(k x_0)\cos(k x_1)\,\mathcal{D}(t), \\
p &= -\tfrac14\rho_0 U^2\,[\cos(2 k x_0) + \cos(2 k x_1)]\,\mathcal{D}(t)^2.
\end{aligned}
\end{align}
The density is slaved to the dynamic pressure, $\rho_{\text{exact}} = \rho_0 + p/c_s^2$, so the density fluctuation is $O(\text{Ma}^2)$, and the conserved momentum is $(\rho\mathbf{u})_{\text{exact}} = \rho_{\text{exact}}\mathbf{u}$.

\begin{table}[H]
\centering\footnotesize
\caption{Symbol values for Weakly Compressible Navier--Stokes Equations (Incompressible Limit) MMS.}\label{tab:mmsvals_navier_stokes}
\begin{tabular}{@{}lll@{}}
\toprule
Symbol & Value & Quantity \\
\midrule
$\rho_{0}$ & \qty{1.225}{\kilogram\per\cubic\metre} & reference density \\
$c_{s}$ & \qty{340}{\metre\per\second} & sound-speed scale \\
$\mu$ & \qty{4.165}{\pascal\second} & dynamic viscosity \\
$U$ & \qty{34}{\metre\per\second} & velocity amplitude \\
\bottomrule
\end{tabular}
\end{table}

\paragraph{Compiler-expanded fields ($Q_i^{\text{exact}}$)}
{\footnotesize
\begin{align*}
Q_{0}^{\text{exact}} &= - \frac{\rho_{0} (\cos{(4 \pi x_0 )} + \cos{(4 \pi x_1 )}) e^{- \frac{4 \pi^{2} t}{25}}}{400} + \rho_{0} \\
Q_{1}^{\text{exact}} &= - \frac{c_{s} \rho_{0} e^{- \frac{2 \pi^{2} t}{25}} \sin{(2 \pi x_1 )} \cos{(2 \pi x_0 )}}{10} \\
&\quad + \frac{c_{s} \rho_{0} e^{- \frac{6 \pi^{2} t}{25}} \sin{(2 \pi x_1 )} \cos{(2 \pi x_0 )} \cos{(4 \pi x_0 )}}{4000} \\
&\quad + \frac{c_{s} \rho_{0} e^{- \frac{6 \pi^{2} t}{25}} \sin{(2 \pi x_1 )} \cos{(2 \pi x_0 )} \cos{(4 \pi x_1 )}}{4000} \\
Q_{2}^{\text{exact}} &= \frac{c_{s} \rho_{0} e^{- \frac{2 \pi^{2} t}{25}} \sin{(2 \pi x_0 )} \cos{(2 \pi x_1 )}}{10} \\
&\quad - \frac{c_{s} \rho_{0} e^{- \frac{6 \pi^{2} t}{25}} \sin{(2 \pi x_0 )} \cos{(4 \pi x_0 )} \cos{(2 \pi x_1 )}}{4000} \\
&\quad - \frac{c_{s} \rho_{0} e^{- \frac{6 \pi^{2} t}{25}} \sin{(2 \pi x_0 )} \cos{(2 \pi x_1 )} \cos{(4 \pi x_1 )}}{4000}
\end{align*}
}

\subsection{Compressible Navier--Stokes--Fourier Equations MMS}\label{sec:appendix_mms_compressible_navier_stokes}
\paragraph{Manufactured solution (physical form)}
The manufactured solution models a supersonic viscous flow with an anisotropic mean velocity at per-axis Mach numbers $3.0$ and $2.0$, at Prandtl number $\text{Pr} = 0.71$ (air). The reference state has density $\rho_0$, gas constant $R$, and sound speed $c_s = \sqrt{\gamma R T_0}$ at temperature $T_0 = \qty{288}{\kelvin}$, with reference pressure $p_0 = \rho_0 R T_0$ and background velocity $u_0 = 3\,c_s$, $v_0 = 2\,c_s$. With wavenumber $k = 2\pi$ the fluctuations use a distinct amplitude and spatial structure per field,
\begin{align}
\begin{aligned}
\Delta\rho &= 0.20\,\rho_0\,\sin(k x_0)\cos(k x_1)\cos t, \\
\Delta u   &= 0.20\,c_s\,\cos(k x_0)\sin(k x_1)\sin t, \\
\Delta v   &= 0.15\,c_s\,\sin(k x_0)\sin(k x_1)\cos t, \\
\Delta p   &= c_s^2\,\Delta\rho + 0.20\,p_0\,\cos(k x_0)\sin(k x_1)\sin t,
\end{aligned}
\end{align}
so the pressure carries an independent mode beyond the linear response. The conserved state is $\rho_{\text{exact}} = \rho_0 + \Delta\rho$, $(\rho\mathbf{u})_{\text{exact}} = \rho_{\text{exact}}(\mathbf{u}_0 + \Delta\mathbf{u})$, and $E_{\text{exact}} = p_{\text{exact}}/(\gamma-1) + \tfrac12\rho_{\text{exact}}|\mathbf{u}_{\text{exact}}|^2$ with $p_{\text{exact}} = p_0 + \Delta p$. The thermal conductivity follows from the Prandtl number, $K = \mu\,c_p/\text{Pr}$ with $c_p = \gamma R/(\gamma-1)$, placing the flow in a resolved viscous regime ($\text{Re} = \rho_0 u_0 L/\mu \approx 3.5\times10^2$ on the unit domain $L = 1$).

\begin{table}[H]
\centering\footnotesize
\caption{Symbol values for Compressible Navier--Stokes--Fourier Equations MMS.}\label{tab:mmsvals_compressible_navier_stokes}
\begin{tabular}{@{}lll@{}}
\toprule
Symbol & Value & Quantity \\
\midrule
$\rho_{0}$ & \qty{1.225}{\kilogram\per\cubic\metre} & reference density \\
$c_{s}$ & \qty{340.2}{\metre\per\second} & sound-speed scale \\
$\gamma$ & \num{1.4} & adiabatic index \\
$R$ & \qty[per-mode=power]{287}{\joule\per\kilogram\per\kelvin} & specific gas constant \\
$\mu$ & \qty{3.52}{\pascal\second} & dynamic viscosity \\
$K$ & \qty[per-mode=power]{4980}{\watt\per\metre\per\kelvin} & thermal conductivity \\
\bottomrule
\end{tabular}
\end{table}

\paragraph{Compiler-expanded fields ($Q_i^{\text{exact}}$)}
{\footnotesize
\begin{align*}
Q_{0}^{\text{exact}} &= \frac{\rho_{0} \sin{(2 \pi x_0 )} \cos{(t )} \cos{(2 \pi x_1 )}}{5} + \rho_{0} \\
Q_{1}^{\text{exact}} &= (\frac{c_{s} \sin{(t )} \sin{(2 \pi x_1 )} \cos{(2 \pi x_0 )}}{5} \\
&\quad + 3 c_{s}) (\frac{\rho_{0} \sin{(2 \pi x_0 )} \cos{(t )} \cos{(2 \pi x_1 )}}{5} + \rho_{0}) \\
Q_{2}^{\text{exact}} &= (\frac{3 c_{s} \sin{(2 \pi x_0 )} \sin{(2 \pi x_1 )} \cos{(t )}}{20} \\
&\quad + 2 c_{s}) (\frac{\rho_{0} \sin{(2 \pi x_0 )} \cos{(t )} \cos{(2 \pi x_1 )}}{5} + \rho_{0}) \\
Q_{3}^{\text{exact}} &= \frac{5 c_{s}^{2} \rho_{0} \sin{(t )} \sin{(2 \pi x_1 )} \cos{(2 \pi x_0 )}}{14} \\
&\quad + \frac{c_{s}^{2} \rho_{0} \sin{(2 \pi x_0 )} \cos{(t )} \cos{(2 \pi x_1 )}}{2} + \frac{25 c_{s}^{2} \rho_{0}}{14} \\
&\quad + (\frac{\rho_{0} \sin{(2 \pi x_0 )} \cos{(t )} \cos{(2 \pi x_1 )}}{10} \\
&\quad + \frac{\rho_{0}}{2}) ((\frac{c_{s} \sin{(t )} \sin{(2 \pi x_1 )} \cos{(2 \pi x_0 )}}{5} + 3 c_{s})^{2} \\
&\quad + (\frac{3 c_{s} \sin{(2 \pi x_0 )} \sin{(2 \pi x_1 )} \cos{(t )}}{20} + 2 c_{s})^{2})
\end{align*}
}

\subsection{Resistive Compressible Magnetohydrodynamics MMS}\label{sec:appendix_mms_mhd}
\paragraph{Manufactured solution (physical form)}
The manufactured solution is a supersonic magnetized flow at per-axis Mach number $3.0$ with smooth fluctuations about a uniform background. The magnetic field is carried in Alfven (velocity) units. The reference state has density $\rho_0$ and sound speed $c_s = \sqrt{\gamma p_0/\rho_0}$ (reference pressure $p_0 = \rho_0 c_s^2/\gamma$), with background velocity $u_0 = v_0 = 3\,c_s$ and background field $B_{x,0} = B_{y,0} = 0.5\,c_s$ (plasma $\beta = 4/\gamma \approx 2.9$). With wavenumber $k = 2\pi$ the fluctuations are
\begin{align}
\begin{aligned}
\Delta\rho &= 0.3\,\rho_0\,\sin(k x_0)\cos(k x_1)\cos t, \\
\Delta p   &= c_s^2\,\Delta\rho + 0.2\,p_0\,\cos(k x_0)\sin(k x_1)\sin t, \\
\Delta u   &= 0.3\,c_s\,\cos(k x_0)\sin(k x_1)\sin t, \\
\Delta v   &= 0.3\,c_s\,\sin(k x_0)\sin(k x_1)\sin t, \\
\Delta B_x &= -0.3\,c_s\,\cos(k x_0)\sin(k x_1)\sin t, \\
\Delta B_y &=  0.3\,c_s\,\sin(k x_0)\cos(k x_1)\sin t,
\end{aligned}
\end{align}
with the magnetic perturbation divergence-free by construction ($\partial_{x_0}\Delta B_x + \partial_{x_1}\Delta B_y = 0$), and the pressure carrying an independent mode beyond the linear response $c_s^2\Delta\rho$. The conserved energy is $E_{\text{exact}} = p_{\text{exact}}/(\gamma-1) + \tfrac12\rho_{\text{exact}}|\mathbf{u}_{\text{exact}}|^2 + \tfrac12|\mathbf{B}_{\text{exact}}|^2$, and a Powell source proportional to $\nabla\cdot\mathbf{B}$ controls the solenoidal error. The magnetic diffusivity, dynamic viscosity, and heat-flux coefficient are set to equal diffusivities.

\begin{table}[H]
\centering\footnotesize
\caption{Symbol values for Resistive Compressible Magnetohydrodynamics MMS.}\label{tab:mmsvals_mhd}
\begin{tabular}{@{}lll@{}}
\toprule
Symbol & Value & Quantity \\
\midrule
$\rho_{0}$ & \qty{1.225}{\kilogram\per\cubic\metre} & reference density \\
$c_{s}$ & \qty{340}{\metre\per\second} & sound-speed scale \\
$\gamma$ & \num{1.4} & adiabatic index \\
$\eta$ & \qty{3.4}{\metre\squared\per\second} & magnetic diffusivity \\
$\mu$ & \qty{4.165}{\pascal\second} & dynamic viscosity \\
$K$ & \qty{4.165}{\pascal\second} & heat-flux coefficient ($T = p/\rho$) \\
\bottomrule
\end{tabular}
\end{table}

\paragraph{Compiler-expanded fields ($Q_i^{\text{exact}}$)}
{\footnotesize
\begin{align*}
Q_{0}^{\text{exact}} &= \frac{3 \rho_{0} \sin{(2 \pi x_0 )} \cos{(t )} \cos{(2 \pi x_1 )}}{10} + \rho_{0} \\
Q_{1}^{\text{exact}} &= (\frac{3 c_{s} \sin{(t )} \sin{(2 \pi x_1 )} \cos{(2 \pi x_0 )}}{10} \\
&\quad + 3 c_{s}) (\frac{3 \rho_{0} \sin{(2 \pi x_0 )} \cos{(t )} \cos{(2 \pi x_1 )}}{10} + \rho_{0}) \\
Q_{2}^{\text{exact}} &= (\frac{3 c_{s} \sin{(t )} \sin{(2 \pi x_0 )} \sin{(2 \pi x_1 )}}{10} \\
&\quad + 3 c_{s}) (\frac{3 \rho_{0} \sin{(2 \pi x_0 )} \cos{(t )} \cos{(2 \pi x_1 )}}{10} + \rho_{0}) \\
Q_{3}^{\text{exact}} &= \frac{5 c_{s}^{2} \rho_{0} \sin{(t )} \sin{(2 \pi x_1 )} \cos{(2 \pi x_0 )}}{14} \\
&\quad + \frac{3 c_{s}^{2} \rho_{0} \sin{(2 \pi x_0 )} \cos{(t )} \cos{(2 \pi x_1 )}}{4} + \frac{25 c_{s}^{2} \rho_{0}}{14} \\
&\quad + \frac{(\frac{3 c_{s} \sin{(t )} \sin{(2 \pi x_0 )} \cos{(2 \pi x_1 )}}{10} + \frac{c_{s}}{2})^{2}}{2} \\
&\quad + \frac{(- \frac{3 c_{s} \sin{(t )} \sin{(2 \pi x_1 )} \cos{(2 \pi x_0 )}}{10} + \frac{c_{s}}{2})^{2}}{2} \\
&\quad + (\frac{3 \rho_{0} \sin{(2 \pi x_0 )} \cos{(t )} \cos{(2 \pi x_1 )}}{20} \\
&\quad + \frac{\rho_{0}}{2}) ((\frac{3 c_{s} \sin{(t )} \sin{(2 \pi x_0 )} \sin{(2 \pi x_1 )}}{10} + 3 c_{s})^{2} \\
&\quad + (\frac{3 c_{s} \sin{(t )} \sin{(2 \pi x_1 )} \cos{(2 \pi x_0 )}}{10} + 3 c_{s})^{2}) \\
Q_{4}^{\text{exact}} &= - \frac{3 c_{s} \sin{(t )} \sin{(2 \pi x_1 )} \cos{(2 \pi x_0 )}}{10} + \frac{c_{s}}{2} \\
Q_{5}^{\text{exact}} &= \frac{3 c_{s} \sin{(t )} \sin{(2 \pi x_0 )} \cos{(2 \pi x_1 )}}{10} + \frac{c_{s}}{2}
\end{align*}
}

\subsection{3D Homogenized Compressible Navier--Stokes--Fourier Equations MMS}\label{sec:appendix_mms_hcfnse}
\paragraph{Manufactured solution (physical form)}
The manufactured solution is fully three-dimensional with active compressibility ($\nabla \rho \neq 0$), evolved as fluctuations of amplitude $A = 0.2$ about a subsonic mean state at per-axis Mach number $\text{Ma} = 0.5$ (air at standard conditions; reference density $\rho_0$, isothermal sound-speed scale $c_s = \sqrt{p_0/\rho_0}$, reference pressure $p_0 = \rho_0 c_s^2$, mean (adiabatic) sound speed $\bar{c} = \sqrt{\gamma}\,c_s$, drift $U_0 = \text{Ma}\,\bar{c}$). The fluid primitives ($k = 2\pi$) are
\begin{align}
\begin{aligned}
\rho &= \rho_0 + A\,\sin(k x_0)\cos(k x_1)\sin(k x_2)\cos t, \\
u_0 &= U_0 + A\,\cos(k x_0)\sin(k x_1)\sin(k x_2)\sin t, \\
u_1 &= U_0 + A\,\sin(k x_0)\cos(k x_1)\sin(k x_2)\sin t, \\
u_2 &= U_0 + A\,\sin(k x_0)\sin(k x_1)\cos(k x_2)\sin t, \\
p   &= p_0 + \bar{c}^2(\rho - \rho_0) + A\,p_0\,\cos(k x_0)\sin(k x_1)\cos(k x_2)\sin t,
\end{aligned}
\end{align}
so the pressure combines the linear response $\bar{c}^2(\rho-\rho_0)$ with an independent mode, and the conserved state is assembled algebraically as $\mathbf{Q} = (\rho,\ \rho\mathbf{u},\ E)$ with $E = p/(\gamma-1) + \tfrac{1}{2}\rho|\mathbf{u}|^2$. The prescribed solid-velocity input field superimposes a rotation (amplitude $0.2$), a uniform translation $\mathbf{c} = (0.1,\,-0.1,\,0.05)$, and an independent oscillation (amplitude $0.1$),
\begin{align}
\begin{aligned}
u_{s,0} &= 0.2\,\sin(k x_0)\cos(k x_1)\cos(k x_2)\cos t + c_0 + 0.1\,\sin t\,\cos(k x_2), \\
u_{s,1} &= -0.2\,\cos(k x_0)\sin(k x_1)\cos(k x_2)\cos t + c_1 + 0.1\,\cos t\,\cos(k x_0), \\
u_{s,2} &= 0.2\,\cos(k x_0)\cos(k x_1)\sin(k x_2)\sin t + c_2 + 0.1\,\sin t\,\cos(k x_1),
\end{aligned}
\end{align}
and the porosity is a smooth, fully three-dimensional and non-separable sinusoid advected along the solid displacement $\boldsymbol{\xi} = \mathbf{x} - \mathbf{d}(\mathbf{x},t)$,
\begin{align}
\alpha &= \frac{1}{2} + \frac{1}{10}\Big[\sin(k\xi_0) + \sin(k\xi_1) + \sin(k\xi_2) \nonumber \\
       &\qquad\quad + 2\sin(k\xi_0)\sin(k\xi_1)\sin(k\xi_2)\Big] \in [0,\,1],
\end{align}
with $d_i = (\text{rotation})\,\sin t + c_i\,t$ a smooth solid displacement, so $\alpha$ spans the full solid-to-fluid range and is transported.

\begin{table}[H]
\centering\footnotesize
\caption{Symbol values for 3D Homogenized Compressible Navier--Stokes--Fourier Equations MMS.}\label{tab:mmsvals_hcfnse}
\begin{tabular}{@{}lll@{}}
\toprule
Symbol & Value & Quantity \\
\midrule
$\rho_{0}$ & \qty{1.225}{\kilogram\per\cubic\metre} & reference density \\
$c_{s}$ & \qty{287.4}{\metre\per\second} & sound-speed scale \\
$\gamma$ & \num{1.4} & adiabatic index \\
$\mu$ & \qty{3.52}{\pascal\second} & dynamic viscosity \\
$\kappa$ & \qty{3.52}{\pascal\second} & heat-flux coefficient ($T = p/\rho$) \\
$\eta$ & \qty{0.000348}{\second} & drag relaxation time \\
\bottomrule
\end{tabular}
\end{table}

\paragraph{Compiler-expanded fields ($Q_i^{\text{exact}}$)}
{\footnotesize
\begin{align*}
Q_{0}^{\text{exact}} &= \rho_{0} (\frac{\sin{(2 \pi x_0 )} \sin{(2 \pi x_2 )} \cos{(t )} \cos{(2 \pi x_1 )}}{5} + 1) \\
Q_{1}^{\text{exact}} &= c_{s} \rho_{0} (\frac{\sin{(t )} \sin{(2 \pi x_1 )} \sin{(2 \pi x_2 )} \cos{(2 \pi x_0 )}}{5} \\
&\quad + 0.591607978309962) (\frac{\sin{(2 \pi x_0 )} \sin{(2 \pi x_2 )} \cos{(t )} \cos{(2 \pi x_1 )}}{5} + 1) \\
Q_{2}^{\text{exact}} &= c_{s} \rho_{0} (\frac{\sin{(t )} \sin{(2 \pi x_0 )} \sin{(2 \pi x_2 )} \cos{(2 \pi x_1 )}}{5} \\
&\quad + 0.591607978309962) (\frac{\sin{(2 \pi x_0 )} \sin{(2 \pi x_2 )} \cos{(t )} \cos{(2 \pi x_1 )}}{5} + 1) \\
Q_{3}^{\text{exact}} &= c_{s} \rho_{0} (\frac{\sin{(t )} \sin{(2 \pi x_0 )} \sin{(2 \pi x_1 )} \cos{(2 \pi x_2 )}}{5} \\
&\quad + 0.591607978309962) (\frac{\sin{(2 \pi x_0 )} \sin{(2 \pi x_2 )} \cos{(t )} \cos{(2 \pi x_1 )}}{5} + 1)
\end{align*}
}
{\footnotesize\noindent\adjustbox{max width=0.99\linewidth}{$\displaystyle\begin{aligned}Q_{4}^{\text{exact}} &= \frac{c_{s}^{2} \rho_{0} \sin^{2}{(t )} \sin^{3}{(2 \pi x_0 )} \sin^{2}{(2 \pi x_1 )} \sin{(2 \pi x_2 )} \cos{(t )} \cos{(2 \pi x_1 )} \cos^{2}{(2 \pi x_2 )}}{250} \\
&\quad + \frac{c_{s}^{2} \rho_{0} \sin^{2}{(t )} \sin^{3}{(2 \pi x_0 )} \sin^{3}{(2 \pi x_2 )} \cos{(t )} \cos^{3}{(2 \pi x_1 )}}{250} \\
&\quad + \frac{c_{s}^{2} \rho_{0} \sin^{2}{(t )} \sin^{2}{(2 \pi x_0 )} \sin^{2}{(2 \pi x_1 )} \cos^{2}{(2 \pi x_2 )}}{50} \\
&\quad + \frac{c_{s}^{2} \rho_{0} \sin^{2}{(t )} \sin^{2}{(2 \pi x_0 )} \sin^{2}{(2 \pi x_2 )} \cos^{2}{(2 \pi x_1 )}}{50} \\
&\quad + \frac{c_{s}^{2} \rho_{0} \sin^{2}{(t )} \sin{(2 \pi x_0 )} \sin^{2}{(2 \pi x_1 )} \sin^{3}{(2 \pi x_2 )} \cos{(t )} \cos^{2}{(2 \pi x_0 )} \cos{(2 \pi x_1 )}}{250} \\
&\quad + \frac{c_{s}^{2} \rho_{0} \sin^{2}{(t )} \sin^{2}{(2 \pi x_1 )} \sin^{2}{(2 \pi x_2 )} \cos^{2}{(2 \pi x_0 )}}{50} \\
&\quad + 0.0236643191323985 c_{s}^{2} \rho_{0} \sin{(t )} \sin^{2}{(2 \pi x_0 )} \sin{(2 \pi x_1 )} \sin{(2 \pi x_2 )} \cos{(t )} \cos{(2 \pi x_1 )} \cos{(2 \pi x_2 )} \\
&\quad + 0.0236643191323985 c_{s}^{2} \rho_{0} \sin{(t )} \sin^{2}{(2 \pi x_0 )} \sin^{2}{(2 \pi x_2 )} \cos{(t )} \cos^{2}{(2 \pi x_1 )} \\
&\quad + 0.0236643191323985 c_{s}^{2} \rho_{0} \sin{(t )} \sin{(2 \pi x_0 )} \sin{(2 \pi x_1 )} \sin^{2}{(2 \pi x_2 )} \cos{(t )} \cos{(2 \pi x_0 )} \cos{(2 \pi x_1 )} \\
&\quad + 0.118321595661992 c_{s}^{2} \rho_{0} \sin{(t )} \sin{(2 \pi x_0 )} \sin{(2 \pi x_1 )} \cos{(2 \pi x_2 )} \\
&\quad + 0.118321595661992 c_{s}^{2} \rho_{0} \sin{(t )} \sin{(2 \pi x_0 )} \sin{(2 \pi x_2 )} \cos{(2 \pi x_1 )} \\
&\quad + 0.118321595661992 c_{s}^{2} \rho_{0} \sin{(t )} \sin{(2 \pi x_1 )} \sin{(2 \pi x_2 )} \cos{(2 \pi x_0 )} \\
&\quad + \frac{c_{s}^{2} \rho_{0} \sin{(t )} \sin{(2 \pi x_1 )} \cos{(2 \pi x_0 )} \cos{(2 \pi x_2 )}}{2} \\
&\quad + \frac{161 c_{s}^{2} \rho_{0} \sin{(2 \pi x_0 )} \sin{(2 \pi x_2 )} \cos{(t )} \cos{(2 \pi x_1 )}}{200} \\
&\quad + \frac{121 c_{s}^{2} \rho_{0}}{40}\end{aligned}$}}\par\addvspace{2pt}

\end{document}